\documentclass[preprint,english,preprintnumbers,amsmath,amssymb,floatfix]{revtex4}
\usepackage[T1]{fontenc}
\usepackage[latin1]{inputenc}
\usepackage{booktabs}
\usepackage{color}
\usepackage{array}
\usepackage{amstext}
\usepackage{graphicx}
\usepackage{esint}
\usepackage{rotating}

\makeatletter

\@ifundefined{textcolor}{}
{%
 \definecolor{BLACK}{gray}{0}
 \definecolor{WHITE}{gray}{1}
 \definecolor{RED}{rgb}{1,0,0}
 \definecolor{GREEN}{rgb}{0,1,0}
 \definecolor{BLUE}{rgb}{0,0,1}
 \definecolor{CYAN}{cmyk}{1,0,0,0}
 \definecolor{MAGENTA}{cmyk}{0,1,0,0}
 \definecolor{YELLOW}{cmyk}{0,0,1,0}
 }

\@ifundefined{definecolor}
 {\usepackage{color}}{}
\@ifundefined{definecolor}
 {\usepackage{color}}{}
\makeatother

\makeatother

\usepackage{babel}

\begin{document}

\title{QCD analysis of structure functions in deep inelastic neutrino-nucleon scattering without using the orthogonal polynomials approach}

\author{A. Ghaffari Tooran}   
\email[]{A.Ghafary@semnan.ac.ir}
   \author{A. Khorramian} 
   \email[]{Khorramiana@semnan.ac.ir}
\author{H. Abdolmaleki}  
   \email[]{Abdolmaleki@semnan.ac.ir}
\affiliation{Faculty of Physics, Semnan University, P. O. Box 35131-19111, Semnan, Iran}

\date{\today}

\begin{abstract}
A nonsinglet QCD analysis of neutrino-nucleon structure function is performed based on all the data for charged current neutrino-nucleon deep inelastic scattering (DIS) corresponds to NLO and NNLO approximations, with taking into account the nuclear and higher twist corrections. In this analysis, we extract  $xu_v(x,Q^2)$ and $xd_v(x,Q^2)$ valence parton distribution functions (PDFs) in a wide range of $x$ and $Q^2$, and determine their parametrization with the correlated errors using the xFitter framework. Our results regarding valence-quark densities with their uncertainties are compared to the prediction extracted using other PDF sets from different groups. We determine $\alpha_{s}(M_Z^2)$= 0.1199 $\pm$ 0.0031  and 0.1185 $\pm$ 0.0023 with considering the nuclear and higher twist corrections at the NLO and NNLO, respectively,  and perform a comparison with other reported results. The extracted results regarding valence-quark distributions and the value of $\alpha_{s}(M_Z^2)$ are in good agreement with available theoretical models.

\end{abstract}


\maketitle

\section{Introduction}

An extremely extensive range of deep inelastic lepton-nucleon (nucleus) scattering, structure functions, and cross sections data are successfully explained in terms of universal parton densities which satisfy in the Dokshitzer-Gribov-Lipatov-Altarelli-Parisi (DGLAP) evolution equations \cite{DGLAP}.
 The electron/neutrino-nucleon deep inelastic scattering (DIS) data have permitted a detailed information of parton densities at small and large values of parton momentum fraction $x$ of the nucleon.  Quantum chromodynamics (QCD) global fits to 
 these data can be used to constrain the
parton distribution functions (PDFs). DIS has been successful at investigating features of QCD such as unpolarized and polarized PDFs within a hadron~\cite{Alekhin:2012ig,Gao:2013xoa,CooperSarkar:2011aa,Martin:2009iq,Ball:2012wy,Jimenez-Delgado:2014xza,Jimenez-Delgado:2013boa,Ball:2013lla,Leader:2010rb,Blumlein:2010rn,Sato:2016tuz, Arbabifar:2013tma,Khanpour:2017cha,Monfared:2014nta,Khorramian:2010qa,AtashbarTehrani:2007odq,Khorramian:2004ih,Soleymaninia:2013cxa}. Precise information
about the structure of the proton plays an important role
in understanding interactions observed at high-energy proton
colliders. Without precise extracted PDFs from the global fits,  each calculation in QCD may be limited by uncertainties in the high energy physics.

The nonsinglet structure function
$xF_3(x,Q^2)$, where mainly information comes from deep inelastic neutrino-nucleon scattering, is the important input to the QCD global analysis of parton distribution
function, especially at large-$x$, where valence quark distributions are dominant.  The neutrino structure function
$xF_3(x,Q^2)$ experimental data are the first experimental source to extract the valence quark densities $xu_v(x,Q^2)$ and $xd_v(x,Q^2)$ of the nucleon in charged current (CC) neutrino nucleon deep inelastic scattering. Because of the absence of gluonic efficacy in the nonsinglet QCD evolution equation, this nonsinglet structure function gives us an opportunity for clear measurements of the strong coupling constant $\alpha_s$. Having the unique feature to discriminate valence quarks from other partons, is a special characteristic of neutrino probes.  Therefore, neutrino DIS measurements are important to determine valence quark distributions in the nucleon due to the parity violating neutrino $xF_3(x,Q^2)$ structure function probes the valence quark densities directly. Essentially, the treatment of $xF_3$ structure function of deep inelastic neutrino-nucleon scattering is similar to nonsinglet part of $F_2$ in deep inelastic electron-nucleon scattering.

The $xF_3$ structure functions of deep inelastic neutrino-nucleon scattering have been measured by different experimental groups, such as the Chicago-Columbia-Fermilab-Rochester collaboration (CCFR) \cite{ccfr:1977}, Neutrinos at the Tevatron (NuTeV) \cite{Tzanov:2005kr}, CERN Hybrid Oscillation Research ApparatUS (CHORUS) collaboration at CERN \cite{Onengut:2005kv}, and CERN-Dortmund-Heidelberg-
Saclay-Warsaw collaboration (CDHSW) \cite{Berge:1989hr}. These experimental data have prepared an accurate experimental origin for the valence quark densities and strong coupling constant determination. The present neutral-current deep inelastic neutrino-nucleon scattering data which assigned for the $xF_3(x,Q^2)$ structure function has not yet reached the level of precision of 1$ - $2 \%. However, more precise charged-current neutrino-nucleon DIS data will be available at the neutrino factories planned \cite{Bonesini:2016nrv,
 Kaplan:2014xda,Banerjee:2015gca,Geer:2009zz}. It is expected that
new nonsinglet experimental data at future facilities, such as the Large Hadron Electron Collider  (LHeC) \cite{AbelleiraFernandez:2012cc,AbelleiraFernandez:2012ni}  and Electron Ion Collider (EIC) \cite{Aschenauer:2016our,Deshpande:2016goi,Accardi:2012qut},  will  improve further the knowledge of the nonsinglet distribution functions and strong coupling constant. 
 
 An accurate information of parton densities at large $x$ is worth to obtain the precise aims of the CERN Large Hadron Collider (LHC), Tevatron, and other high energy accelerators \cite{Lees:2015ymt}. For example, a precise measurement of nucleon structure function at CEBAF accelerator at Jefferson Laboratory has allowed an investigation of partonic landscape mapping at large $x$ region. While much of the works have been done to use the highest energy colliders data at the small value of $x$ to extract the quark and gluon structure of the nucleon in the form of PDF structures information, less effort has been directed and focused on the region of large $x$. However, nonperturbative QCD effects play an important role in the large momentum fractions. As an example, an effort has been performed in the CTEQ-Jefferson Lab collaboration  \cite{CTEQ:CJ} which reported a different series of QCD fit analysis of PDFs with considering perturbative QCD treatment at the high values of
 $x$ \cite{Accardi:2009br,Accardi:2011fa,Owens:2012bv,Accardi:2016qay}.

Previous nonsinglet analyses for CCFR data were performed  based on orthogonal polynomials expansions methods (as approximation methods),  such as  Jacobi polynomials \cite{Kataev:1994rj,Kataev:1996vu,Kataev:1997nc,Kataev:1997vv,Alekhin:1998df,Alekhin:1999af,Kataev:1999bp,Kataev:2001kk,Kataev:2002wr,Sidorov:2013aza,Khorramian:2009xz,AtashbarTehrani:2009zz,Khorramian:2008yh}, Bernstein polynomials \cite{Khorramian:2006wg,Santiago:2001mh}, and Laguerre polynomials  \cite{GhasempourNesheli:2015tva}. The first results of our nonsinglet $F_2$ and $xF_3$ analysis based on Jacobi polynomials and Bernstein polynomials approach, were reported in Refs.~\cite{Khorramian:2006wg,Khorramian:2008yh,Khorramian:2009xz}. Recent NLO and NNLO QCD analysis of nonsinglet $xF_3$ structure function based on the Laplace transform approach
 is reported in Ref. \cite{MoosaviNejad:2016ebo}  without taking into account the nuclear and higher twist corrections. In  Ref. \cite{MoosaviNejad:2016ebo}, the reported results based on  the Laplace transform approach have been performed without including CDHSW data and also without applying the $W^2$ cuts to the data. We will discuss later in Sec.~IV the impact of cuts on the data to obtain the significant improvement in $\chi^2$ per degree of freedom.

In the present paper, the valence quark distribution functions were determined using the available neutrino structure function $xF_3(x,Q^2)$ world data at NLO and NNLO with taking into account the nuclear and higher twist corrections. This QCD analysis is performed to extract the less number of parameters explaining the valence distributions and it is independent of sea-quarks and gluon distributions. The advantage of neutrino structure function
world data is to deal with a restricted set of valence parton densities, and therefore this analysis is free of the correlation between strong coupling constant $\alpha_s$ and the sea-quarks and gluon distributions. The nonsinglet QCD analysis will provide us the $xu_v(x,Q^2)$ and  $xd_v(x,Q^2)$
distribution functions, $\alpha_s$  and their corresponding errors as well.

In our previous work, we used only the deep inelastic neutrino scattering CCFR data to determine valence quark densities at the NNLO level of accuracy using  Mellin-moment space and without taking into consideration the nuclear and higher twist corrections. In this work, to have precise valence PDFs, the number of experimental data increase in comparison with our previous QCD analysis based on the orthogonal polynomials approach \cite{Khorramian:2006wg}. As an important modification, we provide one set of fits corresponding to our parametrization of nonsinglet parton distributions in $x$ space considering the nuclear and higher twist corrections. The valence PDFs, their uncertainties, and also the strong coupling constant central value $\alpha_s(M^2_Z)$ have changed as a consequence of new data and  QCD calculation in $x$ space without applying any orthogonal polynomial approximation methods in $n$ space.  We can also compare the QCD nonsinglet results without and with using the orthogonal polynomial methods. Very recently, we have reported a new QCD  analysis on polarized PDFs without using the orthogonal polynomial methods \cite{Salimi-Amiri:2018had}. 

In this article, we perform our nonsinglet QCD  analysis based on xFitter  open source framework \cite{xFitter,Alekhin:2014irh}, which previously was known as HERAfitter \cite{Sapronov}. 
In this regard, we need to add the neutrino-nucleon experimental data and other necessary modifications, such as nuclear and higher twist effects, which are not included in the main xFitter package. In Refs.~\cite{Vafaee:2017nze,Abdolmaleki:2017wlg,Vafaee:2017jnt,Rostami:2015iva,Azizi:2018iiq,Aleedaneshvar:2017bgs}, we have used the xFitter package for different QCD analyses. Very recently, we presented also a new set of PDFs considering the intrinsic charm content of the
proton using this package \cite{Hamed:2018new}.

The plan of the paper is to give a brief review of basic formalism for neutrino structure function in deep inelastic scattering in Sec. II. 
In this section, we introduce the nuclear and higher twist corrections for neutrino-nucleon structure functions.
In Sec. III, we present the theoretical and experimental inputs of the fit, the parametrization for valence quark densities, and experimental data sets which we apply in the present QCD analysis. In Sec. IV, the fit results for the
valence distribution functions, their evolution, corresponding errors, and our results on $\alpha_{s}(M_Z^2)$ at the NLO and NNLO  are given and compared with other theoretical results. Our discussion and conclusion are given in Sec. V.

\section{Neutrino-nucleon cross sections and  parton distributions}
\subsection{Basic formalism}
The charged-current  (CC)  deep inelastic neutrino (antineutrino)-nucleon scattering differential cross sections are given  by a combination of three structure functions $F_1$, $F_2$, and $F_3$  as \cite{Eisele:1986uz,Diemoz:1986kt}
\begin{eqnarray}
\frac{d^2\sigma ^{\nu ,\bar \nu }}{dxdy}&=& \frac{{G_F^2 M_N E}}{\pi}\left(\frac{M^2_W}{M^2_W+Q^2}\right)^2 \left[F_1^{\nu ,\bar \nu } xy^2 \right.\nonumber \\ 
&+&  F_2^{\nu ,\bar \nu }(1 - y - \frac{M_N xy}{2E}) \pm xF_3^{\nu ,\bar \nu } (y - \frac{y^2}{2})]\;,
\end{eqnarray}
where  $Q^2$ is negative four-momentum transfer squared, and 
$x$ is the Bjorken scaling variable. Here, $y$ is inelasticity which is defined by $y=Q^2/(sx)$, $E$ is the neutrino-beam energy,
and $M_N$ is the nucleon mass. In the above,  $\pm$ indicates $+$ for $\nu$ and $-$ for $\bar \nu$, $G_F=1.16638\times 10^{-5} $~GeV$^{-2}$ is the Fermi constant,
and $M_W$=80.385 GeV is the $W$ boson mass \cite{Kramer:2000hn}.

According to quark-parton model (QPM), the above CC neutrino (antineutrino)-nucleon differential cross section at the leading order of the running
coupling constant $\alpha_s$,   is related to the structure functions  in terms of PDFs. By considering $F_2   = 2xF_1$,  one can have 
the above $F_2^{\nu p}$ and  $F_2^{\bar \nu p}$ structure functions in terms of PDFs, $F_2^{\nu p}   = 2x(d + s + \bar u + \bar c)$,  and $F_2^{\bar \nu p} =2x(u + c + \bar d + \bar s)$.  By changing the signs of $\bar u$, $\bar d$, $\bar s$, and $\bar c$, the structure functions of $xF_3^{\nu p}$ and $xF_3^{\bar \nu p}$  can be written as 
\begin{eqnarray}
xF_3^{\nu p} & =& 2x(d + s - \bar u - \bar c)~,   \nonumber   \\ 
xF_3^{\bar \nu p} & =& 2x(u + c - \bar d - \bar s)~.   
\end{eqnarray}
By considering $u\equiv u_v+\bar u$ and $d\equiv d_v+\bar d$ and combining the above equations,  the structure function $xF_3$ is as follows:
\begin{equation}
xF_3^{{({\nu}+{\bar \nu})}p}=xF_3^{\nu p}  + xF_3^{\bar \nu p}  
= 2x(u_v  + d_v ) + 2x(s - \bar s) + 2x(c - \bar c)~.
\end{equation}
So, one can have the average of the neutrino and antineutrino nucleon structure function as follows:
\begin{eqnarray}
xF_3^{N}(x,Q^2)&=&\frac{1}{2}\left(xF_3^{\nu N}+ xF_3^{\bar \nu N} \right) (x,Q^2)\nonumber \\
&=&\frac{1}{2}\left([xF_3^{(\nu+\bar \nu) p}+xF_3^{(\nu+\bar \nu) n}]/2 \right)(x,Q^2)~.
\end{eqnarray}
However, due to the isospin symmetry, $xF_3^{{({\nu}+{\bar \nu})}p}=xF_3^{{({\nu}+{\bar \nu})}n}$,  the  average of the neutrino and antineutrino nucleon structure is  
\begin{eqnarray}
xF_3^{N}(x,Q^2)&=&\frac{1}{2}~xF_3^{(\nu+\bar \nu) p}(x,Q^2)\nonumber \\
&&= [x(u_v  + d_v ) + x(s - \bar s) + x(c - \bar c)](x,Q^2)~.
\end{eqnarray}

It should be noted that $s-\bar s$ and $c-\bar c$ are considered to be very small. 
Therefore, the average of the neutrino and antineutrino nucleon structure is only related to valence quark distribution as
\begin{eqnarray}
xF_3^{N}(x,Q^2)&= (xu_v + xd_v) (x,Q^2)~.
\end{eqnarray}

According to QPM, the above equation explicitly demonstrates that the $xF_3$ structure function is related to the valence quark distributions. Therefore, DIS neutrino-nucleon scattering $xF_3$ measurements are needed to  determine
the valence-quark densities in the nucleon. 
Also, according to isoscalar correction of the $xF_3$ structure function, these data are only sensitive to the sum of $xu_v + xd_v$. In the next sections, we will discuss how we are able to have a reliable separation of the two contributions. 

\subsection{Nuclear neutrino structure function}

Since the detection of neutrinos always involves the heavy nuclear targets, so the nuclear effect is needed to study the DIS neutrino (antineutrino)-nucleus $xF_3$ structure function. The nuclear targets are used by different neutrino experiments, such as CCFR, NuTeV, and CDHSW with the same iron target, and  CHORUS with a lead target. To have the average of the neutrino and antineutrino nucleus structure functions, we require to have the nuclear PDFs.

We need to discuss which combination of PDFs is related to neutrino-nucleus structure function, for simplicity at leading order. 
If we assume $s^A=\bar s^A$ and $c^A=\bar c^A$, the neutrino and antineutrino-nucleus structure functions $xF_3^{\nu A}$ and $xF_3^{\bar \nu A}$ are given by
\begin{eqnarray}
xF_3^{\nu A} & =& 2x(d^A + s^A - \bar u^A - \bar c^A)~,   \nonumber   \\ 
xF_3^{\bar \nu A} & =& 2x(u^A + c^A - \bar d^A - \bar s^A)~.  
\end{eqnarray}
Here,  $xF_3^{\bar \nu A}$ structure functions for antineutrino  nucleus scattering are obtained by exchanging the quark and antiquark PDFs
in the corresponding  $xF_3^{\nu A}$ neutrino nucleus structure functions, i.~e., $xF_3^{\bar \nu A}=-xF_3^{\nu A}[q\leftrightarrow \bar q] $. Therefore, the average of the neutrino and antineutrino nucleus structure function is as
\begin{eqnarray}
xF_3^{A}(x,Q^2)&=&\frac{1}{2}\left(xF_3^{\nu A}(x,Q^2)+ xF_3^{\bar \nu A}(x,Q^2) \right)\nonumber \\
&=& xu_v^A(x,Q^2)  + xd_v^A(x,Q^2) ~.
\end{eqnarray}

As we mentioned before, DIS neutrino $xF_3$  experiments have used the iron or lead targets, so performing the nuclear corrections \cite{deFlorian:2011fp} in the present analysis would be necessary. In Ref.~\cite{Schienbein:2007fs}, the nuclear effects in charged current DIS neutrino-iron data are studied to determine the iron PDFs.

To include the nuclear effects for neutrino DIS structure functions, we need to have the nuclear parton distribution functions (nPDFs).
The nPDFs are introduced by a number of parameters which appeared in nuclear modification and also by simple summation of free proton and neutron contributions.  For example in Ref.~\cite{Kumano:2016jsb}, the parameters in the nuclear modification are determined by $\chi^2$  analysis of world data on nuclear structure function ratios. 

Basically, the valence nPDFs for a nucleus  can be expressed as 
\begin{equation}
xq_{v}^A(x,Q_0^2)=\frac{Z}{A}\;xq_{v}^{p/A}(x,Q_0^2) +\frac{A-Z}{A}\;xq_{v}^{n/A}(x,Q_0^2)\;,
\label{eq:pdfsnucleus}
\end{equation}
where  $A$ and $Z$ are mass number and atomic number, respectively,  and $p$ and $n$ indicate proton and neutron. In the above, $xq_{v}^{p/A}$  and $xq_{v}^{n/A}$ denote valence PDFs of bound
protons and neutrons in the nucleus $A$. By assuming isospin symmetry, the valence distributions inside a bound neutron, $xq_{v}^{n/A}$, are related 
to the ones in a bound proton, $xq_{v}^{p/A}$. If there are no nuclear modification, the valence nPDFs, $xq_{v}^A$, are expressed by a simple summation of free proton and neutron contributions.

The nuclear effects in hadron production may arise from the nuclear modifications of PDFs. For nonsinglet QCD analysis, this modification create a connection between the bounded valence PDFs in the nucleus $A$ and free valence PDFs  in the proton as  
\begin{equation}
xq_{v}^{p/A}(x,Q_0^2)= R_{{v}}(x,A,Z)\;xq_v(x,Q_0^2)~,
\label{eq:nuclear}
\end{equation} 
with $q_v=u_v, d_v$. In the above, $R_{{v}}(x,A,Z)$  is the nuclear modification which is dependent on the nucleus, and   $xq_v(x,Q_0^2)$ is  the valence PDFs in the free proton. For  $R_{{v}}(x,A,Z)$, we can use the different available parametrizations, such as DSSZ parametrization~\cite{deFlorian:2011fp}. As we mentioned before, in the absence of the nuclear modification $xq_{v}^{{p}/A}=xq_{v}$, it corresponds to $R_{{v}}(x,A,Z)$=1 in Eq.~(\ref{eq:nuclear}). The input scale of $Q_0^2$ is a fixed $Q^2$ value in valence PDFs parametrization, and the $Q^2$ QCD evolution of PDFs for $Q^2>Q_0^2$ can be obtained by the DGLAP evolution equations.

So,  by having the nuclear modification, the valence nPDFs are expressed by a number of the unknown parameters which appeared in \textit{$xq_v(x,Q_0^2)$}. In the present analysis, these parameters can be determined by QCD fits of the neutrino DIS structure function data.

\subsection{Higher twist effects}
In this subsection, we discuss the role of higher twist (HT) effects in the present QCD analysis on the neutrino-nucleus measurements. 

In the standard analysis of DIS neutrino-nucleus $xF_3$ data and to widely eliminate the nonperturbative effects, it is necessary to apply the appropriate
cuts for the invariant-mass squared $W^2=Q^2(1/x-1)+m_N^2$  and the virtual photon $Q^2$ at NLO and NNLO. 
In fact, choosing the appropriate $W^2$ cut value on the neutrino-nucleon data is required to ignore the nonperturbative effects. 

In this article, we use the neutrino-nucleus data, especially in the region of deep inelastic DIS for determination of valence PDFs and the strong coupling constant in the scale of $M_Z^2$. In neutrino-nucleus scattering and in the DIS region, we have to choose  $W^2$ and  $Q^2$ high enough. To determine PDFs using the DGLAP evolution equations, the running coupling constant should be small enough. In this case, it requires that
$Q^2$ should be large, typically a few GeV$^2$. Although, in the region of $W^2\geq 4$ GeV$^2$, the nucleon is broken and it is the deep-inelastic-scattering region, but if we want to eliminate HT  effects from the data, we should choose the standard $W^2\geq$12.5 GeV$^2$ cut on the data. 
In this regard, it seems that the study of valence PDFs and the strong coupling constant value in the scale of $M_Z^2$, with and without taking into account HT effects would be worthy.

As the first step, we used the standard cuts in $Q^2$ and invariant-mass squared $W^2$, $Q^2\geq4$ GeV$^2$,  and $W^2\geq$12.5 GeV$^2$ to eliminate HT effects from the data. In this case,  we can extract the unknown parameters using QCD fits on the data.  

To find the impact of the HT contribution,  we used all data in the $Q^2\geq4$ GeV$^2$ region without any cut on $W^2$, where the experimental data are located in the DIS region in our QCD fits, as the second step. In Refs. \cite{Kataev:1997vv,Sidorov:1996wb,Sidorov:1996if,Virchaux:1991jc}, a lot of efforts have been made in this regard.

To include the HT contribution, the average of the neutrino and antineutrino  structure function may be explained as
\begin{equation}
         xF_3(x,Q^2) = xF_3^{QCD}(x,Q^2)+\frac{h(x)}{Q^2} ~.
 \label{eq:HT}
\end{equation}
Here, the $Q^2$ dependence of the first term is obtained by perturbative QCD and the HT correction term is \cite{Tokarev:1997vc}
\begin{equation}
h(x)=\sum_{k=0}^3 {\cal D}_k z^k,\;\;\;\;  z=log(x) ~.
 \label{eq:HTform}
\end{equation}
The unknown parameters of ${\cal D}_k$  and their uncertainties for the function $h(x)$  can be extracted simultaneously with other unknown parameters which appeared in the valence PDFs and the strong coupling constant by fitting the experimental data. Note that, in the main xFitter package, we need to add the nuclear and higher twist effects modifications, which are not generally included in this package.

\section{Theoretical and experimental inputs of the fit}
In this section, we  introduce the $xu_v$ and  $xd_v$ parametrizations at the input scale of $Q_0^2$. 
Also, we will consider 
$\alpha_{\rm s}(M_Z^2)$, as another fitting parameter, using the nonsinglet QCD analysis of neutrino-nucleon scattering data. A detailed discussion of various combinations of data sets will be presented for neutrino DIS data obtained by CCFR, NuTeV, CHORUS, and CDHSW experiments, which can be used for determination of $xu_v$ and  $xd_v$  distributions and $\alpha_{\rm s}(M_Z^2)$ as well.

\subsection{Nonsinglet parametrization}

In this analysis, we choose the following valence quark densities according to our previous nonsinglet QCD analysis \cite{Khorramian:2006wg} at the input scale of $Q_0^2$:
 \begin{equation}
     xu_{v}(x,Q_0^2)={N_{u}}x^{a_{u}}(1-x)^{b_{u}}(1+c_{u}x+d_{u}\sqrt{x})\;,\\
     \label{eq:parm1}
     \end{equation}
      \begin{equation}
     xd_{v}(x,Q_0^2)=\frac {N_{d}}{N_{u}}(1-x)^ {b_{d}}  xu_{v}(x,Q_0^2)\;.\\
     \label{eq:parm2}
     \end{equation}
       
In this parametrization, $xd_{v}(x,Q_0^2)$ distribution depends on $xu_{v}(x,Q_0^2)$ \cite{Diemoz:1987xu,Gluck:1989ze,Gluck:1998xa}, and such the above parametrization does not exist in the xFitter framework. In fact, to separate the $xu_{v}(x,Q_0^2)$ and $xd_{v}(x,Q_0^2)$ in deep inelastic neutrino nucleon scattering, we assume a valid above relation, which we used in our previous QCD analysis \cite{Khorramian:2006wg}. The terms of $x^{a_{i}}$ and  $(1-x)^{b_{i}}$  control the 
   low and large $x$ region, respectively, and other polynomial terms are important for additional medium-$x$ values.  
 Thus, in the limit as $x\rightarrow1$, the ratio of $xd_{v}(x,Q_0^2)/xu_{v}(x,Q_0^2)$ goes to zero if $b_d>0$, and infinity if $b_d<0$.
   As we will see, the data at intermediate values of $x$ require $b_d>0$.

   Also the normalization constants $N_{u}$ and $N_{d}$ 
  can be obtained from the other parameters, using conservation
of the fermion number by
\begin{equation}
\int_0^1u_{v}dx=2\;,\nonumber \\
\int_0^1d_{v}dx=1\;.
\end{equation}
So the normalization constants $N_{u}$ and $N_{d}$ are 
\begin{eqnarray}
\label{eq:normuv}
N_{u}&=&2/\left[B(a_{u},1+b_{u})+c_{u}B(1/2+a_{u},1+b_{u})\right. \nonumber \\
&&\left. +d_{u}B(1+a_{u},1+b_{u})\right]\;,
 \\ 
N_{d}&=&1/\left[B(a_{u},1+b_{u}+b_{d})\right. \nonumber \\
&&\left. +c_{u}B(1/2+a_{u},1+b_{u}+b_{d})\right.\nonumber \\
&&\left. +d_{u} B(1+a_{u},1+b_{u}+b_{d})\right]\;,
\end{eqnarray}
where $B(a,b)$ is the Euler $\beta$ function. In above parametrization, the normalization constants $N_{u}$ and $N_{d}$ are very effective to determine unknown parameters via the QCD fitting procedure. 

According to the above parametrization, we have five free valence parameters,  which can be extracted from the QCD fits. In the next section, we will see some parameters should be fixed after the first minimization due to DIS neutrino (antineutrino) $xF_3$ data will not constrain some of the parameters in Eqs.~(\ref{eq:parm1}), (\ref{eq:parm2}) well enough. Since the errors of some parameters turn out to be rather large compared to the central values, we should keep fixed these parameters after the first minimization, as  it has  been done even in nonsinglet QCD analyses of $F_2(x,Q^2)$ \cite{Khorramian:2009xz,Khorramian:2008yh,Blumlein:2006be}.  

Generally,  the coupling constant $\alpha_{\rm s}(M_Z^2)$ can be extracted from the global QCD fit to hadronic processes. In this nonsinglet QCD analysis, the strong coupling constant in the scale of $M_Z^2$ is another QCD free parameter and can be  determined  using  deep inelastic neutrino-nucleon scattering data.  The strong coupling constant as an important parameter displays a remarkable
correlation with the nonsinglet PDFs. Since this parameter is correlated with other nonsinglet quark density uncertainties, the determination of  $\alpha_{\rm s}(M_Z^2)$   uncertainty  would also be important. We can compare this fit parameter to the  world average of ${\alpha_s(M^2_Z)}=0.1181\pm0.0011$  which is reported in Ref.~\cite{Tanabashi:2018oca}.

The xFitter  package employs the useful QCDNUM  evolution program \cite{Botje:2010ay,Bertone:2017kne} to determine the $Q^2$ evolution of PDFs and also the coupling constant. Our previous unpolarized and polarized QCD calculations \cite{Arbabifar:2013tma,Monfared:2014nta,Khorramian:2010qa,AtashbarTehrani:2007odq,Khorramian:2004ih,Khorramian:2009xz,Khorramian:2008yh,Khorramian:2006wg,Salimi-Amiri:2018had} are performed in the $n$ space, based on the QCD-PEGASUS package \cite{Vogt:2004ns}.

\subsection{Experimental data sets}
In this analysis, we include the recent DIS neutrino (antineutrino) $xF_3$ measurements from CCFR [with an iron target and $30\leq E_\nu$(GeV)$ \leq 360$] \cite{ccfr:1977}, NuTeV [with an iron target and $30\leq E_\nu$(GeV)$ \leq 500$] \cite{Tzanov:2005kr}, CHORUS  [with a lead target and $10\leq E_\nu$(GeV)$ \leq 200$] \cite{Onengut:2005kv}, and CDHSW [with an iron target and $20\leq E_\nu$(GeV)$ \leq 212$] \cite{Berge:1989hr}.
These experimental data have prepared an accurate experimental origin for the valence-quark densities and  $\alpha_{s}(M_Z^2)$ determination. We need to add the mentioned data in the xFitter because these experimental data are not included in this package. 

The $x$ range for NuTeV and CHORUS are almost the same considering two different targets in these data sets. At moderate $x$, the CDHSW measurement result agrees well with CCFR.  Also, there are differences in NuTeV and CCFR at $x>0.4$ where CCFR measurements
for neutrino and antineutrino-nucleon cross sections are consistently below the NuTeV energies \cite{Tzanov:2005kr}. The NuTeV data are newer and have improved energy scale calibration. It also uses a better theory treatment of  heavy quarks and updated higher twist corrections.

Also, CCFR and CDHSW measurements cover higher $Q^2$ values. The upper limit of the $x$ value for all different data sets is almost the same (i.e., $ $x$\sim$0.7), where it would be very important for valence quark densities. However, NuTeV and CCFR measurements use the same target in their DIS neutrino (antineutrino)-nucleus processes and have the same kinematic upper range of $x$, NuTeV data seem to be more precise than other measurements.

In the DIS region, the experimental data which are used in the QCD analysis may be expected to be somewhat free of nuclear corrections as nonperturbative effects. Since DIS neutrino-nucleon $xF_3$  experiments have used high $Z$, $A$ targets such as iron or lead, so performing the nuclear correction~\cite{deFlorian:2011fp} in the present analysis should be considered. 
In some of the  QCD analyses of DIS neutrino-nucleon $xF_3$ data,  different groups avoid the nuclear corrections in their calculations \cite{Sidorov:2013aza,Khorramian:2009xz,Khorramian:2008yh,Khorramian:2006wg,
 Santiago:2001mh,GhasempourNesheli:2015tva,MoosaviNejad:2016ebo}. 
  In Ref.~\cite{Tzanov:2005kr}, it is mentioned that neutrino (and also antineutrino)-nucleon scattering favors smaller nuclear effects in high-$x$ region that are found in charged-lepton DIS measurements.  To obtain very precise valence quark distributions and also  $\alpha_{s}(M_Z^2)$ in the nonsinglet DIS neutrino-nucleon $xF_3$ analysis, we need to include the nuclear corrections in the present QCD analysis. 

In order to make a more systematic investigation of the stability of the QCD fit,
 a series of analyses were performed to study an improvement in $\chi^2$ with different $W^2$ and $Q^2$ cut values for neutrino-nucleon $xF_3$ data. To include the DIS neutrino-nucleon data in the present analysis, we found that considering  $Q^2\geq4$ GeV$^2$ and $W^2\geq$12.5 GeV$^2$ on the data are the appropriate choices to have the best fit quality. We believe that without attention to the suitable cut values on the neutrino  $xF_3$ data, not only we will get unreliable $\chi^2$ per degree of freedom, but also we cannot ignore the HT effects. Making these cuts on  $Q^2$  and $W^2$, we are able to do a QCD fit procedure for nonsinglet PDFs by 263 and 287 data points considering  $Q^2\geq4$ GeV$^2$ with and without $W^2\geq$12.5 GeV$^2$ cuts, respectively.

 In Fig.~1, we plot all the experimental data which we used in this analysis, such as  CCFR, NuTeV, CHORUS, and CDHSW in the $x$ and $Q^2$ plan with considering  $W^2$ and $Q^2$ cuts on the data ($Q^2\geq$4 GeV$^2$ and $W^2\geq$12.5 GeV$^2$). As we mentioned before,  due to the existence of the differences in NuTeV and CCFR at $x>0.4$, one can exclude the CCFR data in this region. The data points lying above these lines are only included in the present QCD fits. Note that the kinematic cuts on the data depend on the kind of QCD analysis. For example, in a global analysis of $F_2$ in presence of neutrino-nucleon $xF_3$ data to determine valence, sea, and gluon PDFs, one can choose  $W^2\geq$25 GeV$^2$ on the data,  as reported in Ref.~\cite{Martin:2009iq}.

Different combinations of the subset of DIS neutrino-nucleon $xF_3$ data, with the
corresponding $x$ and $Q^2$ ranges, and the number of individual data points before and after  cuts for each data set are listed in Table ~\ref{tab:first}.

\subsection{$\chi^2$ minimization and treatment of experimental systematic uncertainties}
\label{sec:systematic_errors}

Basically, the general ansatz applied in QCD fits is the parametrization of parton densities at the input scale of $Q^2_0$, using an appropriate functional form, as we chose in Eqs.~(\ref{eq:parm1}), (\ref{eq:parm2}). Several QCD analyses have been done to assess the uncertainties on parton distribution functions obtained from the QCD fits. 

The nonsinglet DGLAP evolution equations are used to obtain the valence parton distribution $xu_v(x, Q^2)$ and $xd_v(x, Q^2)$ at any $Q^2$ from the valence parton distribution at $Q^2_0$.
This allows the theoretical structure functions of the neutrino-nucleon $xF_3$ data to be computed. The parameters that define the valence distributions at the input scale (e.g., $a_u, b_u, c_u, d_u, b_d$) in Eqs.~(\ref{eq:parm1}), (\ref{eq:parm2}) can then be extracted by fitting these theoretical predictions to the  neutrino-nucleon experimental measurements. This is performed by minimizing a $\chi^2$ function as \cite{Pascaud:1995qs,PDFUncertainties}
\begin{equation}
         \chi^2 = \sum_i \frac{ [d_i -  t_i(1-\sum_j \beta_{j}^{i} s_j)]^2 }{\delta^2_{i,unc} t^2_i +\delta^2_{i,stat} d_i t_i } + \sum_j s^2_j ~,
 \label{eq:chi2}
\end{equation}
where  $t_i$ is the corresponding theoretical prediction, $d_i$ the measured value of the $i$$th$ data point, and  $\delta_{i,stat}$, $\delta_{i,unc}$, and $\beta^i_j$ are the relative statistical, uncorrelated systematic, and correlated systematic uncertainties. 
In the above, $j$ labels the sources of correlated systematic uncertainties and, in the Hessian method, $s_j$ are not fixed. 
When the $s_j$ parameters are fixed to zero, the correlated systematic errors are ignored. In fact, the central fit is performed to the data shifted with the best setting for the systematic error sources.

The correlated piece entries in Table~\ref{tab:first} correspond to the second term in the right-hand side of Eq.~(\ref{eq:chi2}). A reduction of the first term of Eq.~(\ref{eq:chi2}) indicates that the fit does not require the predictions to be shifted so far within the tolerance of the correlated systematic uncertainties, while a reduction of the second term reflects a better agreement of the theoretical predictions with the data.

The Hessian uncertainties on the fitted PDF parameters are obtained from $\Delta \chi^2 = T^2$. A tolerance parameter, $T$, is selected, such that the criterion $\Delta \chi^2 = T^2$ ensures that each data set is described within the desired confidence level.
The correlated statistical error on any given quantity $q_v$  is then obtained from standard error propagation \cite{Perez:2012um}:
	\begin{equation}
	(\sigma_{q_v})^2 = \Delta \chi^2 \left( \sum_{\alpha,\beta} \frac{\partial q_v}{\partial p_{\alpha}}C_{\alpha,\beta} \frac{\partial q_v}{\partial p_{\beta}} \right)~. 
	\label{eqn:error_propagation}
	\end{equation}
	
By considering the Hessian matrix as $ H_{\alpha,\beta} = \frac{1}{2} \partial^2 \chi^2 / \partial p_{\alpha} \partial p_{\beta}$, the covariance matrix $ C = H^{-1}$ is the inverse of the Hessian matrix, evaluated at the $\chi^2$ minimum. In order to be able to calculate the fully correlated  1$\sigma$ error bands corresponding  to  68\% confidence level for PDFs, one can choose $T=1$ in the xFitter package.

   \begin{sidewaystable}[h]
   \caption{Different combinations of the subset of $xF_3$ data, with the
corresponding $x$ and $Q^2$ ranges. The fourth and fifth columns, $xF_3$ and $xF_3$(cuts)  contain the number of individual data points before and after  cuts  for each data set with  considering $Q^2\geq$ 4 GeV$^2$, $W^2\geq$12.5 GeV$^2$ cuts  on the data, respectively. Also, the reduction of the number of  CCFR data points only by the additional cuts on this data ($x>0.4$) due to the disagreement between CCFR and NuTeV in this region are given in these columns.  
The sixth column, $xF_3$(HT) contains the number of experimental data points in the  range of  $Q^2\geq$4 GeV$^2$ used to fit the higher twist corrections. The 7th to 10th  columns contains the $\chi^2$ values for each set for different fits, $i.~e.$ pQCD+nuclear correction and pQCD+nuclear correction+HT at NLO and NNLO approximation. The  correlated $\chi^2$  and total  $\chi^2/$d.o.f.  are shown.}
     \begin{tabular}{*{10}{c}} \toprule
		\toprule
		   &&&& \multicolumn{2}{c}{ \ $\#$Data \ } & \multicolumn{2}{c}{ NLO  \ }  & \multicolumn{2}{c}{  NNLO  \ }    \\
		   \midrule        
			{Experiment} &   $x$   & Q$^2$ &  {$xF_3$} & {$xF_3$(cuts)}  &{$xF_3$(HT)}  &   \multicolumn{2}{l}{ pQCD+NC ~~pQCD+NC+HT}  &  \multicolumn{2}{l}{pQCD+NC ~~pQCD+NC+HT}   \\
				 \midrule  
		{CCFR \cite{ccfr:1977}}	 &  { 0.0075-0.75} & { 1.3 -125.9} & 116 &67(87-20) & 67& 50 & 49 & 47 & 49\\
		{NuTeV \cite{Tzanov:2005kr}}	 &   { 0.015 - 0.75} &{ 1.26 - 50.12}  & 75 & 59 &65&   96 & 77 & 86   & 77  \\
		{CHORUS \cite{Onengut:2005kv} }	  &{ 0.02 -0.65} &{ 0.325 - 81.55}  & 67 &41 &48& 52 & 52 & 51  & 53  \\ 
     	{CDHSW \cite{Berge:1989hr}}	  & { 0.015 - 0.65} & { 0.19 -196.3} &143 & 96 & 107& 193 & 158 & 175 & 155\\

		  Correlated $\chi^2$  &       & & & &&  31  & 24 & 25 & 24  \\ \midrule 
		  { Total $\chi^2$}	&       & & & & &422 & 360 &  384 &  358 \\ 
		  { d.o.f.}	&       & & & & &259 & 279 & 259 &  279 \\ 
		  { Total $\chi^2$/d.o.f.}	&       & & & & &1.629 & 1.290 & 1.482 &  1.283 \\ \bottomrule
     \end{tabular}

\label{tab:first}
\end{sidewaystable}

\section{Fit results}
In this paper, we use four different CCFR, NuTeV, CHORUS, and CDHSW data sets to extract valence PDFs and strong coupling constants at NLO and NNLO taking into account the nuclear and higher twist corrections.

In the presence of CCFR data, if NuTeV data in DIS neutrino-nucleon scattering need to be included in a QCD fit analysis, an exact attention is necessary.
As we mentioned before about the disagreement between CCFR and NuTeV for $x$ values above 0.4, we can have an alternative way to use both CCFR and NuTeV measurements at lower $x$ and only NuTeV for $x>$ 0.4. A cut-study in the $x$ region is performed to ignore the disagreement between CCFR and NuTeV $x>$ 0.4. In this regard, we exclude the CCFR  data only in the above cut.  In the absence of CCFR data in this region, small improvements in the central values and their uncertainties of valence PDFs and strong coupling constant are observed. We obtain 0.1192 $\pm$ 0.0022 and 0.1185 $\pm$ 0.0023 including and excluding  the CCFR data for $x>$0.4 at NNLO.

But according to the relative  $\frac{\bigtriangleup \chi^2}{\chi^2_{\rm}}$ definition with  $\bigtriangleup \chi^2=\chi^2-\chi^{'2}$, where  $\chi^2$ and $\chi^{'2}$ correspond  to exclude and include the CCFR data for $x>$0.4, respectively, we get $\sim 16$\% improvement in the fit quality. Since a significant improvement for fit quality is obtained, we exclude the CCFR  data in this region.

Another kinematic cut-study in $Q^2$ and $W^2$ was performed to isolate the HT contributions to the neutrino-nucleon deep-inelastic structure function $xF_3$ data.  
It seems that HT corrections are not widely available in  $Q^2\geq$4 GeV$^2$ and $W^2\geq$12.5 GeV$^2$ kinematic regions. 
 
In Fig.~\ref{fig:Nucl} we present our results for the $xF_3 (x, Q^2)$ structure function,  as a function of $Q^2$ for different values of $x$ and in the valence quark region. In this figure, the pQCD fits using CCFR, NuTeV, CHORUS, and CDHSW experimental data considering nuclear correction (pQCD+NC) at NLO  and NNLO approximation are shown. As we mentioned before, by including the above cuts which we choose in our analysis, higher twist corrections can be widely eliminated. 

To include HT contributions to the neutrino-nucleon deep-inelastic structure function $xF_3$ data, we need to include the  experimental data points in the  range of  $W^2<$ 12.5 GeV$^2$. To find the impact of this correction, we compare our NNLO fit results for $xF_3 (x, Q^2)$ as a function of $Q^2$ and for fixed $x$ values in Fig.~\ref{fig:QCD-fit-HT}  with and without HT corrections, indicating the above cuts by a dashed-dotted line. The blue lines correspond to NNLO fit results taking into account the HT corrections in the region of $Q^2\geq$ 4 GeV$^2$.

In Table~\ref{tab:first}, the $\chi^2$ values for each data set for different fits, without and with HT, i.e., pQCD+nuclear correction and pQCD+nuclear correction+HT at NLO and NNLO approximation, are shown. We also present the correlated $\chi^2$  and total  $\chi^2/$ d.o.f. in this table.

In Ref.~\cite{MoosaviNejad:2016ebo}, the totalv$\chi^2/$ d.o.f. values are reported as 2.731 and  2.632 for NLO and NNLO, respectively, using the Laplace transform approach without including CDHSW data. By considering the appropriate cuts, and including the nuclear and HT corrections, we obtain totalv$\chi^2/$ d.o.f. = 1.629 and 1.290  without and with higher twist corrections, respectively. We also obtain total $\chi^2/$ d.o.f. = 1.482 and 1.283 at NNLO.  In fact, it seems that attention to appropriate cuts on the data and also taking into account the nuclear and higher twist corrections are necessary.

\begin{table*}[htb]
\caption{The parameters values of the $u$- and $d$-valence quark densities in Eqs.~(\ref{eq:parm1}), (\ref{eq:parm2}) at the input scale of $Q_0^2$  GeV$^2$, obtained from the best fit with CCFR, NuTeV, CHORUS, and CDHSW  considering pQCD and nuclear corrections and also pQCD and nuclear and HT corrections (see text) at NLO and NNLO. The parameter values without error have been fixed after the first minimization in xFitter, due to the fact that the data do not constrain some parameters well enough.}
	\begin{tabular}{|c|cc|cc|}
		\hline \hline 

		    &  \multicolumn{2}{c|}{ \ pQCD+NC \ }  &  \multicolumn{2}{c|}{ \ pQCD+NC+HT \ }    \\    \cline{2-5}
 Parameter  &  NLO &  NNLO  & NLO &  NNLO  \\ \hline
  $N_{u}$ & $ 0.187 $ &  $0.280 $  & $0.208$ & $0.289$ \\ 
  $a_{u}$ & $0.375 \pm 0.021$ & $0.467 \pm 0.029$ & $0.390 \pm 0.038$& $0.455 \pm 0.031$  \\ 
  $b_{u}$ & $2.995 \pm 0.036$ & $3.159 \pm 0.026$ & $3.278\pm 0.068$& $3.384 \pm 0.047$  \\ 
  $c_{u}$ & $35.000 $ & $29.930 $ & $35.000 $ & $ 29.930 $  \\ 
  $d_{u}$ & $ 14.690$ & $11.990 $ & $ 14.690 $& $11.990 $\\  \hline
  $N_{d}$ & $ 0.152 $ & $ 0.246 $ & $0.163$ & $0.238$ \\ 
  $b_{d}$ & $2.590 \pm 0.230 $ & $2.920 \pm 0.280$ & $2.460 \pm 0.360$ & $2.700 \pm 0.240$  \\\hline 
 ${\cal D}_0$ & - & - & $0.970 \pm 0.120$& $0.784 \pm 0.070$  \\ 
 ${\cal D}_1$ & - & - & $1.950 \pm 0.220$& $1.545 \pm 0.059$  \\ 
 ${\cal D}_2$ & - & - & $0.840 \pm 0.110$& $0.672 \pm 0.020$  \\ 
 ${\cal D}_3$ & - & - & $0.100 \pm 0.017$& $0.080 \pm 0.004$  \\ \hline
 $\alpha_{s}(M_Z^2)$ & $0.1235 \pm 0.0015$ & $0.1219 \pm 0.0012$ & $0.1199 \pm 0.0031$ & $0.1185 \pm 0.0023 $    \\ 
  		\hline \hline  \end{tabular}
	
\label{tab:secound}
\end{table*}

According to our parametrization, we have 5+1 free parameters for valence PDFs and the coupling constant $\alpha_{\rm s}(M_Z^2)$,  which can be extracted from the QCD fits.   
Since DIS 
neutrino-nucleon $xF_3$ data  do not constrain 
parameters $c_{u}$ and $d_{u}$ in Eq.~(\ref{eq:parm1}) well enough, we fixed these parameters  after the first minimization. In fact, we should keep fixed the mentioned  parameters after the first minimization, because the errors of these parameters turn out to be rather large compared to the central values, as  it has  been done even in the nonsinglet QCD analyses of $F_2(x,Q^2)$ \cite{Khorramian:2009xz,Khorramian:2008yh,Blumlein:2006be}.

In Table \ref{tab:secound}, we summarize the QCD fit results for the parameters of $xu_v(x,Q_0^2)$ and  $xd_v(x,Q_0^2)$ valence PDFs at NLO and NNLO  for our parametrization which is defined in Eqs.~(\ref{eq:parm1}), (\ref{eq:parm2}) without and with higher twist corrections.
 However, according to Table~\ref{tab:secound} we present our results for  ${\cal D}_k$  parameters for the function $h(x)$ in Eq.~(\ref{eq:HTform}) and 
  $\alpha_{s}(M_Z^2)$  for the NLO and NNLO as well. It seems that by taking into account higher twist corrections we can get the significant improvement for $\alpha_{s}(M_Z^2)$ for both NLO and NNLO as well.

In Fig.~\ref{fig:Q=1(Proton)} we present our NLO and NNLO results for proton $xu_v$ and $xd_v$ valence PDFs  at  $Q^2$ = 1 GeV$^2$ in the region of $x\in[10^{-4},1]$ for pQCD with nuclear corrections (left) and also pQCD with nuclear and HT corrections (right) with their uncertainty bands.

By having the $xu_v$ and $xd_v$ valence PDFs in the free proton at  $Q^2$ = 1 GeV$^2$, it is possible to present ${xu_v }^{p/Fe}$ and ${xd_v} ^{p/Fe}$.  Figure ~\ref{fig:Q=1(Fe)} illustrates our NLO and NNLO results for ${xu_v }^{p/Fe}$ and ${xd_v} ^{p/Fe}$  valence distribution functions  as a function of $x$ at $Q^2$ = 1 GeV$^2$ with nuclear corrections (left) and also with nuclear and HT corrections (right) with their uncertainty bands.

On the other hand, in order to verify the accuracy of the extracted valence PDFs, comparison of the extracted results with other reported ones seems necessary.
In Figs.~\ref{fig:xuv-NNLO} and \ref{fig:xdv-NNLO}, our results for $xu_v(x,Q^2)$ and $xd_v(x,Q^2)$ valence PDFs  with their uncertainties at NNLO compared with the results obtained by CT14 \cite{Dulat:2015mca} and MMHT14 \cite{Harland-Lang:2014zoa} as a function of $x$ at $Q^2 = 4, 100, M_W^2$ and $M_Z^2$ GeV$^2$ (left panel) and the ratio of ${xu_v}/({xu_v})_{ref}$ and ${xd_v}/({xd_v})_{ref}$ (right panel) with respect to our NNLO results.  We show our results only in the range of $x\in[10^{-2},0.8]$, where the data existed and were applied in the present analysis.

It is clear that the results for $xu_v(x,Q^2)$ and $xd_v(x,Q^2)$ valence PDFs are in good agreement with the results of CT14 \cite{Dulat:2015mca} and MMHT14 \cite{Harland-Lang:2014zoa}. We have enough motivation to compare our results to CT14 and MMHT14 analyses because these PDF sets were extracted by including different combinations of data sets for the DIS, especially the neutrino-nucleon data experiments. 

Comparing the central values of  $xu_v(x,Q^2)$ and $xd_v(x,Q^2)$ valence PDFs in the present QCD analysis comes out with almost the same behavior with CT14 \cite{Dulat:2015mca} and MMHT14 \cite{Harland-Lang:2014zoa}, where the effect for $xd_v$ PDF is larger than in the case of $xu_v$ PDF. 

Although the  central values of the PDFs are almost the same, the ratio  of  ${xq_v}/({xq_v})_{ref}$ can  
 illustrate  better the differences of the different analyses. Our results for the $d$-valence PDF ratio ${xd_v}/({xd_v})_{ref}$ are different with the ratio of ${xu_v}/({xu_v})_{ref}$  at large $x$ which are possibly due to the restricted parametrization form of the $d$-valence PDF in the presence of neutrino-nucleon structure function data. Also according to the results for the relative errors on the ratio of ${xu_v}/({xu_v})_{ref}$ and ${xd_v}/({xd_v})_{ref}$, the central values and their uncertainties  on the $u_v$ and $d_v$ with respect to NNLO, for the CT14 and MMHT analyses are different in some regions of $x$. This is reliable since the current QCD nonsinglet analysis is free of the gluonic effects and other PDF sets \cite{Dulat:2015mca,Harland-Lang:2014zoa} are obtained from a global singlet analysis. Note that the CT14 used general-mass variable flavor number (GM-VFN) scheme and its  PDF uncertainties are shown at 68\% C.L. Also, the MMHT14  applied GM-VFN scheme and their parametrizations of the input distributions are based on Chebyshev polynomials and they use the cuts of $Q^2\geq2$ GeV$^2$  and $W^2\geq$ 15 GeV$^2$ on the data.

Another way to compare the QCD fit results consists in forming moments of the valence densities. In Table~\ref{tab:mom}, we present our comparison of low order moments at
 $Q^2$=4 GeV$^2$ from our nonsinglet NNLO QCD analysis
with the NNLO analysis, KT08~\cite{Khorramian:2008yh}, KT07~\cite{Khorramian:2006wg}, MMHT14~\cite{Harland-Lang:2014zoa}, BBG06~\cite{Blumlein:2006be}, A02~\cite{Alekhin:2002fv}, and A06~\cite{Alekhin:2006zm}.

\begin{table*}[htb]
\caption{Comparison of low order moments at
 $Q^2$= 4 GeV$^2$ from our nonsinglet NNLO QCD analysis
with the NNLO analysis, KT08~\cite{Khorramian:2008yh}, KT07~\cite{Khorramian:2006wg}, BBG06~\cite{Blumlein:2006be},
MMHT14~\cite{Harland-Lang:2014zoa}, A02~\cite{Alekhin:2002fv} and
A06~\cite{Alekhin:2006zm}.}
\begin{tabular}{|c|c|c|c|c|c|c|c|c|}
\hline\hline $f$ & $N$ & pQCD+NC+HT& KT08 & KT07 & BBG06 & MMHT14 & A02 & A06 \\
 &  &&(Jacobi poly.)  & (Bernstein poly.) &&&&\\
\hline\hline
$u_{v}$ & 2 & 0.3112
& 0.3056 &0.2934  & 0.2986  & 0.2851 & 0.304 & 0.2947 \\
& 3 & 0.0914
&0.0871 & 0.0825 & 0.0871  & 0.0831 & 0.087 & 0.0843 \\
& 4 & 0.0346
&  0.0330  &0.0311  & 0.0333  & 0.0322 & 0.033 & 0.0319 \\
$d_{v}$ &2& 0.1019
&0.1235 & 0.1143  & 0.1239   & 0.1202 & 0.120 & 0.1129 \\
& 3 & 0.0207
& 0.0298 & 0.0262 & 0.0315  & 0.0305 & 0.028 & 0.0275 \\
& 4 & 0.0058
& 0.0098 &0.0083 & 0.0105  & 0.0106 & 0.010 & 0.0092 \\
\hline\hline
\end{tabular}

\label{tab:mom}
\end{table*}

In Fig.~\ref{fig:uv} our NLO and NNLO results for  ${xu_v}^{p/Fe}$ (left panel) and ${xd_v}^{p/Fe}$ (right panel) valence PDFs with considering nuclear and higher twist effects as a function of $x$ at different values of $Q^2$ = 4, 100, $M_W^2$, and $M_Z^2$ GeV$^2$ with their uncertainty bands are shown.

We extracted the strong coupling constant of $\alpha_{s}(M_Z^2)$ using our different NLO and NNLO QCD fits. We obtained $\alpha_{s}(M_Z^2)=0.1235 \pm 0.0015$ and $0.1219 \pm 0.0012$ in the case of  pQCD and nuclear corrections and 0.1199 $\pm$ 0.0031  and 0.1185 $\pm$ 0.0023 taking into account nuclear  and HT corrections at NLO and NNLO, respectively. 

In Fig.~\ref{fig:Alphas}, we compare our results with the reported results of different NLO and NNLO QCD analyses for $\alpha_{s}(M_Z^2)$  and the world average ${\alpha_s(M^2_Z)}=0.1181\pm0.0011$  which has been reported in Ref.~\cite{Tanabashi:2018oca}. The grey band and dashed line present the world average value of the strong coupling constant $\alpha_{s}(M_{Z}^{2})$ \cite{Tanabashi:2018oca}.  

According to Ref.~\cite{Tanabashi:2018oca},  many experimental observables are used to determine the average value of $\alpha_{\rm s}(M_Z^2)$. In fact, the central value of the world average value is determined as the weighted average of the individual measurements.  In the DIS case and using the global fit to deep inelastic lepton-nucleon scattering data, the average of the results from world data leads to a preaverage value of $\alpha_{s}(M_{Z}^{2})= 0.1156 \pm 0.0021$ \cite{Tanabashi:2018oca}. So it would be worth to show the preaverage value for DIS which reported in Ref.~\cite{Tanabashi:2018oca}. In Fig.~\ref{fig:Alphas} also, the dotted line with yellow band indicates the pre-average results of the strong coupling constant $\alpha_{s}(M_{Z}^{2})$  in the DIS subfield. It should be noted that the preaverage value of $\alpha_{s}(M_{Z}^{2})$ in the DIS process is smaller in comparison to the world average value of $\alpha_{s}(M_{Z}^{2})=0.1181\pm0.0011$. 

The difference of the reported results of $\alpha_{s}(M_{Z}^{2})$ by different groups is due to the fact that this value depends not only on the renormalization scheme, but also on different kinds of measurements in DIS, cuts on the data, and different parametrization and methodology. As we mentioned before the advantage of neutrino structure function
world data is dealing with a restricted set of valence parton densities, and therefore this analysis is free of the correlation between strong coupling constant $\alpha_{s}(M_{Z}^{2})$ and the sea-quarks and gluon distributions. Due to this reason, our results are in good agreement with both pre-average and world average result values of the strong coupling constant.

It is also worth noting that the deep inelastic neutrino-nucleon scattering data are somewhat sensitive to the HT contribution. 
In fact, when HT terms are fitted we obtain about 3\% improvement for the $\alpha_{s}(M_{Z}^{2})$ value in comparison to when the HT terms are set to zero in both NLO and NNLO.

\section{Summary and conclusions}

We perform a QCD analysis of the deep inelastic neutrino-nucleon scattering data from CCFR \cite{ccfr:1977}, NuTeV \cite{Tzanov:2005kr}, CHORUS  \cite{Onengut:2005kv} and CDHSW \cite{Berge:1989hr} without using the orthogonal expansion methods at NLO and NNLO. We determine $xu_v$ and $xd_v$ valence-PDFs and the corresponding errors including the nuclear and higher twist corrections using xFitter framework. 

We studied the CCFR and NuTeV for $x$ values above 0.4, where there is a disagreement between CCFR and NuTeV.
By excluding the CCFR  data only in the above cut, we get $\sim$ 16\% improvement in the fit quality.  We exclude the CCFR  data only at  $x>$0.4 to remove the disagreement between CCFR and NuTeV for $x$ values above 0.4.

 The fit quality for the $xF_3$ structure function for our parametrization is in good consistency with the neutrino-nucleon scattering data without and with HT corrections. In particular, it is interesting to investigate the quality of the fits improvement in some regions of $x$ and $Q^2$. To find the impact of HT corrections in the nonsinglet QCD analysis, we compare the results with and without HT corrections. We found by taking into account the HT corrections we can obtain 20\% and 13\%   improvements of total $\chi^2/$d.o.f. at the NLO and NNLO, respectively.  
 
It is also worth noting that the $\alpha_{s}(M_Z^2)$ value is somewhat sensitive to the HT contribution. In the present analysis we obtain $\alpha_{s}(M_Z^2)=0.1235 \pm 0.0015$ and  0.1199 $\pm$ 0.0031  without and with considering HT corrections at NLO, respectively. Our NNLO results for  $\alpha_{s}(M_Z^2)$  without and with considering HT corrections  are $0.1219 \pm 0.0012$ and  0.1185 $\pm$ 0.0023, respectively.

In the present analysis, the central values and their uncertainties for  $xu_v$ and $xd_v$  valence PDFs of the free proton at NLO and NNLO and also for ${xu_v }^{p/Fe}$ and ${xd_v} ^{p/Fe}$ valence PDFs of the bound proton in iron are reliable in our parametrization.  

To compare our NLO and NNLO QCD fit results, we have shown  $xu_v$ and $xd_v$  valence PDFs  of the free proton with corresponding errors to other theoretical models.   
The present QCD analysis shows that the central values for $xu_v$ and $xd_v$ valence PDFs at low and large $x$ values are good in agreement with obtained results from CT14 and MMHT  models \cite{Dulat:2015mca,Harland-Lang:2014zoa}.  
The discrepancy between CT14 and MMHT and our results for the central value of the valence PDF, their uncertainties or both is due to different kinds of data sets, various cuts on the data, and also different kind of parametrization. In fact, the main reason is due to the fact that we used only $xF_3 $ data rather than being a QCD global fit.

Although there are several analyses for DIS neutrino-nucleon data using different approaches, such as orthogonal polynomial approaches, we have shown our present results taking into account all available deep inelastic neutrino-nucleon scattering data taking into account that the nuclear and higher twist effects can give us a very precise valence quark distribution and also  $\alpha_{s}(M_Z^2)$. 

Another way to compare our QCD fit results with other reported results is the determination of the lowest moments of  valence distributions. Our calculation for the lowest moments of valence PDFs  shows us that the differences between different models are due to the kind of data sets and  theoretical methods, such as the orthogonal polynomials approach.

In this analysis, we also present the strong coupling constant $\alpha_{s}(M_Z^2)$, which was obtained at NLO and NNLO. The corresponding estimates, based on the QCD analysis of deep inelastic neutrino-nucleon scattering data, compare well with the recently reported results and the world average of ${\alpha_s(M^2_Z)}=0.1181\pm0.0011$  which is reported in Ref. \cite{Tanabashi:2018oca}.  As we mentioned before the advantage of neutrino structure function
world data is to deal with a restricted set of valence parton densities, and therefore this analysis is free of the correlation between the strong coupling constant  and the sea-quarks and gluon distributions. Due to this reason, our extracted results  for $\alpha_{s}(M_Z^2)$ are in good agreement with the world average result value of the strong coupling constant.

Undoubtedly, by having the new measurements from the future colliders, such as the Large Hadron Electron Collider  (LHeC),  and Electron Ion Collider (EIC), 
our knowledge of the nonsinglet PDFs and strong coupling constant will improve.

A standard LHAPDF library of this QCD analysis at NLO and NNLO can be obtained via email from the authors.

\section{Acknowledgments}
We gratefully acknowledge A. Glazov, R. Placakyte, Stefan Schmitt, and D. Naples  for help and useful discussions. A.K.  is also grateful to the CERN TH-PH division for the hospitality where a portion of this work
was performed. H.A. acknowledges the financial support from the DESY FH group (xFitter) and is also grateful to the CTEQ-MCnet school organizers for their hospitality and scholarship at DESY.
\section*{References}


\begin{thebibliography}{99}

\bibitem{DGLAP}
 V.~N.~Gribov and L.~N.~Lipatov, Sov.\ J.\ Nucl.\ Phys.\  { 15}, 438 (1972) [Yad.\ Fiz.\  { 15}, 781 (1972)]; L.~N.~Lipatov, Sov.\ J.\ Nucl.\ Phys.\  { 20}, 94 (1975) [Yad.\ Fiz.\  { 20}, 181 (1974)]; Y.~L.~Dokshitzer,  Sov.\ Phys.\ JETP { 46}, 641 (1977) [Zh.\ Eksp.\ Teor.\ Fiz.\  { 73}, 1216 (1977)]; G.~Altarelli and G.~Parisi,
  Nucl.\ Phys.\ B { 126}, 298 (1977).


  


\bibitem{Alekhin:2012ig} 
  S.~Alekhin, J.~Blumlein and S.~Moch,
  Phys.\ Rev.\ D { 86}, 054009 (2012)
  [arXiv:1202.2281 [hep-ph]].
  




\bibitem{Gao:2013xoa} 
  J.~Gao {\it et al.},
  Phys.\ Rev.\ D { 89}, no. 3, 033009 (2014)
[arXiv:1302.6246 [hep-ph]].
  
\bibitem{CooperSarkar:2011aa} 
  A.~M.~Cooper-Sarkar (ZEUS and H1 Collaborations),
  PoS EPS { -HEP2011}, 320 (2011)
  [arXiv:1112.2107 [hep-ph]].
  
\bibitem{Martin:2009iq} 
  A.~D.~Martin, W.~J.~Stirling, R.~S.~Thorne and G.~Watt,
  Eur.\ Phys.\ J.\ C { 63}, 189 (2009)
  [arXiv:0901.0002 [hep-ph]].
  
  
\bibitem{Ball:2012wy} 
  R.~D.~Ball {\it et al.},
 JHEP { 1304}, 125 (2013)
 [arXiv:1211.5142 [hep-ph]].
  
\bibitem{Jimenez-Delgado:2014xza} 
  P.~Jimenez-Delgado {\it et al.} (Jefferson Lab Angular Momentum
(JAM) Collaboration),
  Phys.\ Lett.\ B { 738}, 263 (2014)
  [arXiv:1403.3355 [hep-ph]].
  
  
\bibitem{Jimenez-Delgado:2013boa} 
  P.~Jimenez-Delgado, A.~Accardi and W.~Melnitchouk,
  Phys.\ Rev.\ D { 89}, no. 3, 034025 (2014)
 [arXiv:1310.3734 [hep-ph]].
  
\bibitem{Ball:2013lla} 
  R.~D.~Ball {\it et al.} (NNPDF Collaboration), 
  Nucl.\ Phys.\ B { 874}, 36 (2013)
[arXiv:1303.7236 [hep-ph]].
  

  

  
\bibitem{Leader:2010rb} 
  E.~Leader, A.~V.~Sidorov and D.~B.~Stamenov,
  Phys.\ Rev.\ D { 82}, 114018 (2010)
 [arXiv:1010.0574 [hep-ph]].
  
\bibitem{Blumlein:2010rn} 
  J.~Blumlein and H.~Bottcher,
  Nucl.\ Phys.\ B { 841}, 205 (2010)
 [arXiv:1005.3113 [hep-ph]].
  
\bibitem{Sato:2016tuz} 
  N.~Sato {\it et al.} (Jefferson Lab Angular Momentum Collaboration),
  Phys.\ Rev.\ D { 93}, no. 7, 074005 (2016)
  [arXiv:1601.07782 [hep-ph]].
  
\bibitem{Khanpour:2017cha} 
  H.~Khanpour, S.~T.~Monfared and S.~Atashbar Tehrani,
  Phys.\ Rev.\ D { 95}, no. 7, 074006 (2017)
 [arXiv:1703.09209 [hep-ph]].
  
\bibitem{Arbabifar:2013tma} 
  F.~Arbabifar, A.~N.~Khorramian and M.~Soleymaninia,
  Phys.\ Rev.\ D { 89}, no. 3, 034006 (2014)
  [arXiv:1311.1830 [hep-ph]].
  


\bibitem{Monfared:2014nta} 
  S.~Taheri Monfared, Z.~Haddadi and A.~N.~Khorramian,
  Phys.\ Rev.\ D { 89}, no. 7, 074052 (2014)
  [arXiv:1405.4633 [hep-ph]].
  
  
\bibitem{Khorramian:2010qa} 
  A.~N.~Khorramian, S.~Atashbar Tehrani, S.~Taheri Monfared, F.~Arbabifar and F.~I.~Olness,
  Phys.\ Rev.\ D { 83}, 054017 (2011)
[arXiv:1011.4873 [hep-ph]].
  
\bibitem{AtashbarTehrani:2007odq} 
  S.~Atashbar Tehrani and A.~N.~Khorramian, 
  JHEP { 0707}, 048 (2007)
  [arXiv:0705.2647 [hep-ph]].
  
\bibitem{Khorramian:2004ih} 
  A.~N.~Khorramian, A.~Mirjalili and S.~A.~Tehrani,
 JHEP { 0410}, 062 (2004)
  [hep-ph/0411390].

 
  
  
\bibitem{Soleymaninia:2013cxa} 
  M.~Soleymaninia, A.~N.~Khorramian, S.~M.~Moosavi Nejad and F.~Arbabifar,
  Phys.\ Rev.\ D { 88}, no. 5, 054019 (2013)
  [arXiv:1306.1612 [hep-ph]].
  
 
  
\bibitem{ccfr:1977}  
  W.~G.~Seligman {\it et al.} (CCFR Collaboration),
  Phys.\ Rev.\ Lett.\  { 79}, 1213 (1997).






  
\bibitem{Tzanov:2005kr}
 M.~Tzanov {\it et al.} (NuTeV Collaboration), 
  Phys.\ Rev.\ D { 74}, 012008 (2006)
 [hep-ex/0509010].

  
  
\bibitem{Onengut:2005kv} 
  G.~Onengut {\it et al.} (CHORUS Collaboration),
  Phys.\ Lett.\ B { 632}, 65 (2006).

 
\bibitem{Berge:1989hr} 
  J.~P.~Berge {\it et al.} (CDHSW Collaboration),
  Z.\ Phys.\ C { 49}, 187 (1991).
   
  
\bibitem{Bonesini:2016nrv} 
  M.~Bonesini,
  Frascati Phys.\ Ser.\  { 61}, 11 (2016)
 [arXiv:1606.00765 [physics.acc-ph]].

\bibitem{Kaplan:2014xda} 
  D.~M.~Kaplan (MAP and MICE Collaborations),
  EPJ Web Conf.\  { 95}, 03019 (2015)
  [arXiv:1412.3487 [physics.acc-ph]].

\bibitem{Banerjee:2015gca} 
  S.~Banerjee, P.~S.~B.~Dev, A.~Ibarra, T.~Mandal and M.~Mitra,
  Phys.\ Rev.\ D { 92}, 075002 (2015)
  [arXiv:1503.05491 [hep-ph]].

\bibitem{Geer:2009zz} 
  S.~Geer,
  Ann.\ Rev.\ Nucl.\ Part.\ Sci.\  { 59}, 347 (2009).


  
\bibitem{AbelleiraFernandez:2012cc} 
  J.~L.~Abelleira Fernandez {\it et al.} [LHeC Study Group],
  J.\ Phys.\ G { 39}, 075001 (2012)
[arXiv:1206.2913 [physics.acc-ph]].
  
\bibitem{AbelleiraFernandez:2012ni} 
  J.~L.~Abelleira Fernandez {\it et al.},
  [arXiv:1211.4831 [hep-ex]].


\bibitem{Aschenauer:2016our} 
  E.~C.~Aschenauer {\it et al.},
  [arXiv:1602.03922 [nucl-ex]].
  
\bibitem{Deshpande:2016goi} 
  A.~Deshpande, Z.~E.~Meziani and J.~W.~Qiu,
  EPJ Web Conf.\  { 113}, 05019 (2016).

\bibitem{Accardi:2012qut} 
  A.~Accardi {\it et al.},
  Eur.\ Phys.\ J.\ A { 52}, no. 9, 268 (2016)
  [arXiv:1212.1701 [nucl-ex]].

 
\bibitem{Lees:2015ymt} 
  J.~P.~Lees {\it et al.} (BaBar Collaboration),
  Phys.\ Rev.\ D { 93}, no. 5, 052015 (2016)
  [arXiv:1508.07960 [hep-ex]].
  

\bibitem{CTEQ:CJ} 
 The CTEQ-Jefferson Lab (CJ) Collaboration, http://www.jlab.org/cj. 
 

 
\bibitem{Owens:2012bv} 
  J.~F.~Owens, A.~Accardi and W.~Melnitchouk,
  Phys.\ Rev.\ D { 87}, no. 9, 094012 (2013)
  [arXiv:1212.1702 [hep-ph]].
  
\bibitem{Accardi:2011fa} 
  A.~Accardi, W.~Melnitchouk, J.~F.~Owens, M.~E.~Christy, C.~E.~Keppel, L.~Zhu and J.~G.~Morfin,
  Phys.\ Rev.\ D { 84}, 014008 (2011)
[arXiv:1102.3686 [hep-ph]].
\bibitem{Accardi:2009br} 
  A.~Accardi, M.~E.~Christy, C.~E.~Keppel, P.~Monaghan, W.~Melnitchouk, J.~G.~Morfin and J.~F.~Owens,
  Phys.\ Rev.\ D { 81}, 034016 (2010)
  [arXiv:0911.2254 [hep-ph]].
  

  
\bibitem{Accardi:2016qay} 
  A.~Accardi, L.~T.~Brady, W.~Melnitchouk, J.~F.~Owens and N.~Sato,
  Phys.\ Rev.\ D { 93}, no. 11, 114017 (2016)
 [arXiv:1602.03154 [hep-ph]].
  


  
  
\bibitem{Kataev:1994rj} 
  A.~L.~Kataev and A.~V.~Sidorov,
  Phys.\ Lett.\ B { 331}, 179 (1994)
  [hep-ph/9402342].



\bibitem{Kataev:1996vu} 
  A.~L.~Kataev, A.~V.~Kotikov, G.~Parente and A.~V.~Sidorov,
  Phys.\ Lett.\ B { 388}, 179 (1996)
  [hep-ph/9605367].

\bibitem{Kataev:1997nc} 
  A.~L.~Kataev, A.~V.~Kotikov, G.~Parente and A.~V.~Sidorov,
  Phys.\ Lett.\ B { 417}, 374 (1998)
  [hep-ph/9706534].
%

\bibitem{Kataev:1997vv} 
  A.~L.~Kataev, A.~V.~Kotikov, G.~Parente and A.~V.~Sidorov,
  Nucl.\ Phys.\ Proc.\ Suppl.\  { 64}, 138 (1998)
  [hep-ph/9709509].
  
\bibitem{Alekhin:1998df} 
  S.~I.~Alekhin and A.~L.~Kataev,
  Phys.\ Lett.\ B { 452}, 402 (1999)
  [hep-ph/9812348].


\bibitem{Alekhin:1999af} 
  S.~I.~Alekhin and A.~L.~Kataev,
  Nucl.\ Phys.\ A { 666}, 179 (2000)
 [hep-ph/9908349].
  
  
\bibitem{Kataev:1999bp} 
  A.~L.~Kataev, G.~Parente and A.~V.~Sidorov,
  Nucl.\ Phys.\ B { 573}, 405 (2000)
  [hep-ph/9905310].

\bibitem{Kataev:2001kk} 
  A.~L.~Kataev, G.~Parente and A.~V.~Sidorov,
  Phys.\ Part.\ Nucl.\  { 34}, 20 (2003)
  Fiz.\ Elem.\ Chast.\ Atom.\ Yadra { 34}, 43 (2003); 38, 827 (2007)
  [hep-ph/0106221].
  
\bibitem{Kataev:2002wr} 
  A.~L.~Kataev, G.~Parente and A.~V.~Sidorov,
  J.\ Phys.\ G { 29}, 1985 (2003)
  [hep-ph/0209024].

\bibitem{Sidorov:2013aza} 
  A.~V.~Sidorov and O.~P.~Solovtsova,
  Nonlin.\ Phenom.\ Complex Syst.\  { 16}, 397 (2013)
 [arXiv:1312.3082 [hep-ph]].

\bibitem{Khorramian:2009xz} 
  A.~N.~Khorramian, H.~Khanpour and S.~A.~Tehrani,
  Phys.\ Rev.\ D { 81}, 014013 (2010)
  [arXiv:0909.2665 [hep-ph]].

\bibitem{AtashbarTehrani:2009zz} 
  S.~Atashbar Tehrani and A.~N.~Khorramian,
  Nucl.\ Phys.\ Proc.\ Suppl.\  { 186}, 58 (2009).
 
\bibitem{Khorramian:2008yh} 
  A.~N.~Khorramian and S.~A.~Tehrani,
  Phys.\ Rev.\ D { 78}, 074019 (2008)
  [arXiv:0805.3063 [hep-ph]].

\bibitem{Khorramian:2006wg} 
  A.~N.~Khorramian and S.~Atashbar Tehrani,
  JHEP { 0703}, 051 (2007)
  [hep-ph/0610136].
  
  
\bibitem{Santiago:2001mh} 
  J.~Santiago and F.~J.~Yndurain,
  Nucl.\ Phys.\ B { 611}, 447 (2001)
  [hep-ph/0102247].
  


\bibitem{GhasempourNesheli:2015tva} 
  A.~Ghasempour Nesheli, A.~Mirjalili and M.~M.~Yazdanpanah,
  Eur.\ Phys.\ J.\ Plus { 130}, no. 4, 82 (2015).
  
  
\bibitem{MoosaviNejad:2016ebo} 
  S.~M.~Moosavi Nejad, H.~Khanpour, S.~Atashbar Tehrani and M.~Mahdavi,
  Phys.\ Rev.\ C { 94}, no. 4, 045201 (2016)
 [arXiv:1609.05310 [hep-ph]].
 

\bibitem{Salimi-Amiri:2018had} 
  M.~Salimi-Amiri, A.~Khorramian, H.~Abdolmaleki and F.~I.~Olness,
  Phys.\ Rev.\ D {\bf 98}, no. 5, 056020 (2018)
  [arXiv:1805.02613 [hep-ph]].


\bibitem{xFitter} 
  xFitter, An open source QCD fit framework.  http://xFitter.org
 [arXiv:1410.4412 [hep-ph]].
  
\bibitem{Alekhin:2014irh} 
  S.~Alekhin {\it et al.},
  Eur.\ Phys.\ J.\ C { 75}, no. 7, 304 (2015)
  [arXiv:1410.4412 [hep-ph]].
  
  
\bibitem{Sapronov} 
  A.~Sapronov (HERAFitter Team Collaboration),
  J.\ Phys.\ Conf.\ Ser.\  { 608}, no. 1, 012051 (2015).


  
\bibitem{Vafaee:2017nze} 
  A.~Vafaee and A.~N.~Khorramian,
  Nucl.\ Phys.\ B { 921}, 472 (2017).
  

 
\bibitem{Abdolmaleki:2017wlg} 
  H.~Abdolmaleki, A.~Khorramian and A.~Aleedaneshvar,
  Nucl.\ Part.\ Phys.\ Proc.\  { 282-284}, 27 (2017).
  
\bibitem{Vafaee:2017jnt} 
  A.~Vafaee and A.~Khorramian,
  Nucl.\ Part.\ Phys.\ Proc.\  { 282-284}, 32 (2017).
  
  
\bibitem{Rostami:2015iva} 
  S.~Rostami, A.~Khorramian, A.~Aleedaneshvar,
  J.\ Phys.\ G { 43}, no. 5, 055001 (2016)
  [arXiv:1510.08421 [hep-ph]].
  
\bibitem{Azizi:2018iiq} 
  M.~Azizi, A.~Khorramian, H.~Abdolmaleki and S.~P.~Mehdiabadi,
  Int.\ J.\ Mod.\ Phys.\ A {\bf 33}, no. 24, 1850142 (2018).
  
\bibitem{Aleedaneshvar:2017bgs} 
  A.~Aleedaneshvar and A.~N.~Khorramian,
  Nucl.\ Phys.\ A {\bf 979}, 215 (2018)
  [arXiv:1709.07247 [hep-ph]].
  
  
    \bibitem{Hamed:2018new}
   H.~Abdolmaleki and A.~Khorramian, Phys.\ Rev.\ D (to be published). 


 
  
  
\bibitem{Eisele:1986uz} 
  F.~Eisele,
  Rept.\ Prog.\ Phys.\  { 49}, 233 (1986).

\bibitem{Diemoz:1986kt} 
  M.~Diemoz, F.~Ferroni and E.~Longo,
  Phys.\ Rept.\  { 130}, 293 (1986).
  




\bibitem{Kramer:2000hn} 
  M.~Kramer, F.~I.~Olness and D.~E.~Soper,
  Phys.\ Rev.\ D { 62}, 096007 (2000)
 [hep-ph/0003035].


\bibitem{deFlorian:2011fp} 
  D.~de Florian, R.~Sassot, P.~Zurita and M.~Stratmann,
  Phys.\ Rev.\ D {\bf 85}, 074028 (2012)
 [arXiv:1112.6324 [hep-ph]].
  
\bibitem{Schienbein:2007fs} 
  I.~Schienbein, J.~Y.~Yu, C.~Keppel, J.~G.~Morfin, F.~Olness and J.~F.~Owens,
  Phys.\ Rev.\ D {\bf 77}, 054013 (2008)
 [arXiv:0710.4897 [hep-ph]].
  
\bibitem{Kumano:2016jsb} 
  S.~Kumano,
  JPS Conf.\ Proc.\  {\bf 12}, 010004 (2016)
 [arXiv:1602.02459 [hep-ph]].
  
\bibitem{Sidorov:1996wb} 
  A.~V.~Sidorov,
  Phys.\ Lett.\ B {\bf 389}, 379 (1996)
  [hep-ph/9607275].

\bibitem{Sidorov:1996if} 
  A.~V.~Sidorov,
  JINR Rapid Commun.\  {\bf 80}, 11 (1996)
  [hep-ph/9609345].
  
\bibitem{Virchaux:1991jc} 
  M.~Virchaux and A.~Milsztajn,
  Phys.\ Lett.\ B {\bf 274}, 221 (1992).
  
\bibitem{Tokarev:1997vc} 
  M.~V.~Tokarev and A.~V.~Sidorov,
  Nuovo Cim.\ A {\bf 110}, 1401 (1997)
  [hep-ph/9707438].
  
  
\bibitem{Diemoz:1987xu} 
  M.~Diemoz, F.~Ferroni, E.~Longo and G.~Martinelli,
  Z.\ Phys.\ C {\bf 39}, 21 (1988).
     
\bibitem{Gluck:1989ze} 
  M.~Gluck, E.~Reya and A.~Vogt,
  Z.\ Phys.\ C {\bf 48}, 471 (1990).
  
\bibitem{Gluck:1998xa} 
M.~Gluck, E.~Reya and A.~Vogt,
  Eur.\ Phys.\ J.\ C {\bf 5}, 461 (1998)
 [hep-ph/9806404].


\bibitem{Blumlein:2006be} 
  J.~Blumlein, H.~Bottcher and A.~Guffanti,
  Nucl.\ Phys.\ B { 774}, 182 (2007)
 [hep-ph/0607200].


 
\bibitem{Tanabashi:2018oca} 
  M.~Tanabashi {\it et al.} [Particle Data Group],
  Phys.\ Rev.\ D {\bf 98}, no. 3, 030001 (2018).

 
\bibitem{Botje:2010ay} 
  M.~Botje,
  Comput.\ Phys.\ Commun.\  { 182}, 490 (2011)
  [arXiv:1005.1481 [hep-ph]].
 
\bibitem{Bertone:2017kne} 
  V.~Bertone and M.~Botje,
  [arXiv:1712.08162 [hep-ph]].

 
 
 
 

 
\bibitem{Vogt:2004ns} 
  A.~Vogt,
  Comput.\ Phys.\ Commun.\  { 170}, 65 (2005)
 [hep-ph/0408244].

 
  
  

\bibitem{PDFUncertainties} J.~Pumplin {\it et~al.}, JHEP {\bf 07} 012 (2007);
        J.~Pumplin {\it et~al.}, Phys. Rev. {\bf D65} 014013 (2001);
         J.~Pumplin {\it et~al.}, Phys. Rev. {\bf D65} 014011 (2001).
       

 
\bibitem{Pascaud:1995qs}
  C.~Pascaud and F.~Zomer,
  LAL-95-05.


\bibitem{Perez:2012um} 
  E.~Perez and E.~Rizvi,
  Rept.\ Prog.\ Phys.\  {\bf 76}, 046201 (2013)
 [arXiv:1208.1178 [hep-ex]].

 
\bibitem{Dulat:2015mca} 
  S.~Dulat {\it et al.},
  Phys.\ Rev.\ D { 93}, no. 3, 033006 (2016)
 [arXiv:1506.07443 [hep-ph]].
  
\bibitem{Harland-Lang:2014zoa} 
  L.~A.~Harland-Lang, A.~D.~Martin, P.~Motylinski and R.~S.~Thorne,
  Eur.\ Phys.\ J.\ C { 75}, no. 5, 204 (2015)
  [arXiv:1412.3989 [hep-ph]].
  
  
\bibitem{Alekhin:2002fv} 
  S.~Alekhin,
  Phys.\ Rev.\ D { 68}, 014002 (2003)
 [hep-ph/0211096].


\bibitem{Alekhin:2006zm} 
  S.~Alekhin, K.~Melnikov and F.~Petriello,
  Phys.\ Rev.\ D { 74}, 054033 (2006)
  [hep-ph/0606237].

  


\bibitem{Martin:2003tt} 
  A.~D.~Martin, R.~G.~Roberts, W.~J.~Stirling and R.~S.~Thorne,
 [hep-ph/0307262].

\bibitem{Adloff:2000qk} 
  C.~Adloff {\it et al.} (H1 Collaboration),
  Eur.\ Phys.\ J.\ C {\bf 21}, 33 (2001)
  [hep-ex/0012053].

\bibitem{Chekanov:2001qu} 
  S.~Chekanov {\it et al.} (ZEUS Collaboration),
  Eur.\ Phys.\ J.\ C {\bf 21}, 443 (2001)
 [hep-ex/0105090].


  
   
  
  
\bibitem{Arneodo:1996qe} 
  M.~Arneodo {\it et al.} (New Muon Collaboration),
  Nucl.\ Phys.\ B {\bf 483}, 3 (1997)
  [hep-ph/9610231].
    

\bibitem{Lionetti:2011pw} 
  S.~Lionetti {\it et al.},
  Phys.\ Lett.\ B {\bf 701}, 346 (2011)
 [arXiv:1103.2369 [hep-ph]].
  
  
  
\bibitem{Alekhin:2009ni} 
  S.~Alekhin, J.~Blumlein, S.~Klein and S.~Moch,
  Phys.\ Rev.\ D {\bf 81}, 014032 (2010)
 [arXiv:0908.2766 [hep-ph]].

  
\bibitem{Alekhin:2018pai} 
   S.~Alekhin, J.~Blumlein and S.~Moch,
  Eur.\ Phys.\ J.\ C {\bf 78}, no. 6, 477 (2018)
  [arXiv:1803.07537 [hep-ph]].
 

 
\bibitem{Abramowicz:2015mha} 
  H.~Abramowicz {\it et al.} (H1 and ZEUS Collaborations),
  Eur.\ Phys.\ J.\ C {\bf 75}, no. 12, 580 (2015)
  [arXiv:1506.06042 [hep-ex]].

\bibitem{Ball:2014uwa} 
  R.~D.~Ball {\it et al.} (NNPDF Collaboration),  JHEP {\bf 1504}, 040 (2015)
 [arXiv:1410.8849 [hep-ph]].
 
\bibitem{JimenezDelgado:2008hf} 
  P.~Jimenez-Delgado and E.~Reya,
  Phys.\ Rev.\ D {\bf 79}, 074023 (2009)
  [arXiv:0810.4274 [hep-ph]].
 
\bibitem{Aaron:2009aa} 
  F.~D.~Aaron {\it et al.} (H1 and ZEUS Collaborations), JHEP01(2010)109
 [arXiv:0911.0884 [hep-ex]].
  
 
\end{thebibliography}

\clearpage

\begin{figure}[ht]
\begin{center}
        \resizebox{0.80\textwidth}{!}{\includegraphics{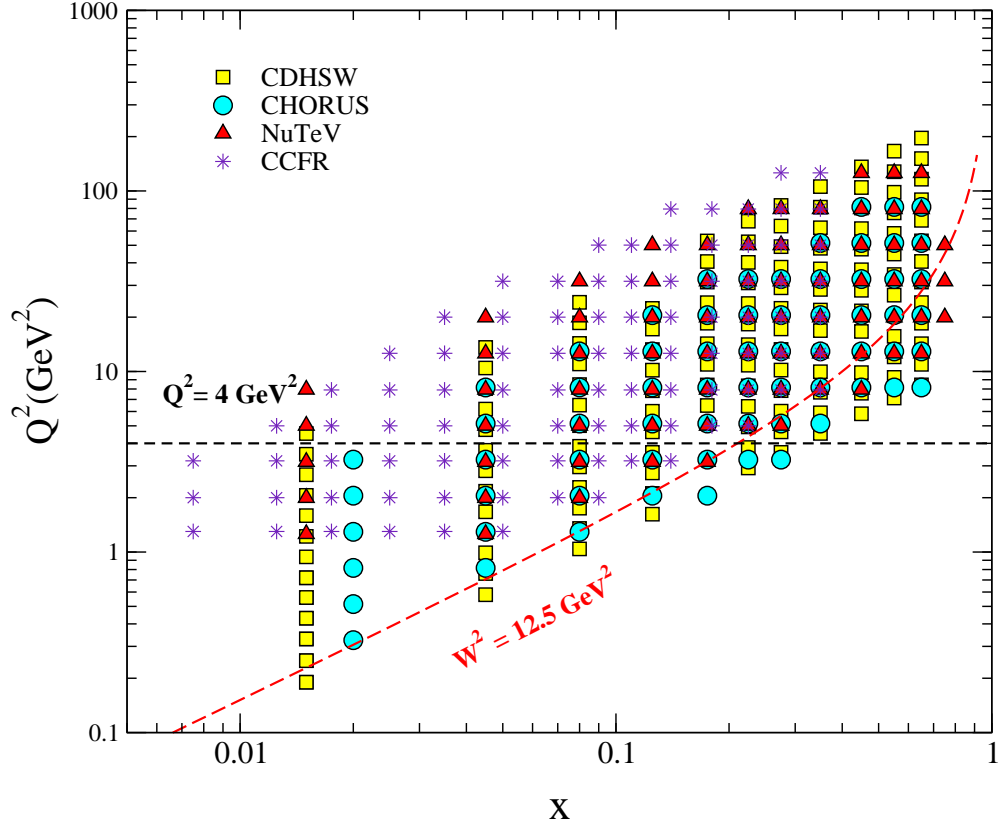}} 
\caption{ Different experiments of  DIS neutrino-nucleon data in the $x$ and $Q^2$ plane. The dashed line represents
the kinematic $W^2$ and $Q^2$ cuts on the data ($Q^2\geq$ 4 GeV$^2$ and $W^2\geq$ 12.5 GeV$^2$) in this analysis. The data points lying below these lines are only excluded in the present QCD fits.}
\label{fig:1}
        \end{center}
\end{figure}

\begin{figure*}[h!]

\includegraphics[width=0.32\textwidth]{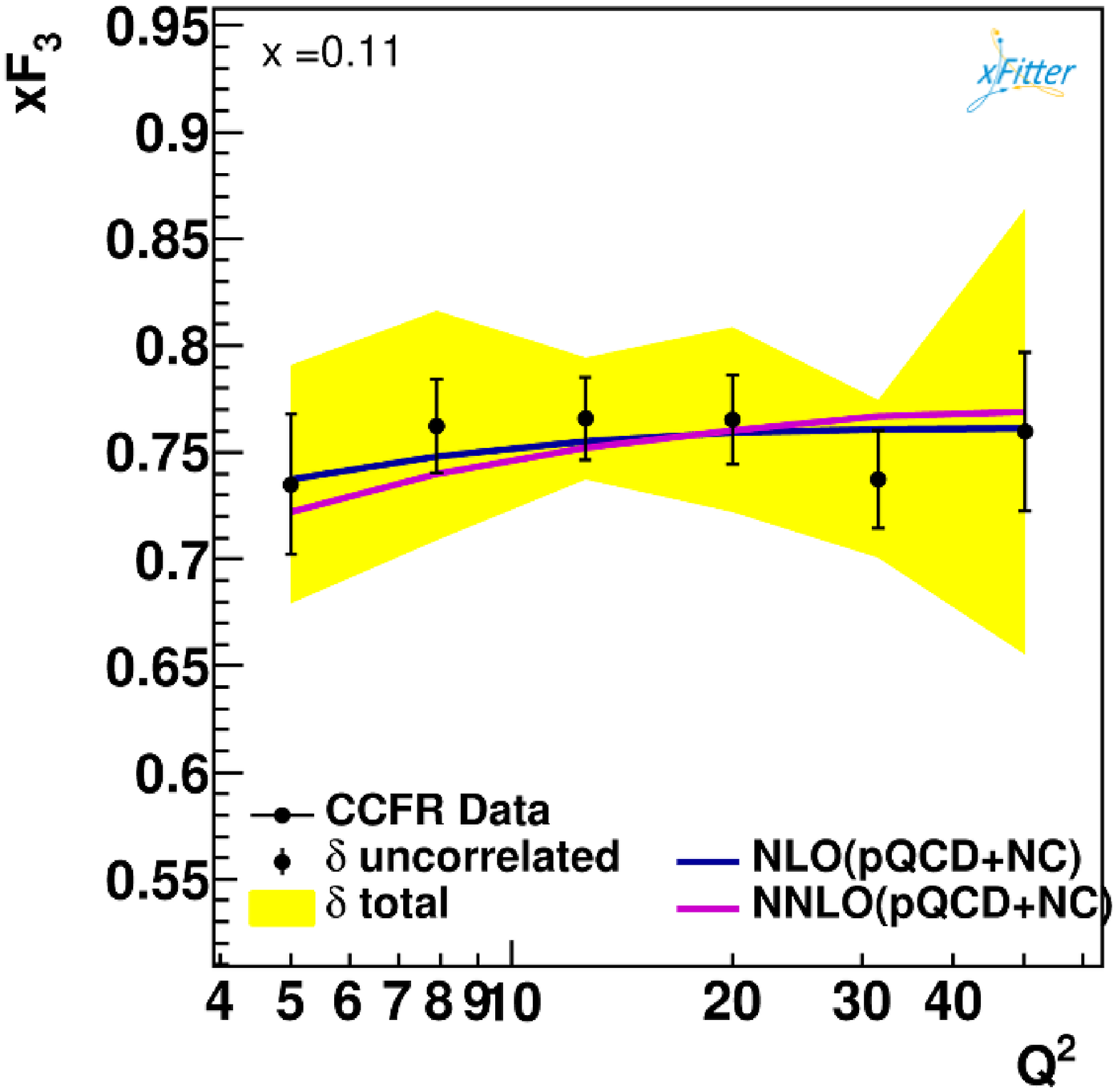}
\includegraphics[width=0.32\textwidth]{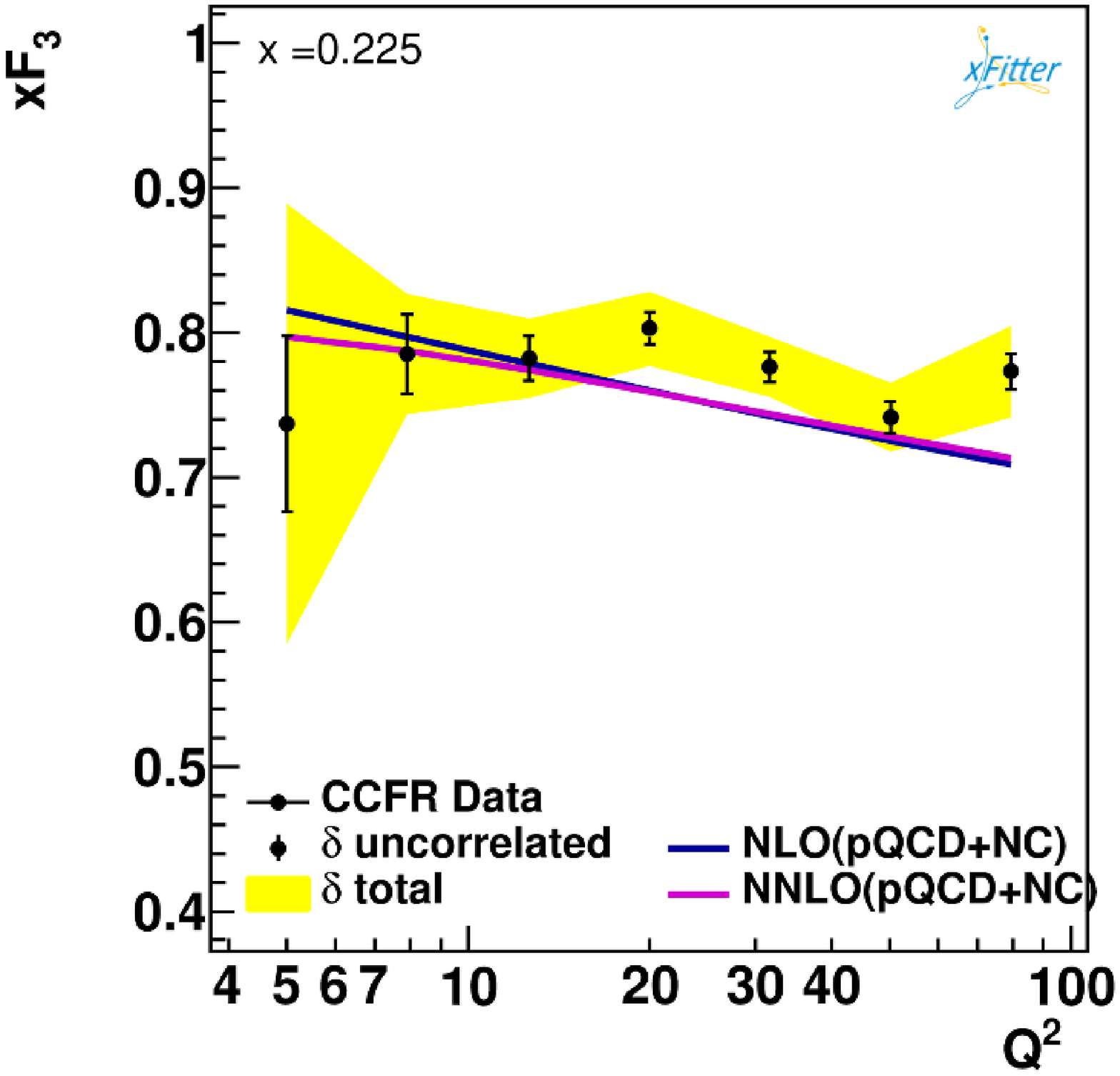}
\includegraphics[width=0.32\textwidth]{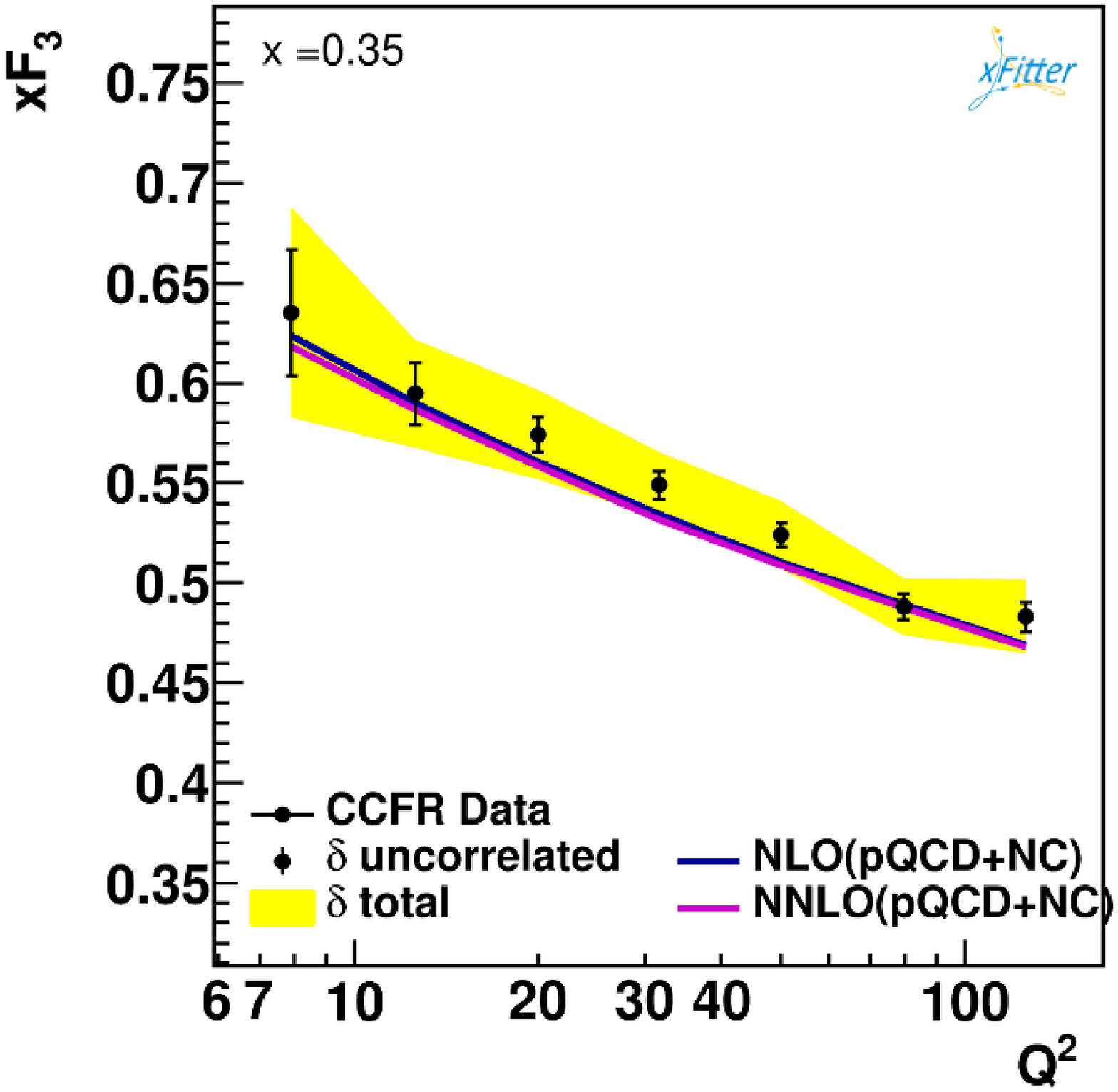}

\includegraphics[width=0.32\textwidth]{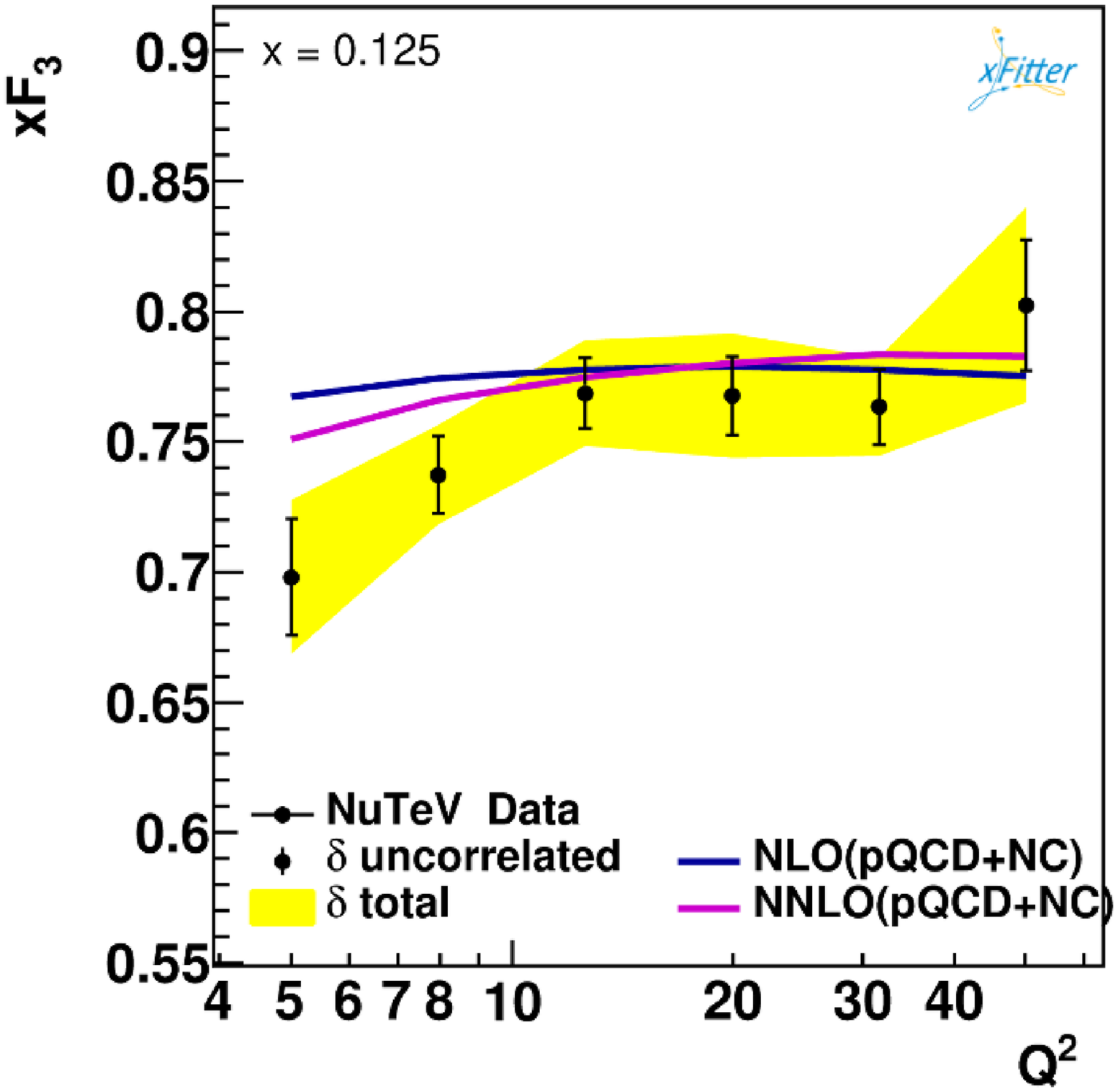}
\includegraphics[width=0.32\textwidth]{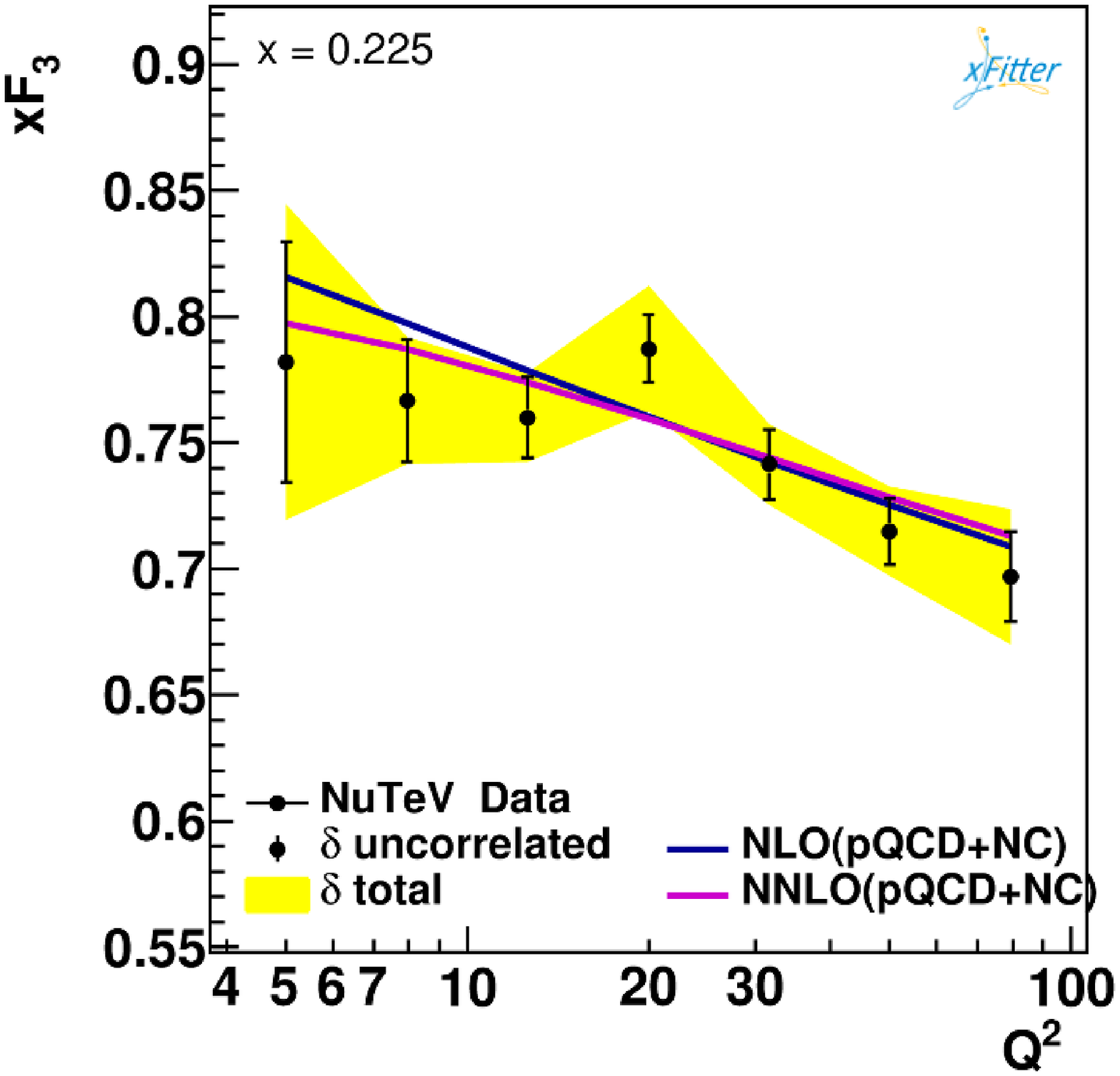}
\includegraphics[width=0.32\textwidth]{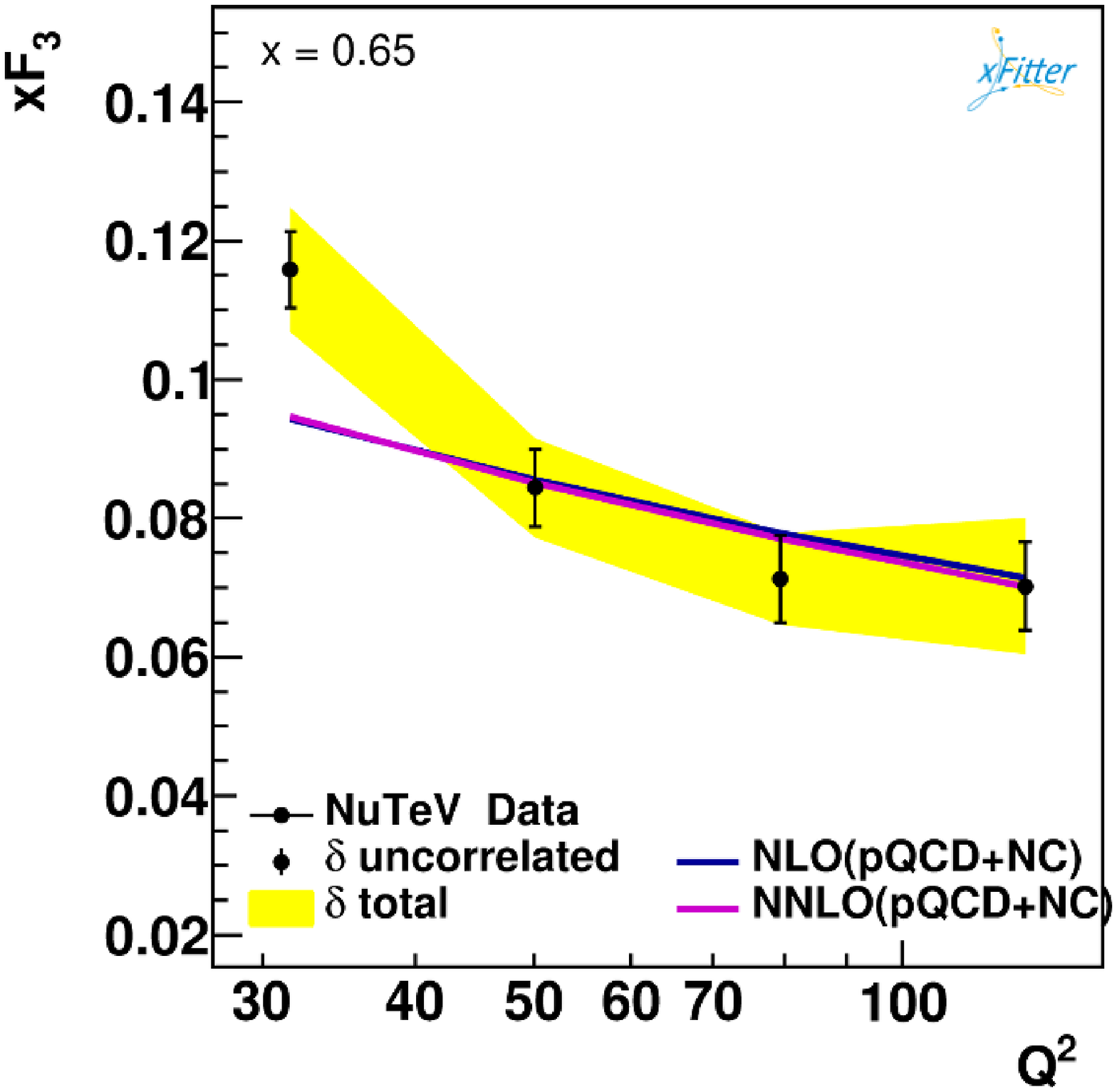}

\includegraphics[width=0.32\textwidth]{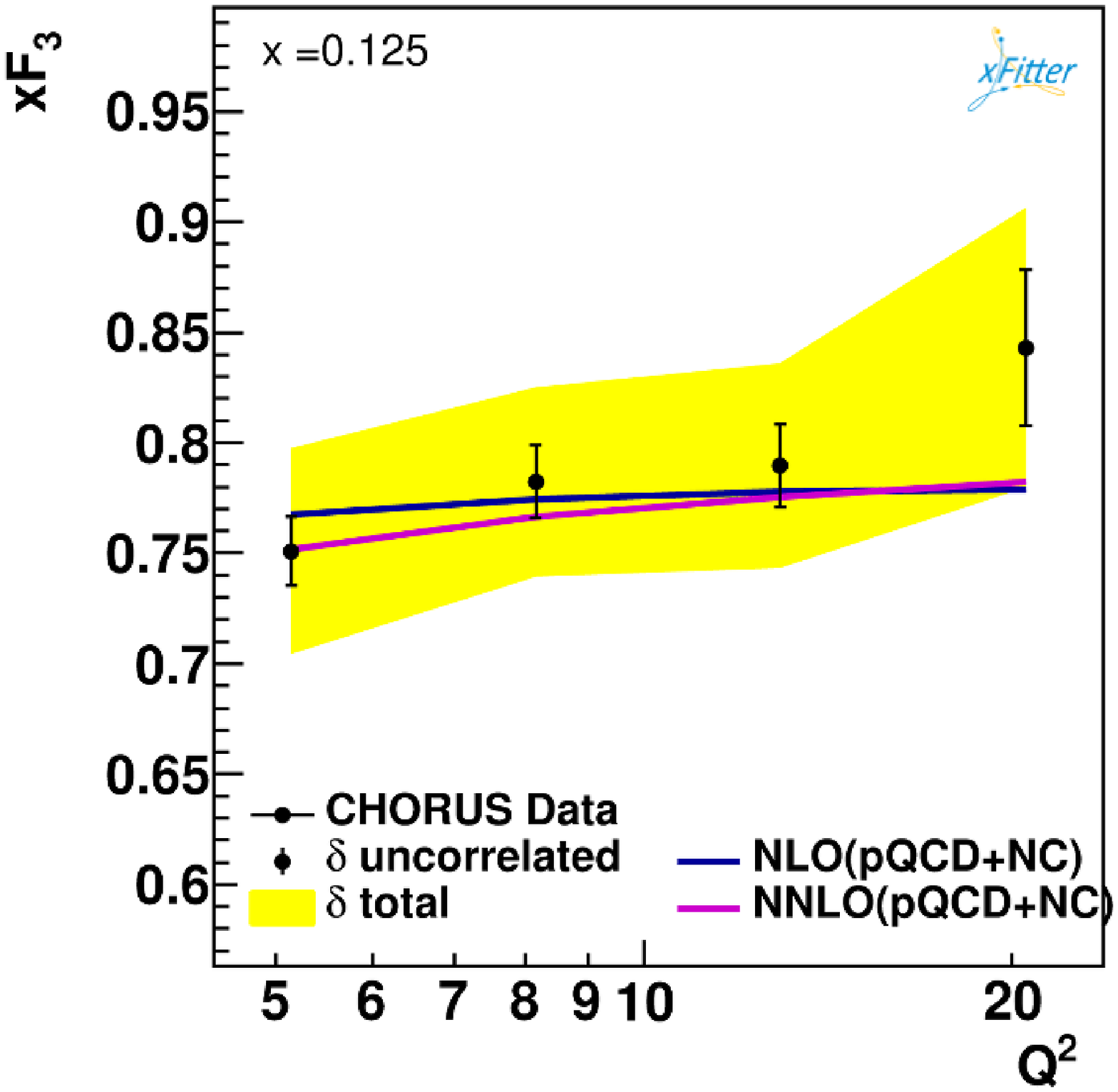}
\includegraphics[width=0.32\textwidth]{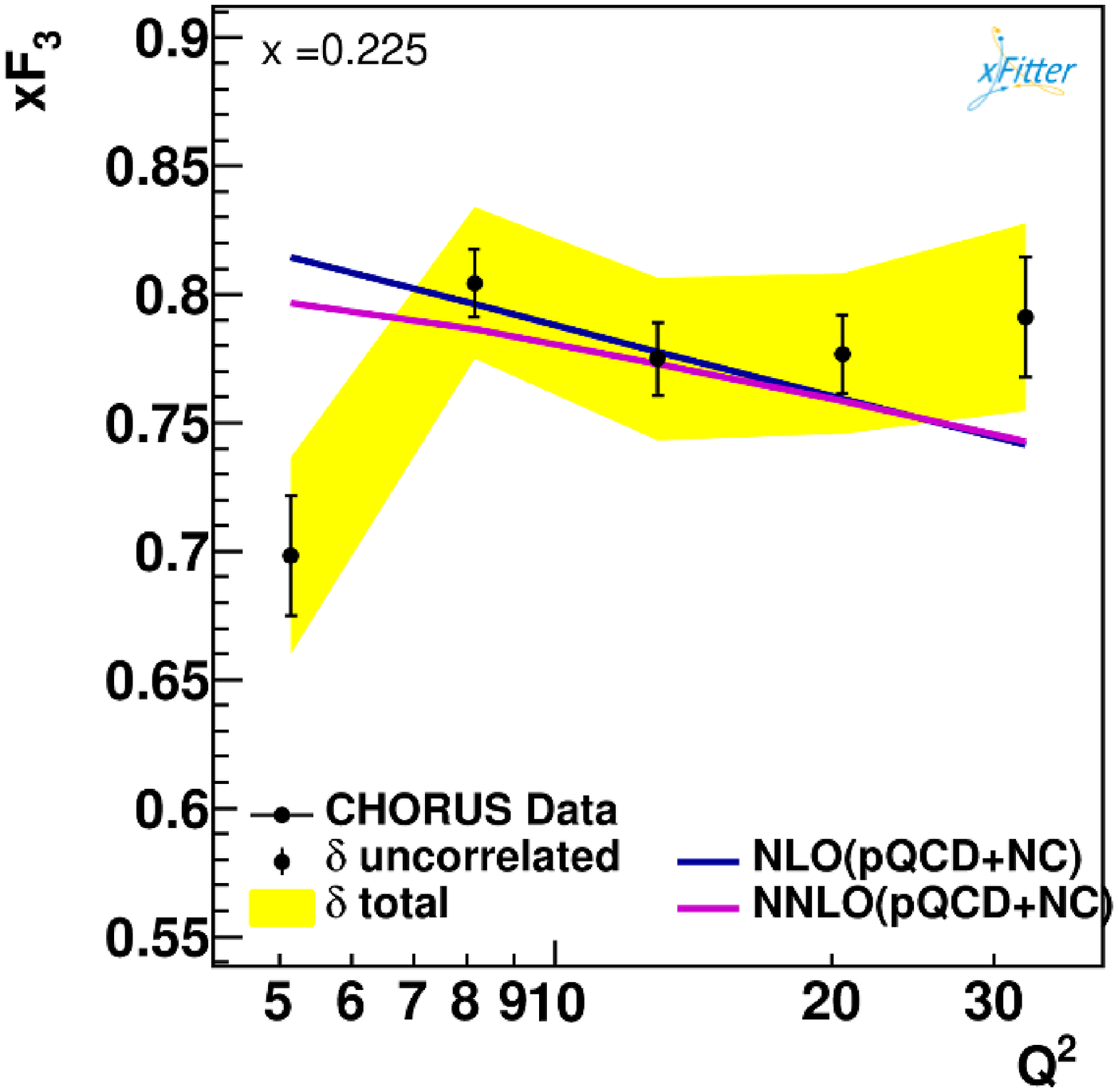}
\includegraphics[width=0.32\textwidth]{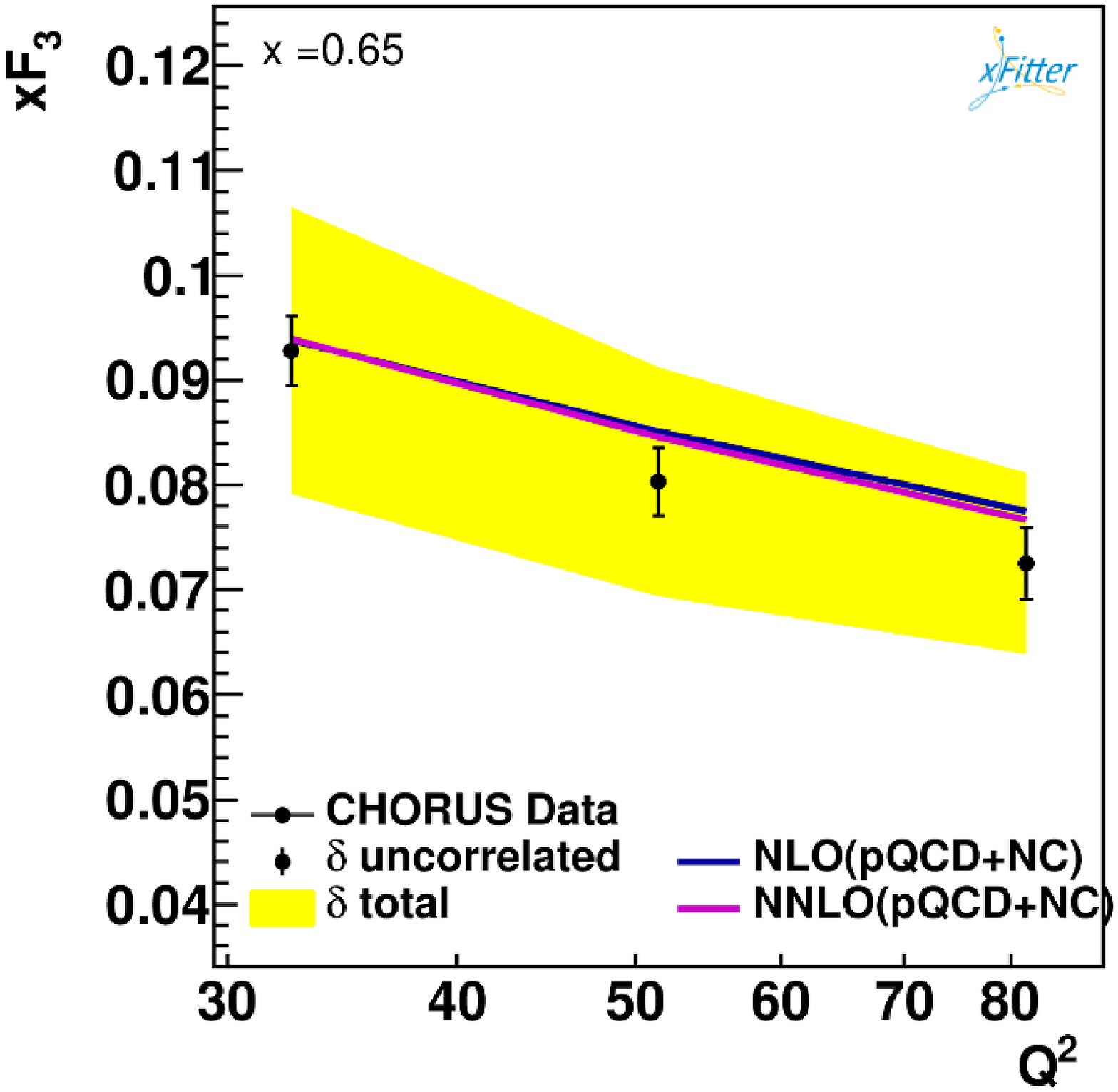}

\includegraphics[width=0.32\textwidth]{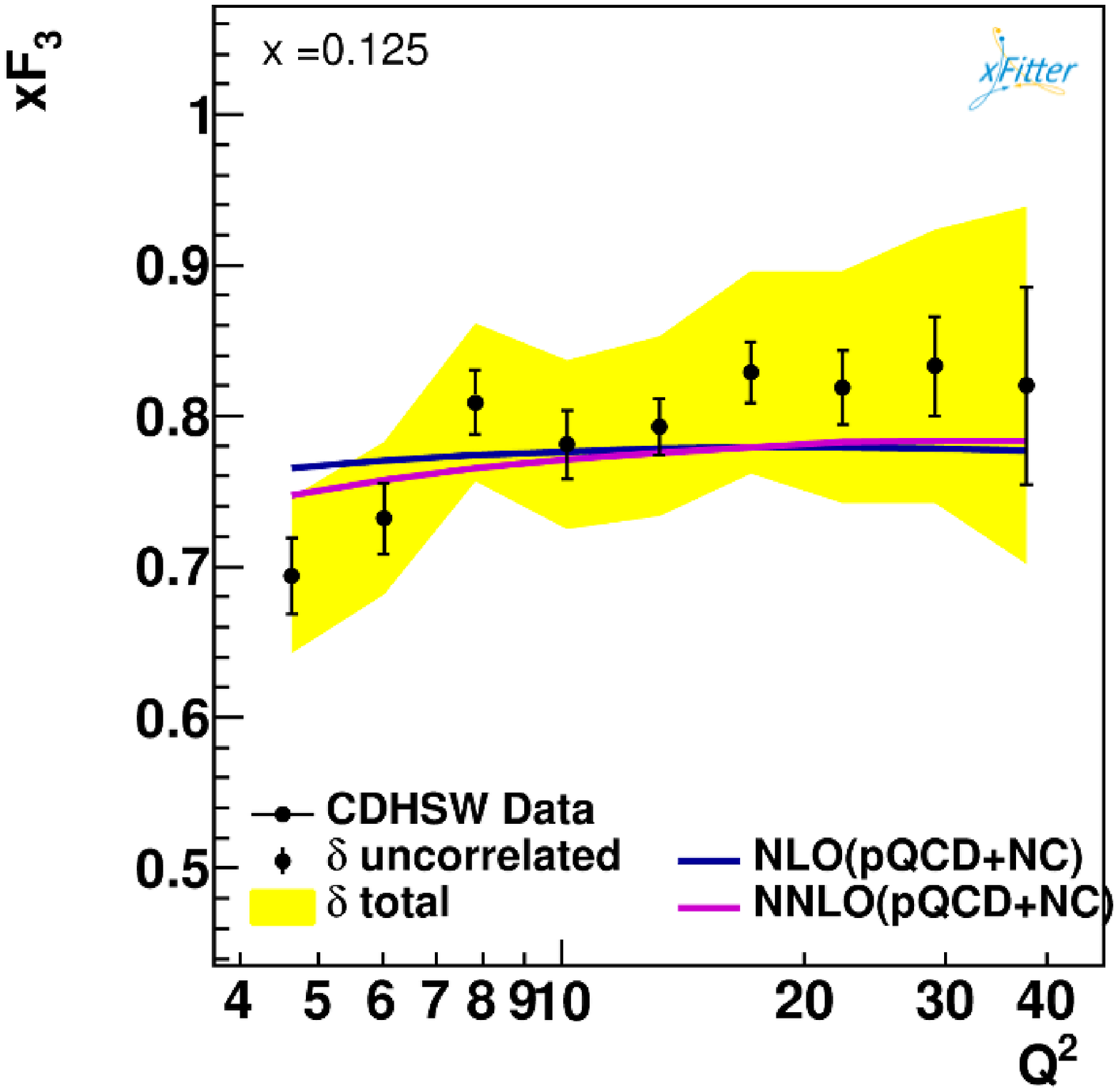}
\includegraphics[width=0.32\textwidth]{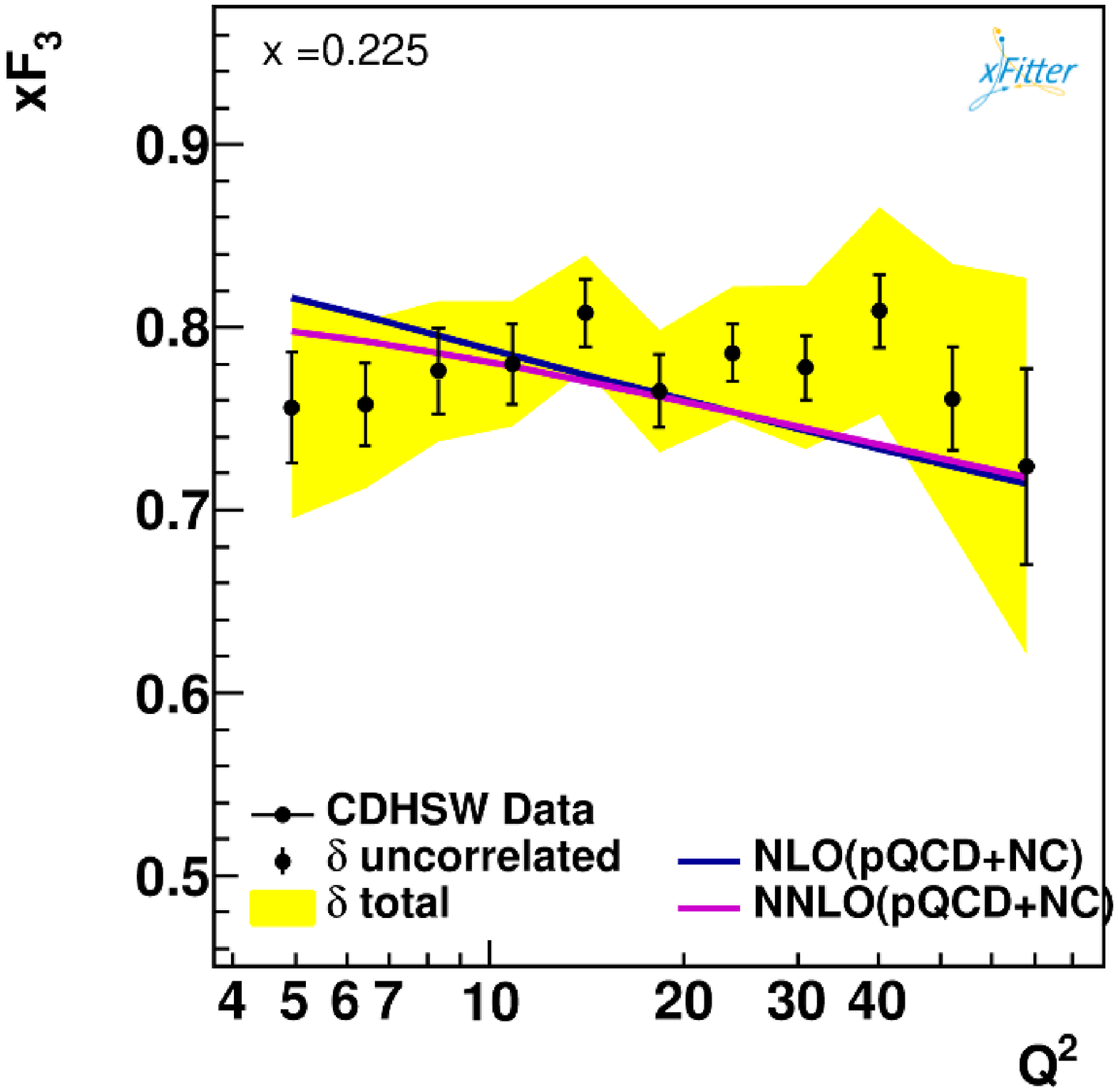}
\includegraphics[width=0.32\textwidth]{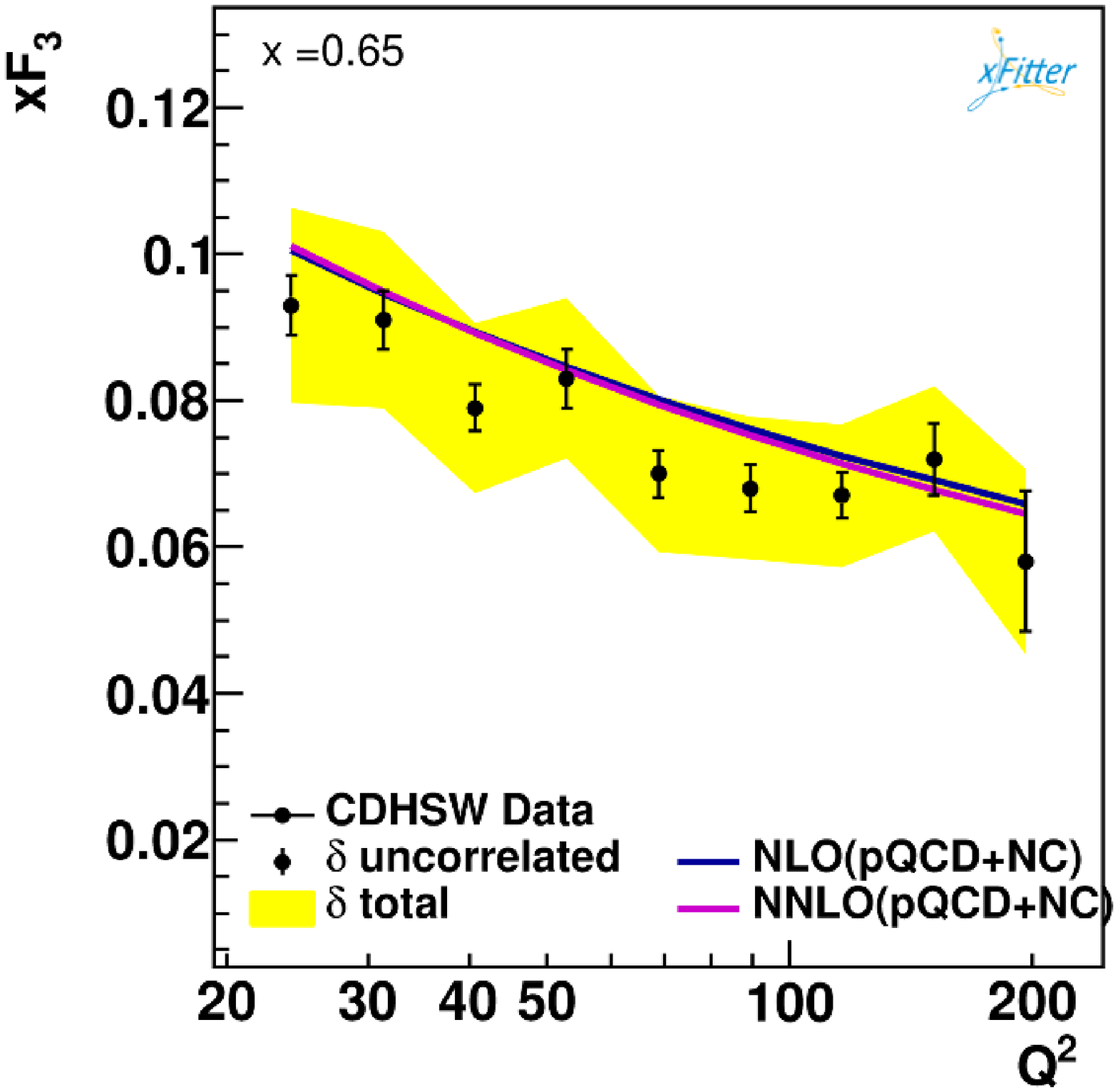}

\caption{The comparison of structure function $xF_3$ obtained from the fit at NLO and NNLO as a function of $Q^2$ in various $x$ with considering nuclear corrections, by using CCFR \cite{ccfr:1977}, NuTeV \cite{Tzanov:2005kr}, CHORUS \cite{Onengut:2005kv}, and CDHSW \cite{Berge:1989hr} data sets.}
\label{fig:Nucl}
\end{figure*}

\begin{figure*}[h!]
\includegraphics[width=0.49\textwidth]{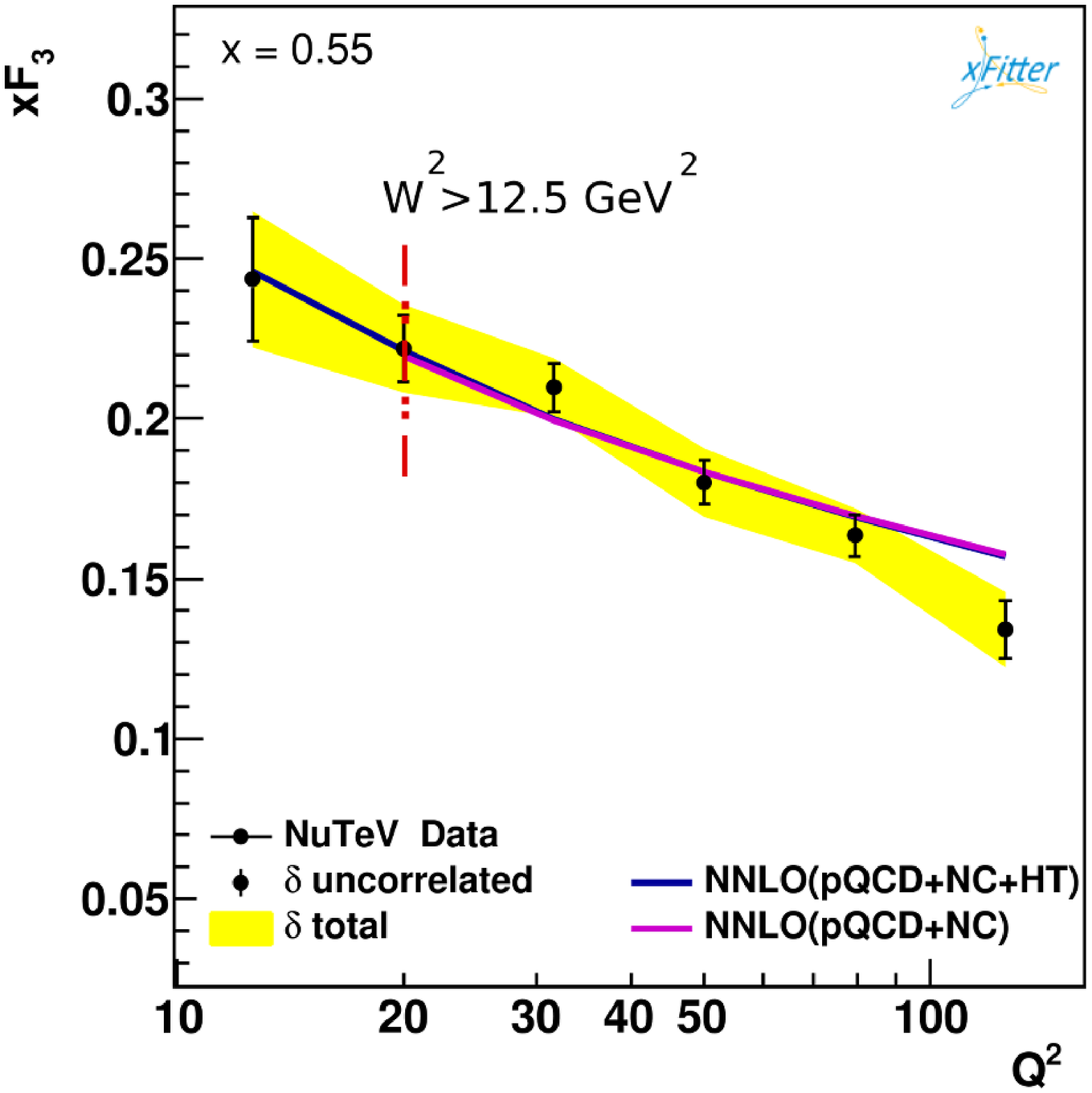}
\includegraphics[width=0.49\textwidth]{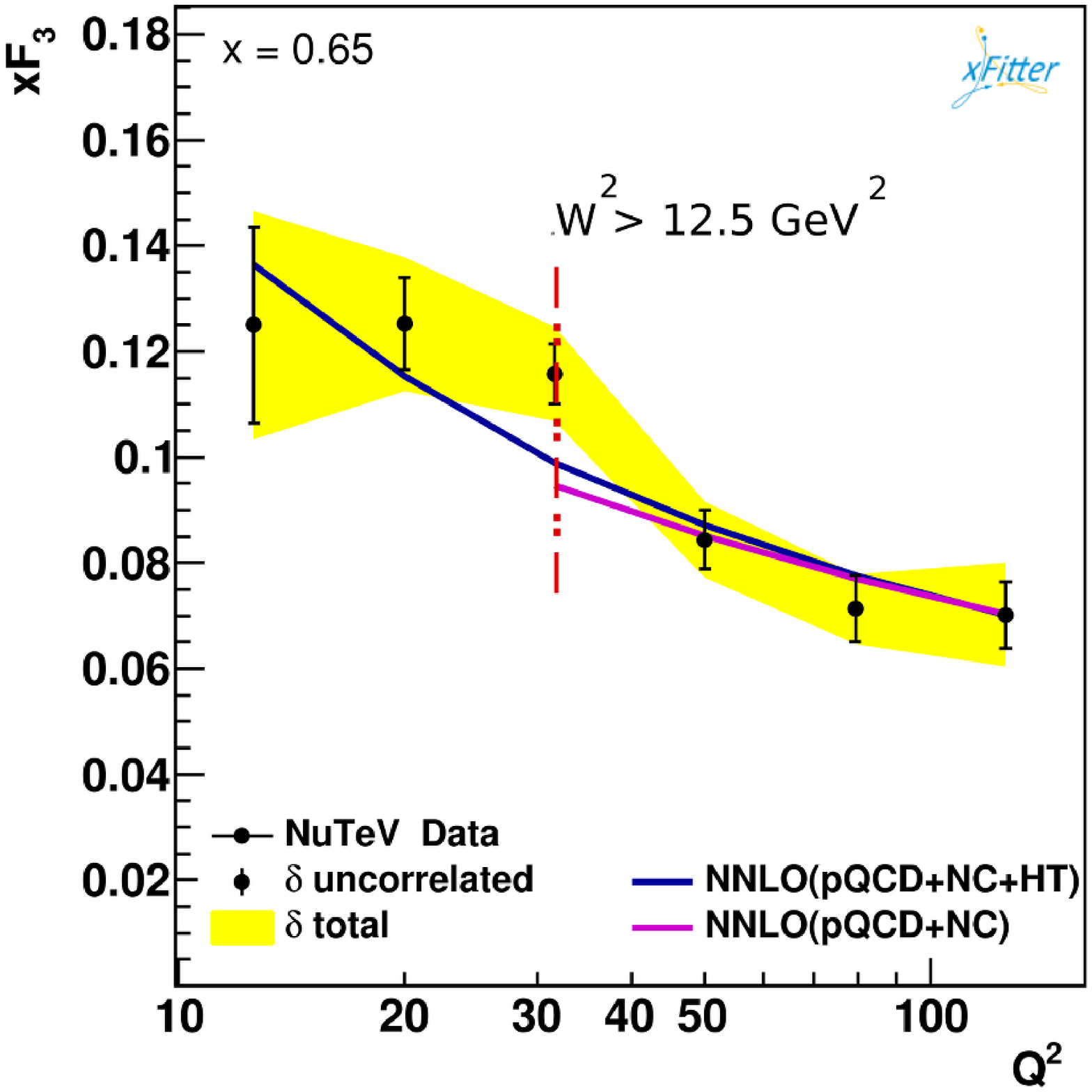}

\includegraphics[width=0.49\textwidth]{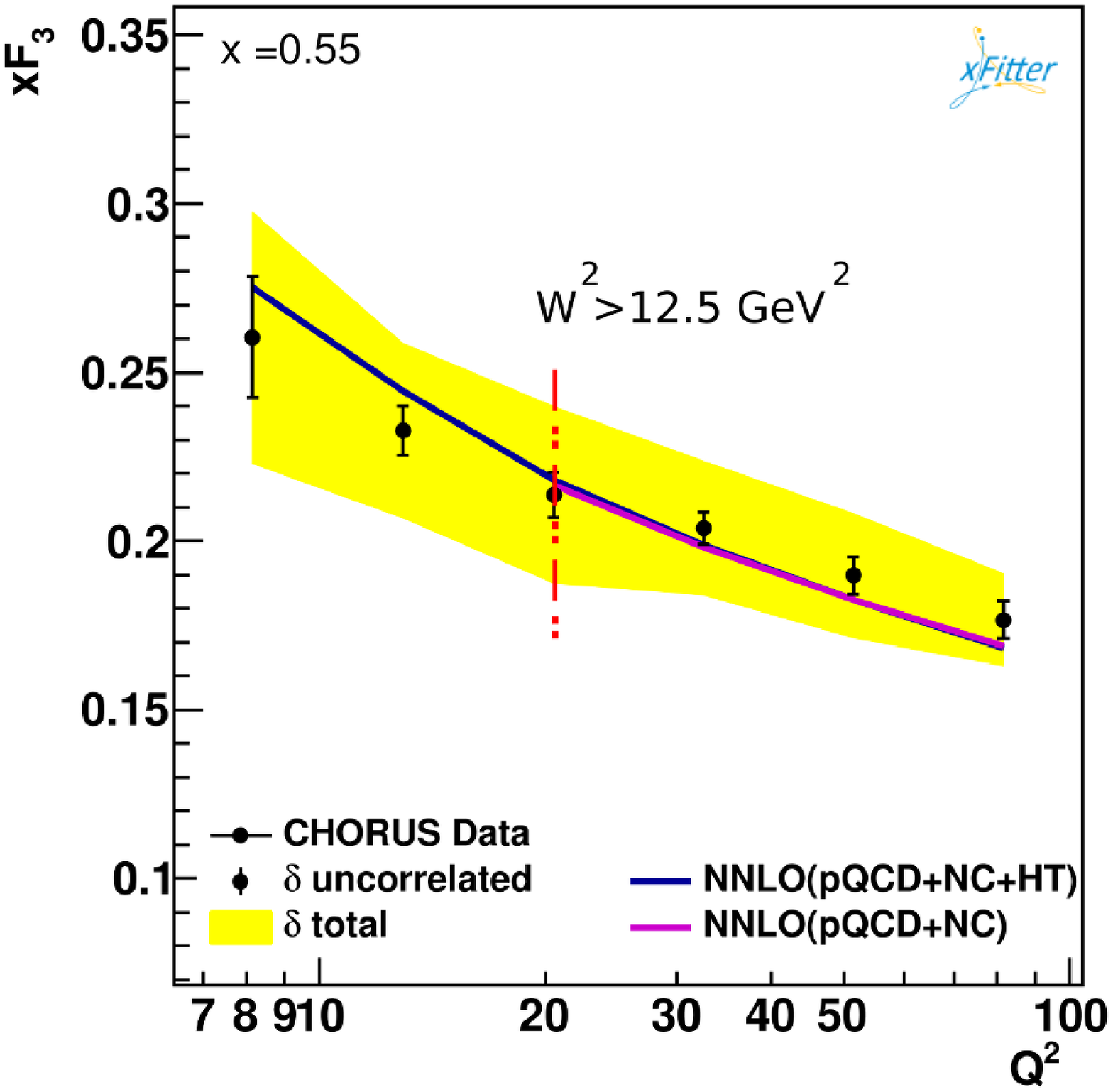}
\includegraphics[width=0.49\textwidth]{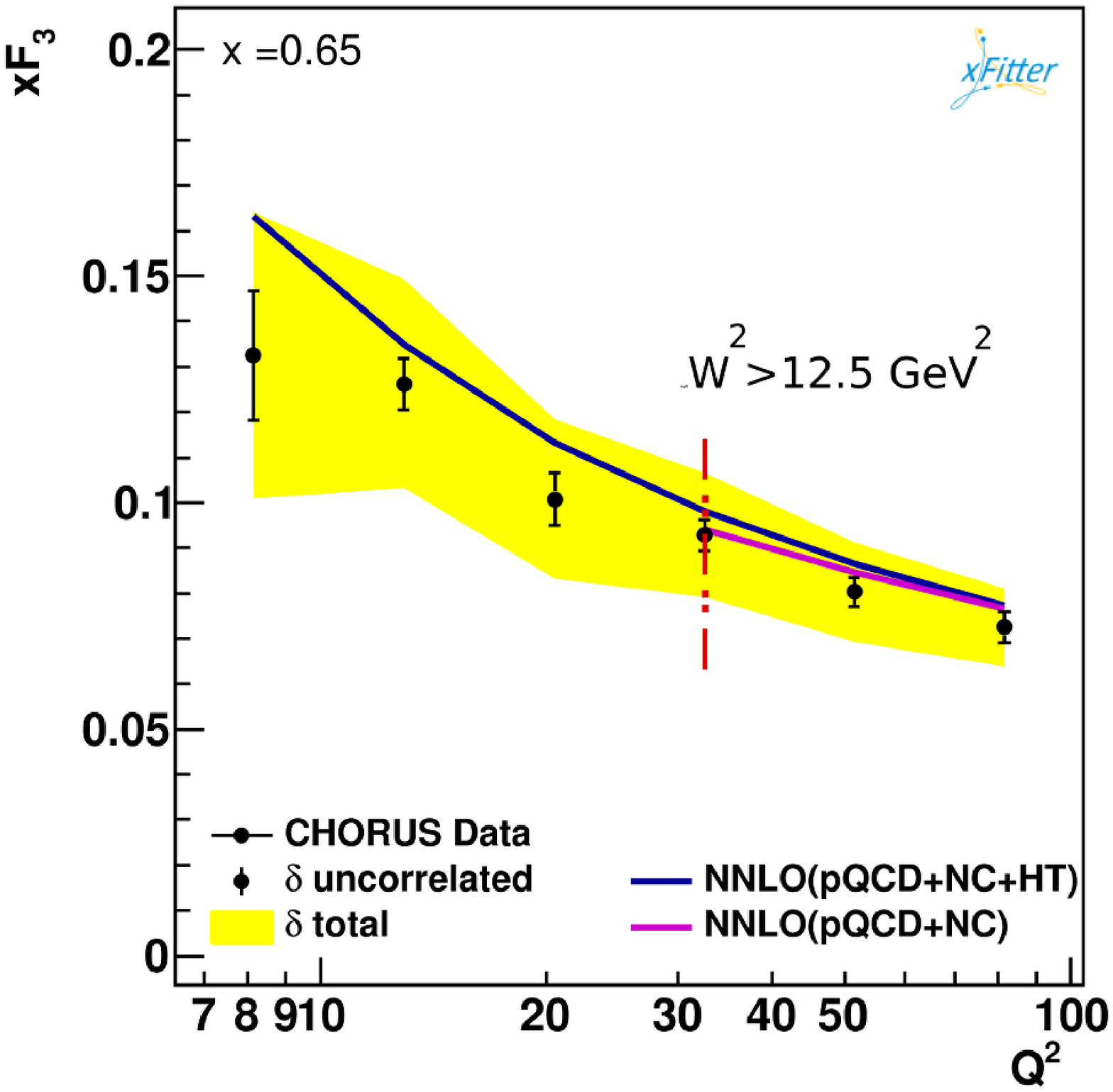}

\includegraphics[width=0.49\textwidth]{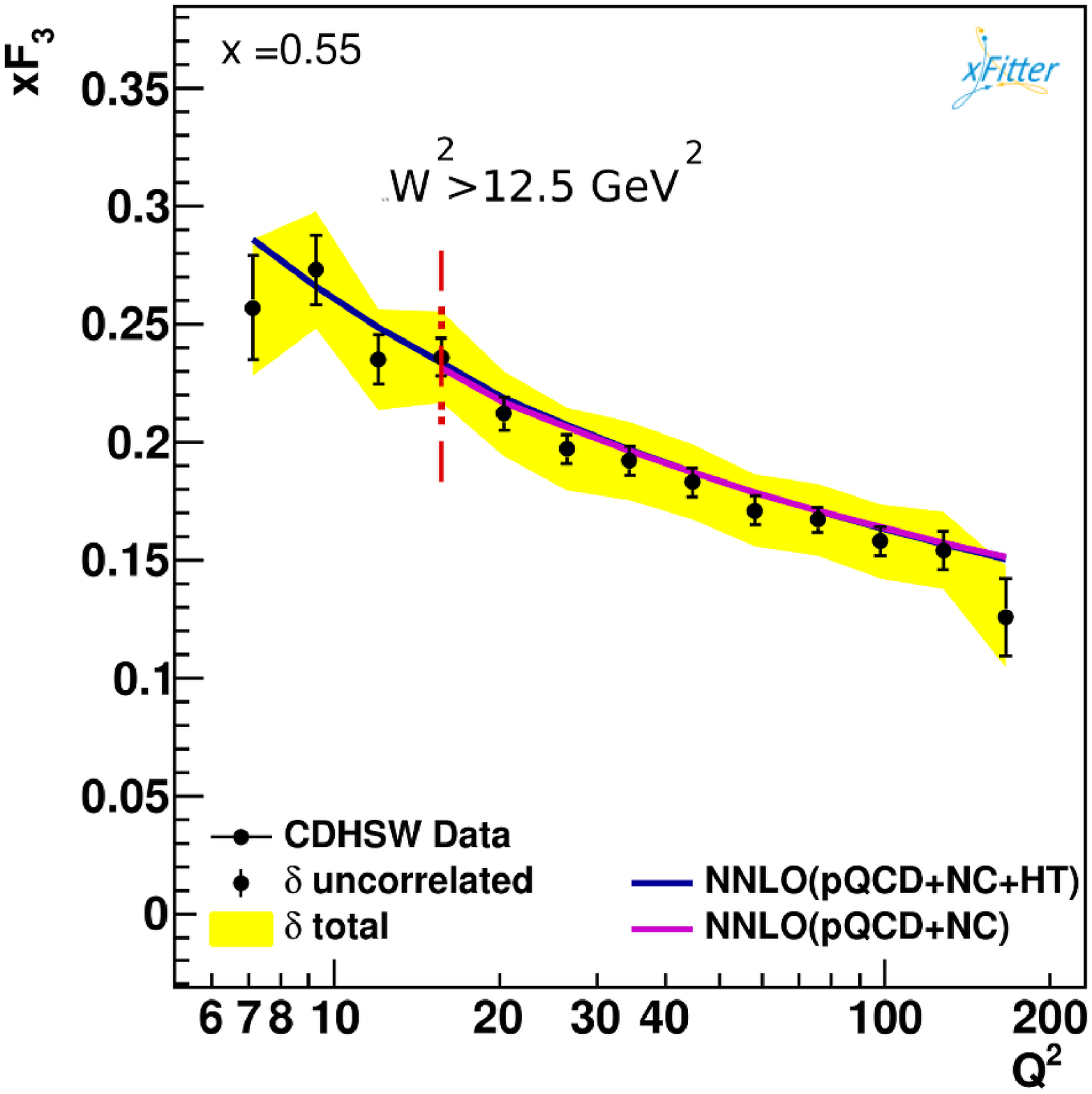}
\includegraphics[width=0.49\textwidth]{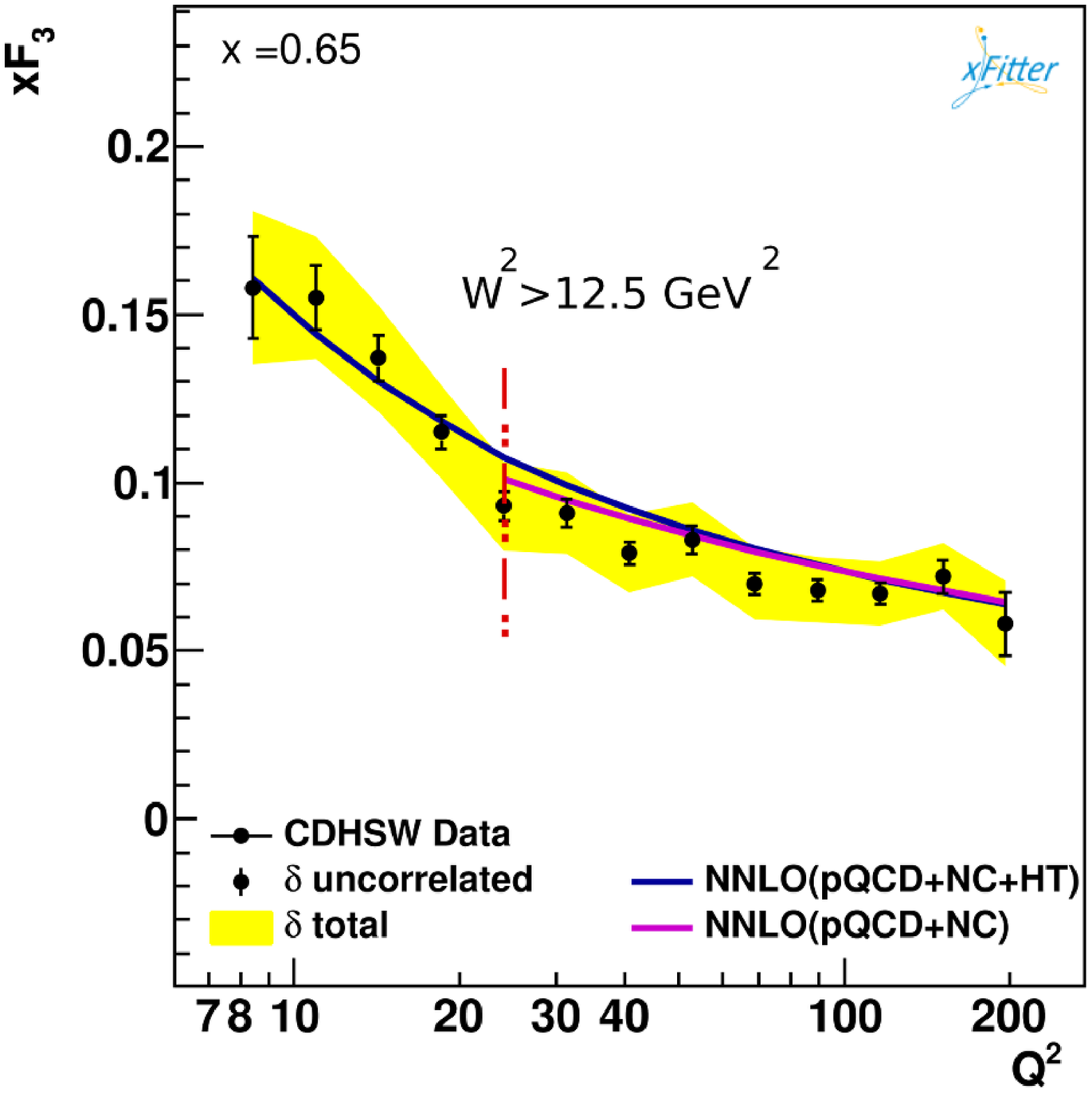}
  \vspace{-0.830cm}

\caption{The comparison of the structure function $xF_3$ obtained from the fit with and without higher twist corrections as a function of $Q^2$ in the various $x$, at NNLO approximation.
 }
\label{fig:QCD-fit-HT}
\end{figure*}

  \begin{figure}
\includegraphics[width=0.49\textwidth]{xuvandxdv-q1NC.eps}
\includegraphics[width=0.49\textwidth]{xuvandxdv-q1HT.eps}

\caption{The comparison of the proton valence ${xu_v}$ and ${xd_v}$ PDFs  as a function of $x$ at $Q^2$ = 1 GeV$^2$ taking into account nuclear corrections (left) and nuclear and higher twist corrections (right), at NLO and NNLO with their uncertainty bands.}
\label{fig:Q=1(Proton)}
\end{figure}

  \begin{figure}
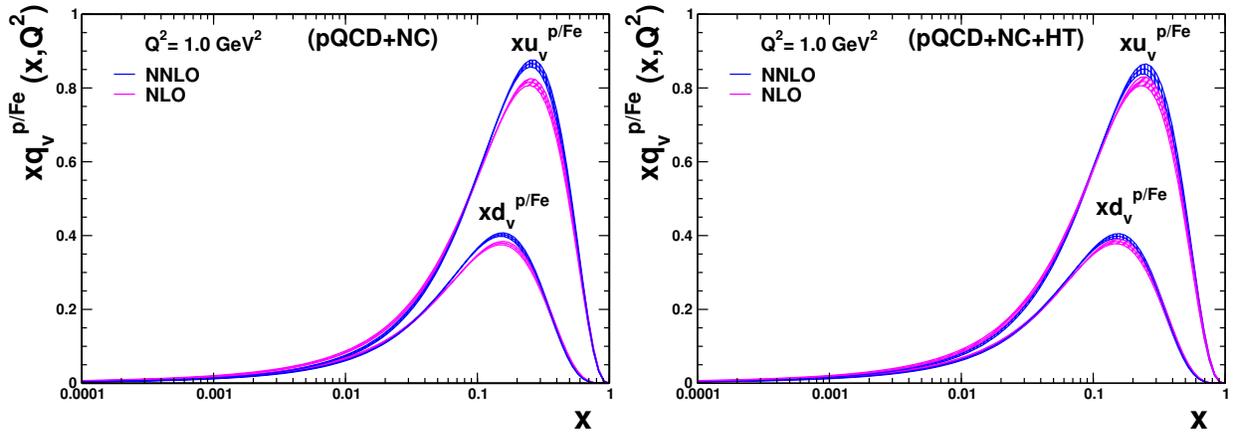

\includegraphics[width=0.49\textwidth]{xuvandxdv-q1NCFe.eps}
\includegraphics[width=0.49\textwidth]{xuvandxdv-q1HTFe.eps}

\caption{The comparison of the iron valence  ${xu_v^{p/Fe}}$ and ${xd_v^{p/Fe}}$  PDFs  as a function of $x$ at $Q^2$ = 1 GeV$^2$ taking into account nuclear corrections (left) and nuclear and higher twist corrections (right), at NLO and NNLO with their uncertainty bands.}
\label{fig:Q=1(Fe)}
\end{figure}

\begin{figure}[h!]
\begin{center}

\includegraphics[width=0.315\textwidth]{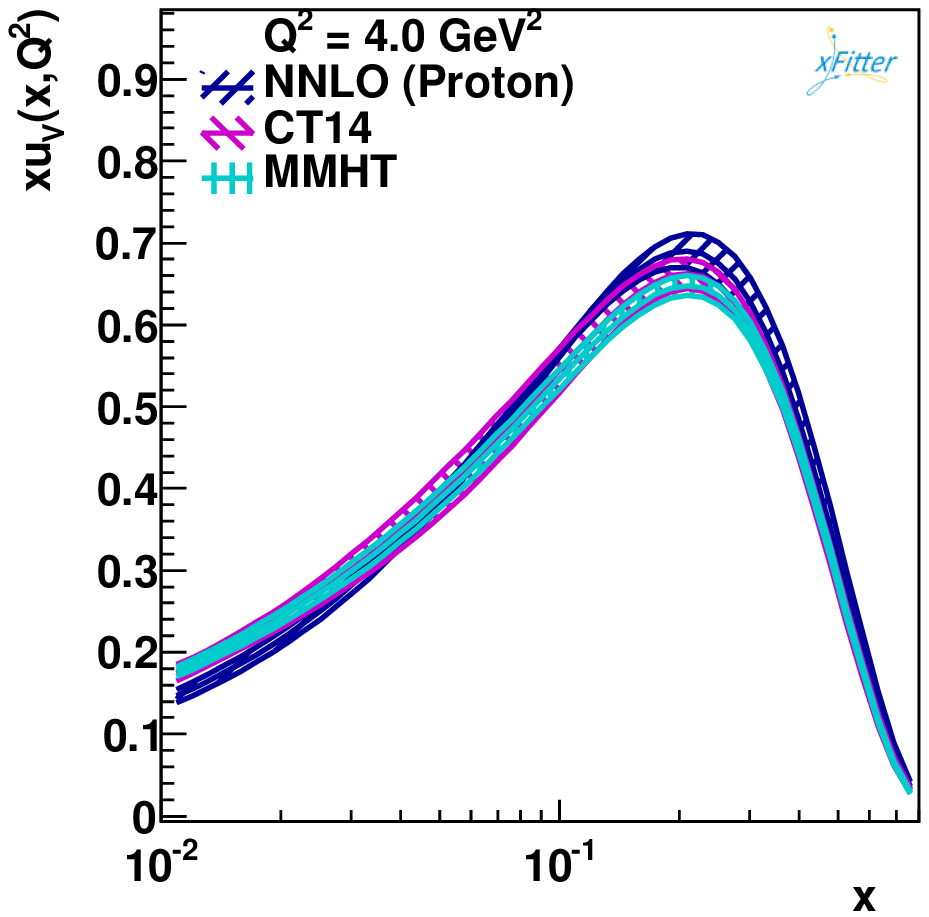}
\includegraphics[width=0.315\textwidth]{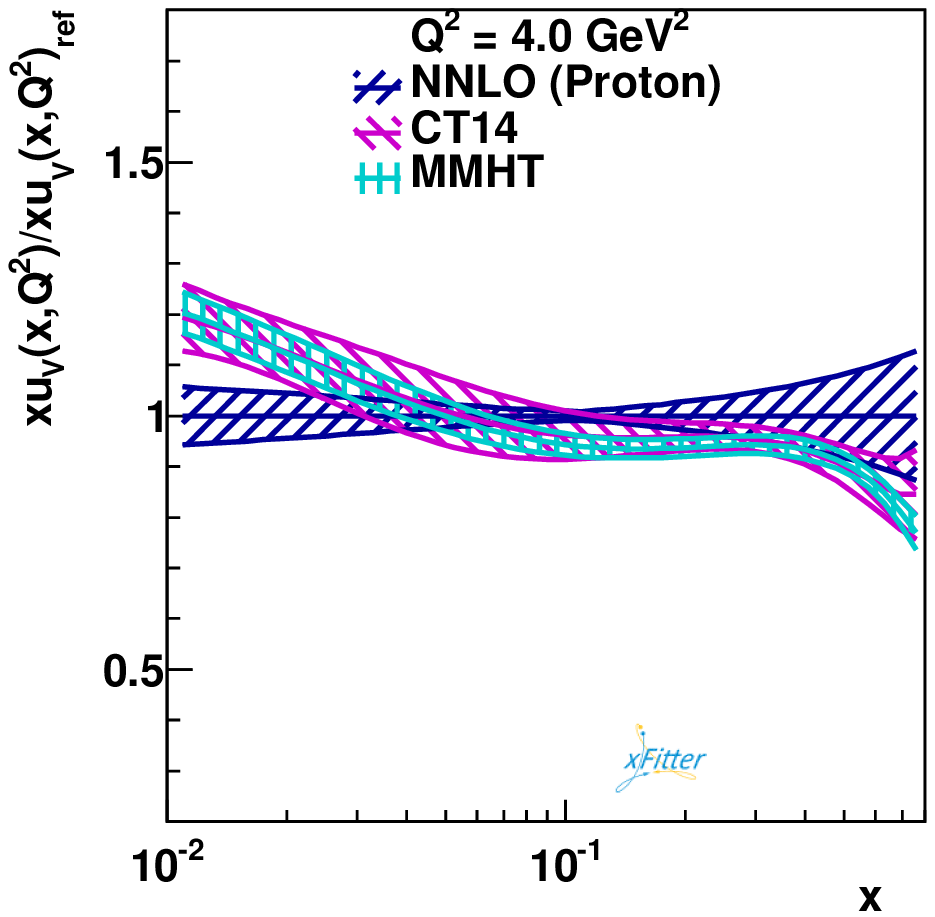}

\includegraphics[width=0.315\textwidth]{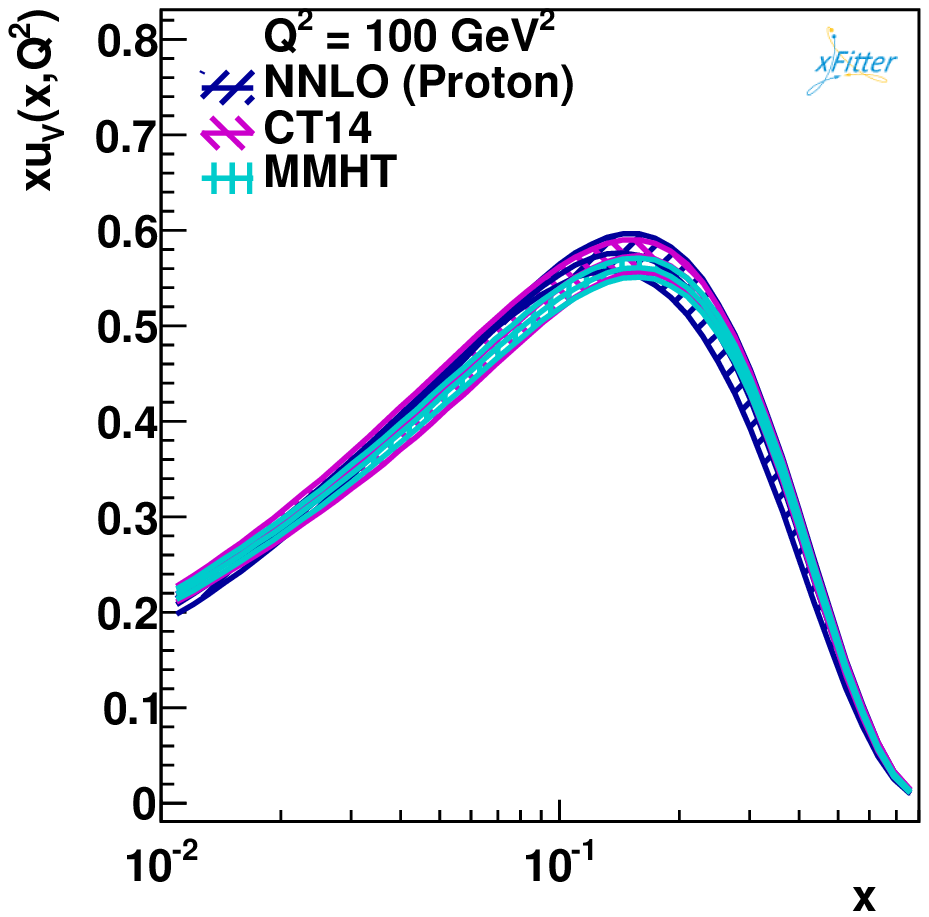}
\includegraphics[width=0.315\textwidth]{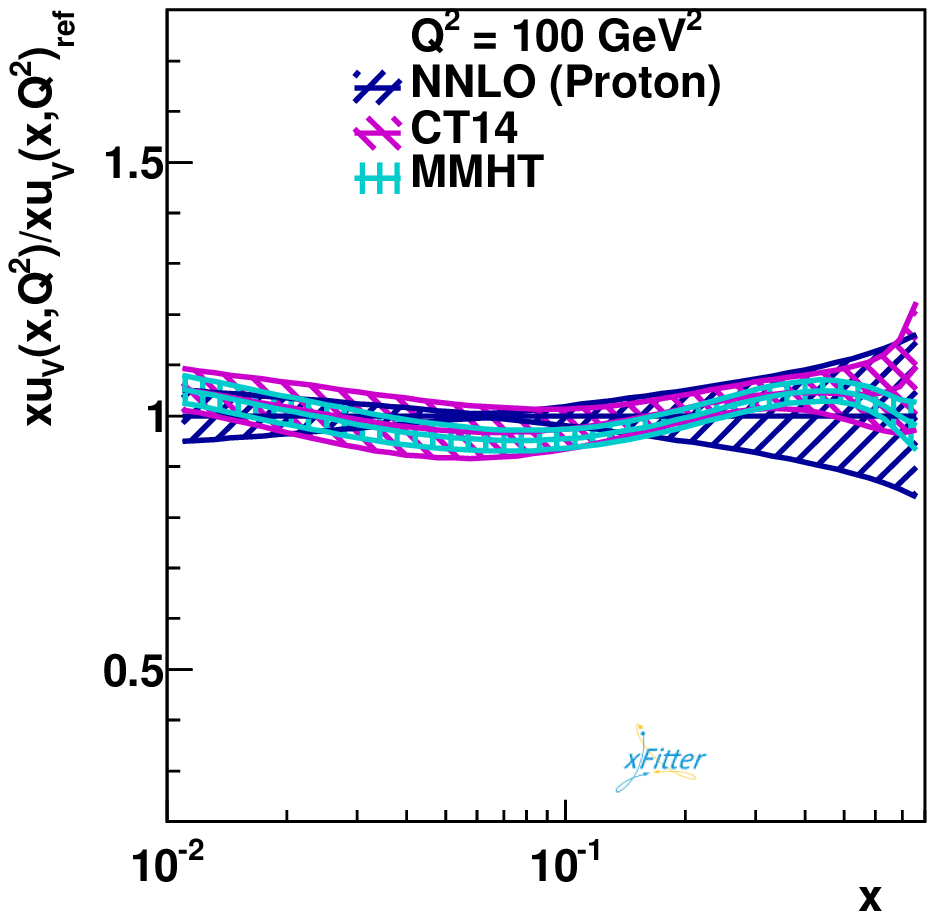}

\includegraphics[width=0.315\textwidth]{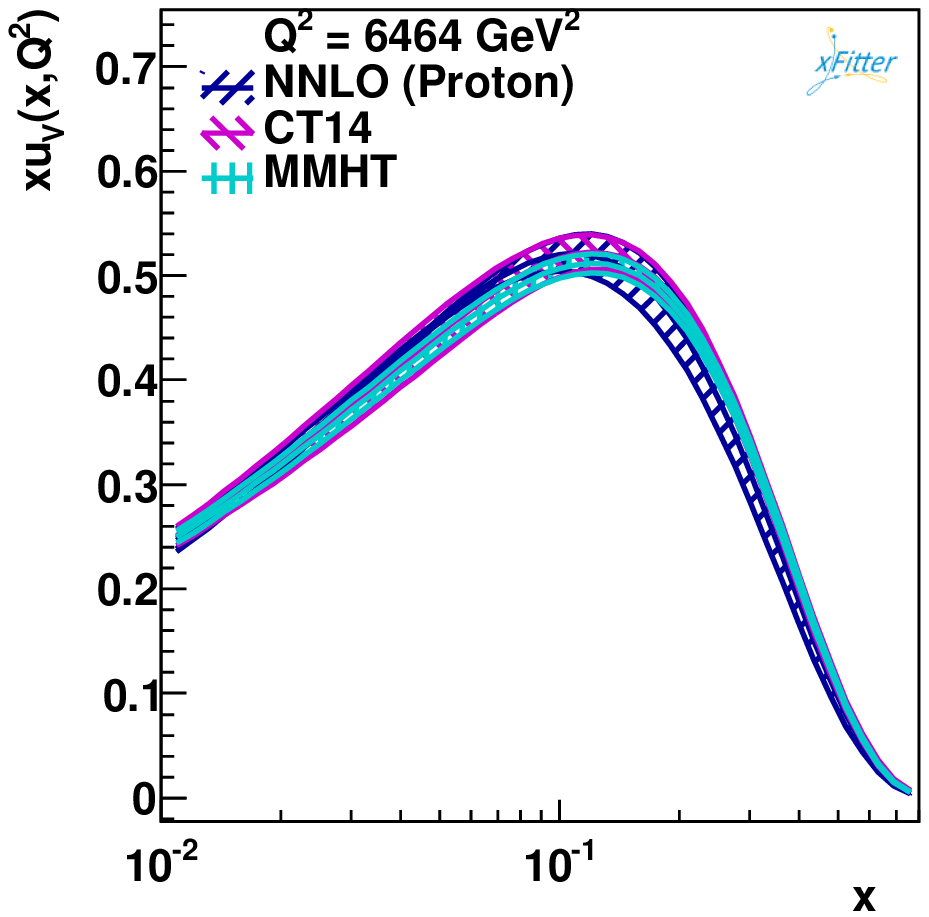}
\includegraphics[width=0.315\textwidth]{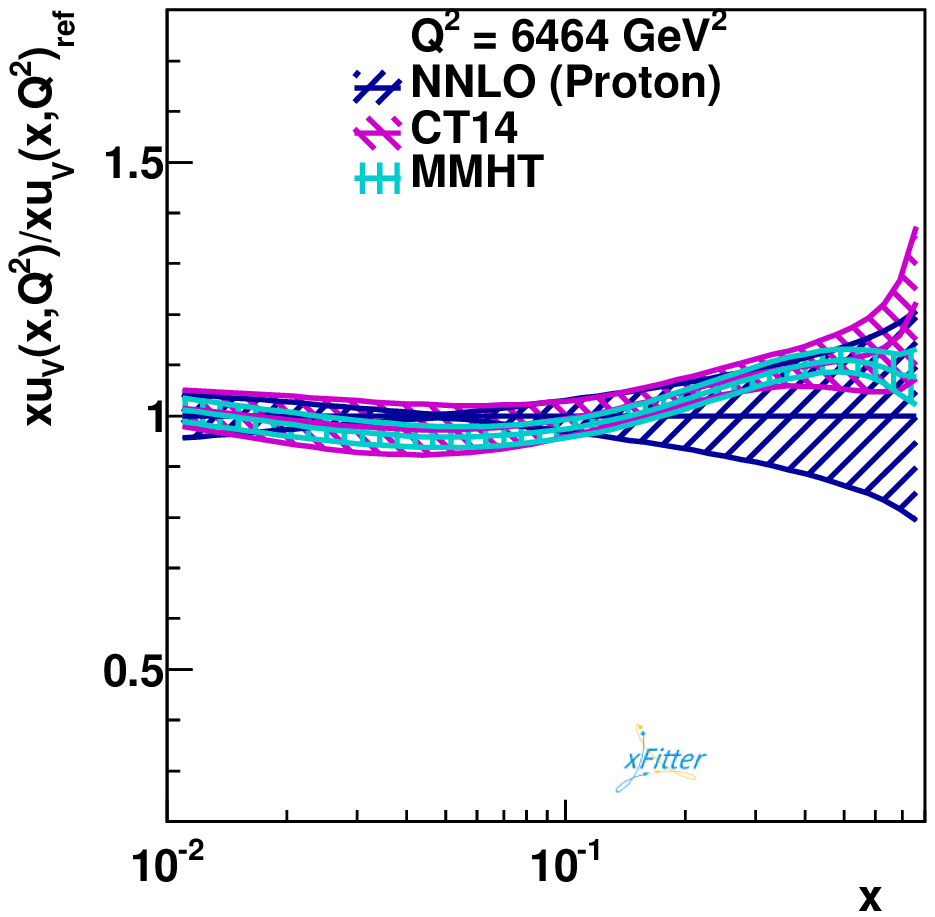}

\includegraphics[width=0.315\textwidth]{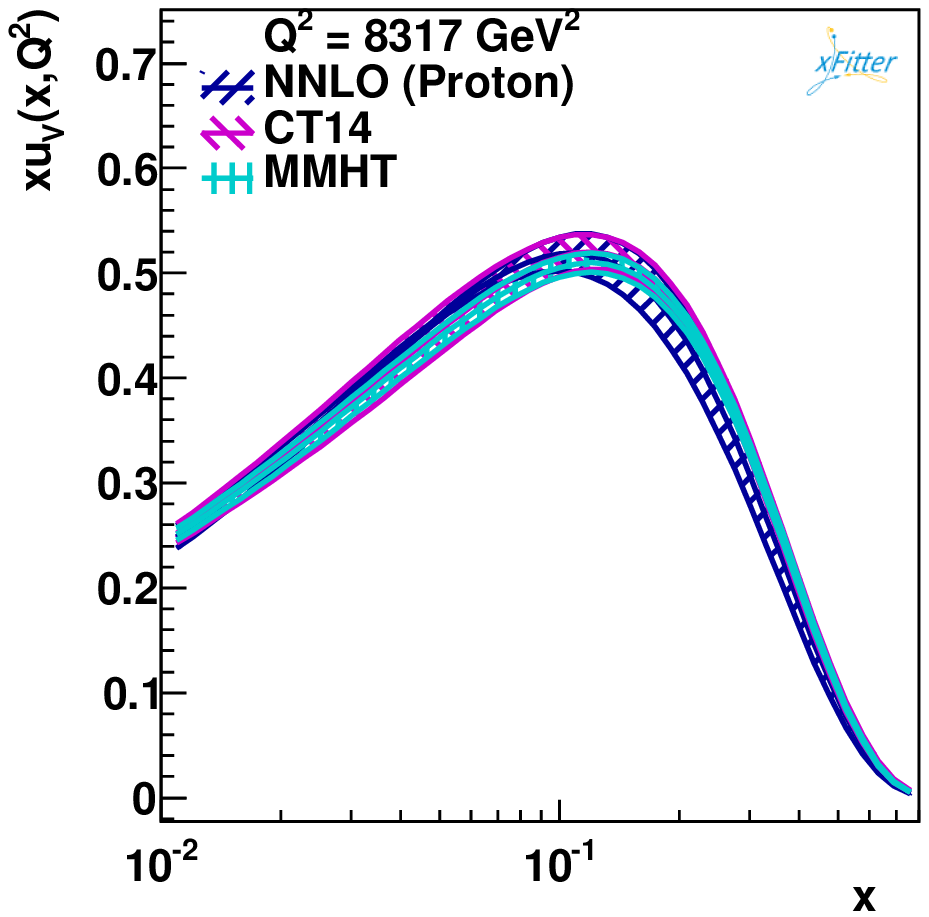}
\includegraphics[width=0.315\textwidth]{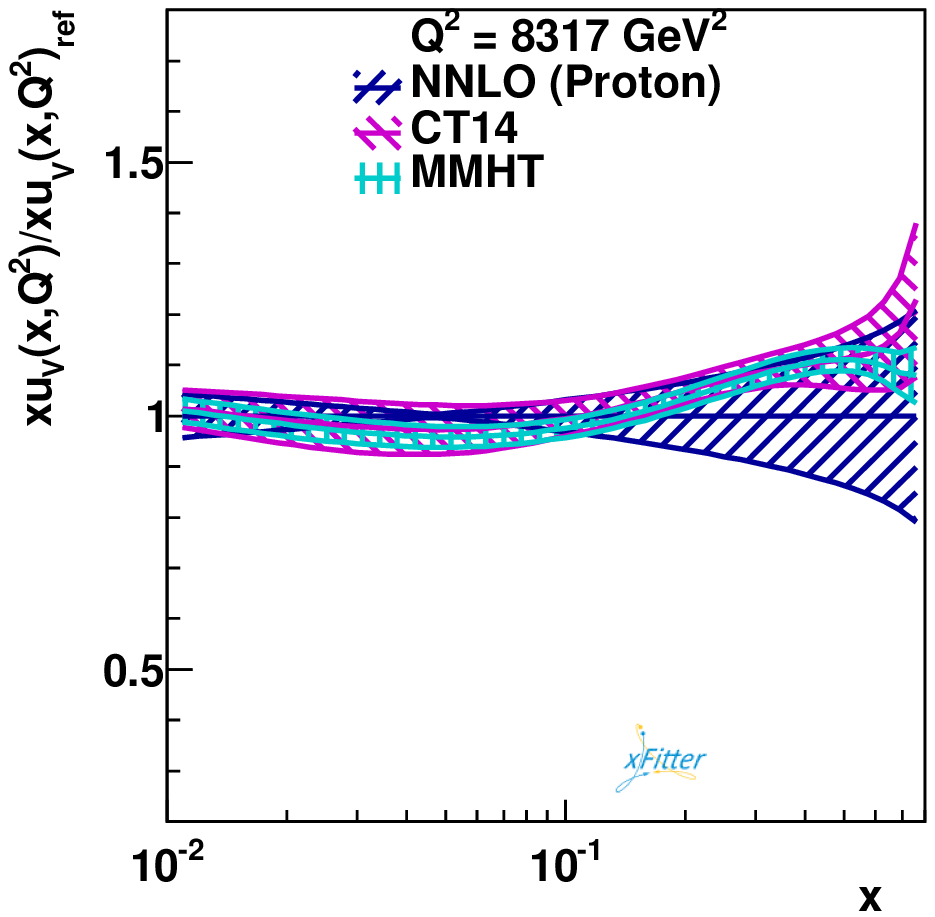}

  \end{center}
   \vspace{-1.15cm} 
\caption{$xu_v$ valence PDF results at different values of $Q^2$ = 4, 100, $M_W^2$, and $M_Z^2$ GeV$^2$ obtained with our QCD fits to the DIS neutrino-nucleon data, which have been compared with the results obtained by CT14 \cite{Dulat:2015mca} and MMHT \cite{Harland-Lang:2014zoa} as a function of $x$ at the NNLO (left panel) and the ratio of ${xu_v}/({xu_v})_{ref}$ (right panel) with respect to NNLO (proton).  We show our results only in the range of $x\in[10^{-2},0.8]$, where the data are existed and were applied in the present analysis.}
\label{fig:xuv-NNLO}
\end{figure}

  \begin{figure}[h!]
  \begin{center}

\includegraphics[width=0.315\textwidth]{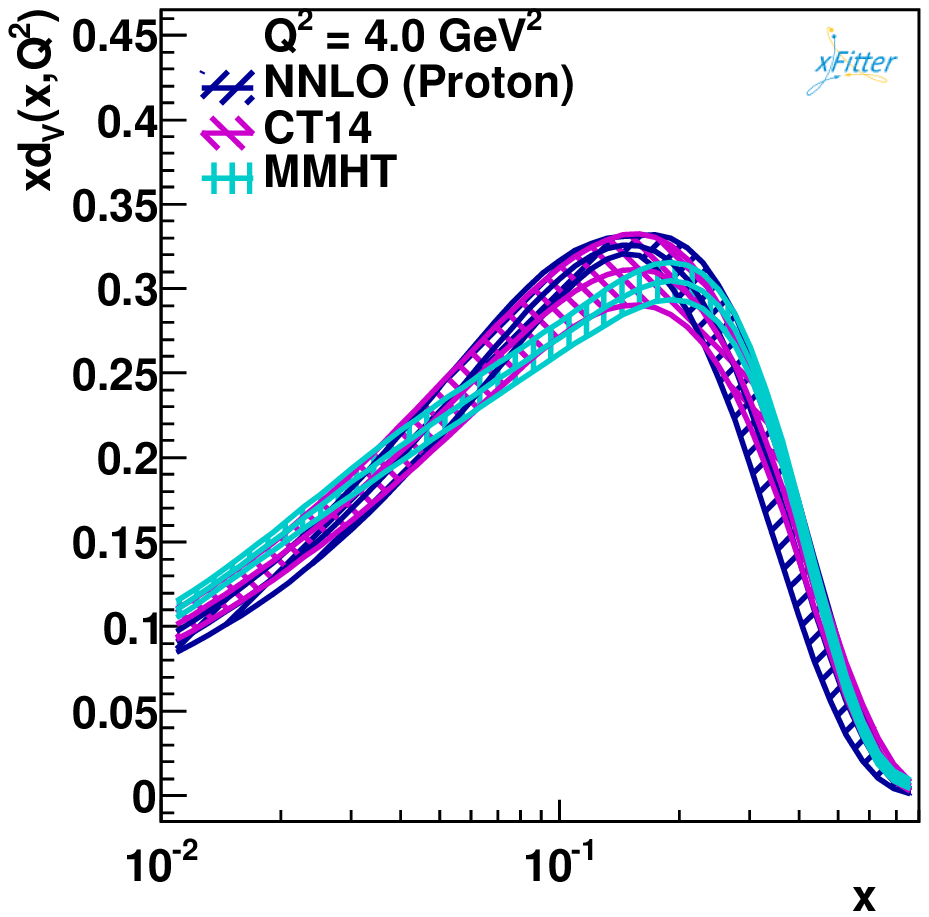}
\includegraphics[width=0.315\textwidth]{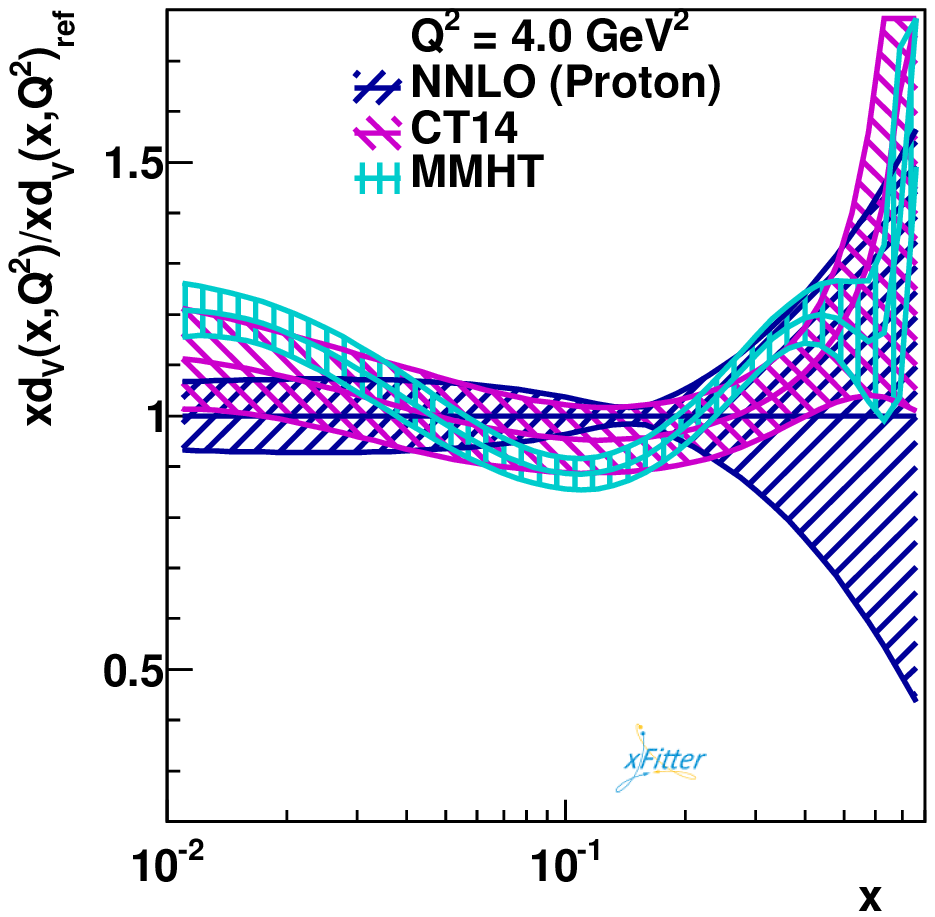}

\includegraphics[width=0.315\textwidth]{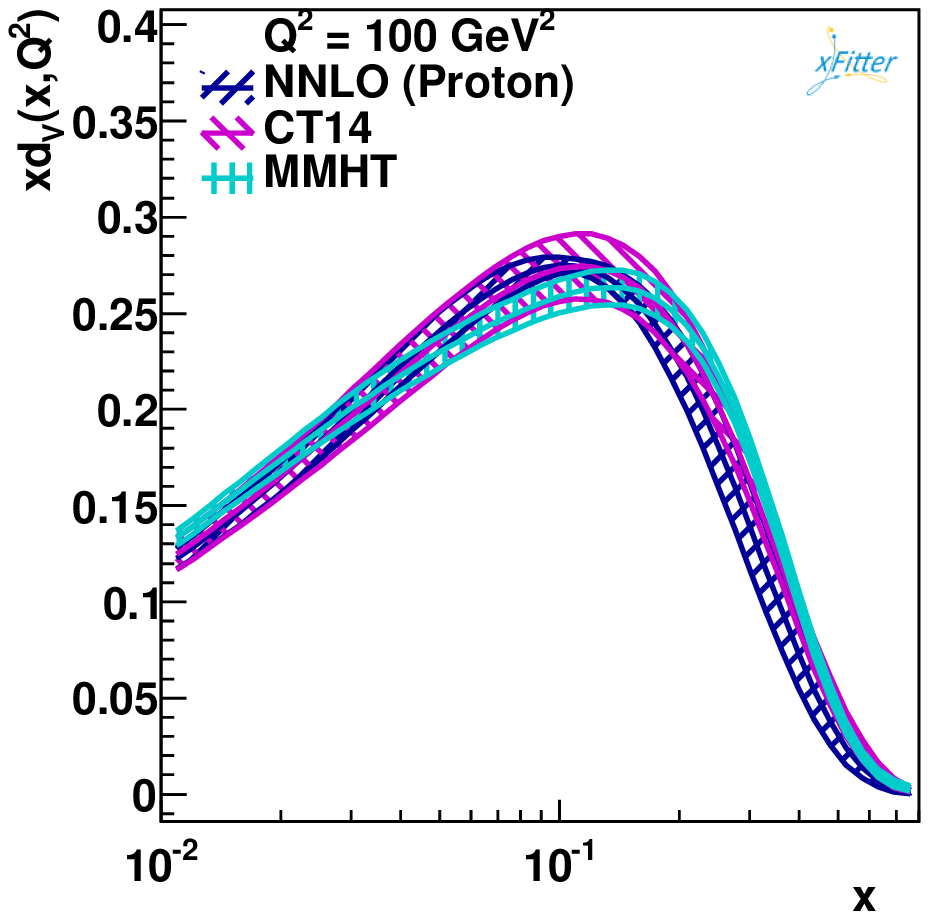}
\includegraphics[width=0.315\textwidth]{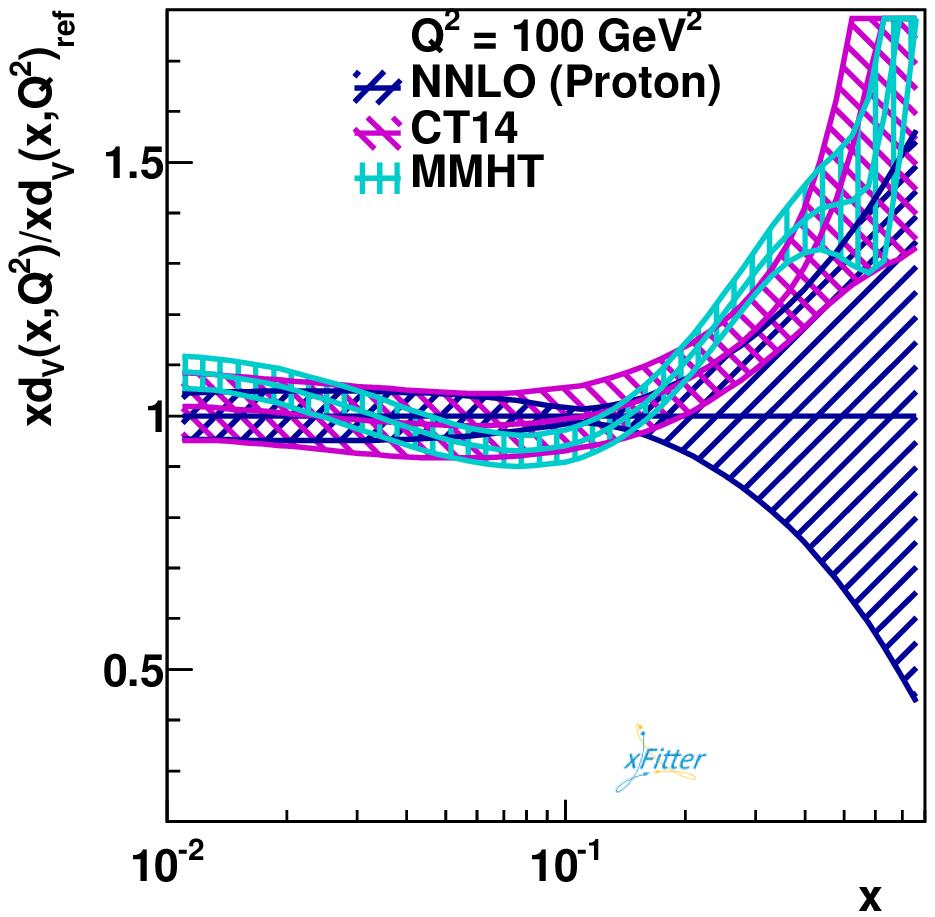}

\includegraphics[width=0.315\textwidth]{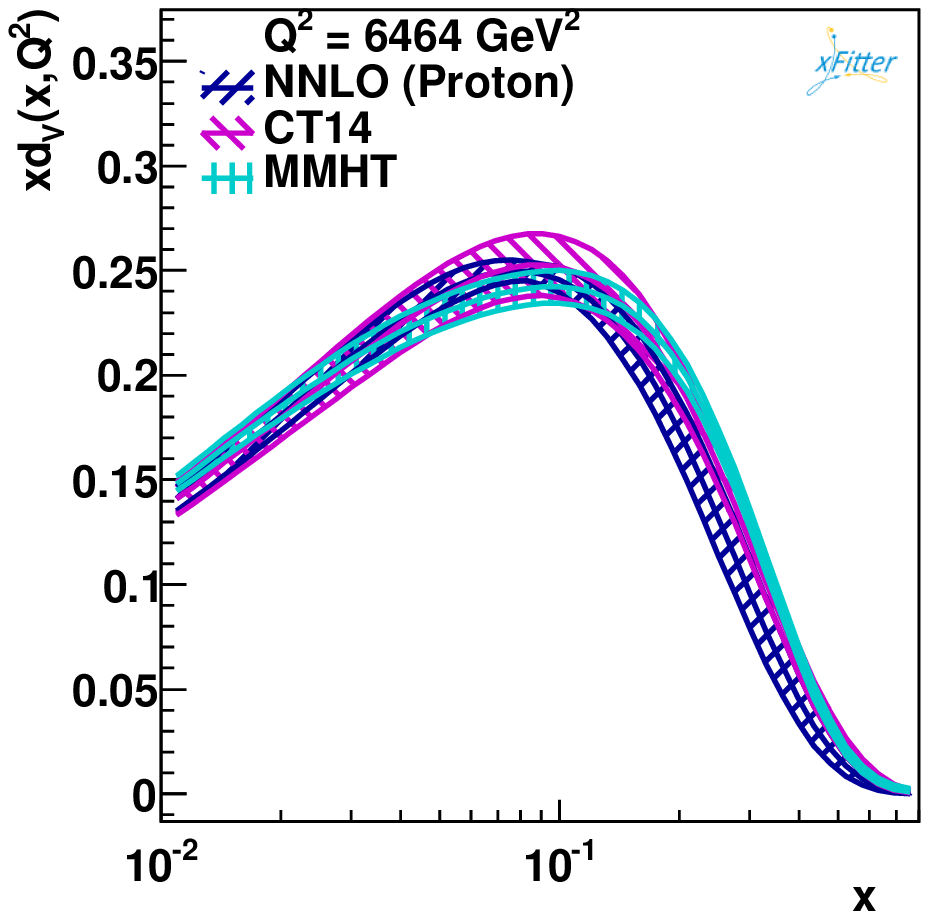}
\includegraphics[width=0.315\textwidth]{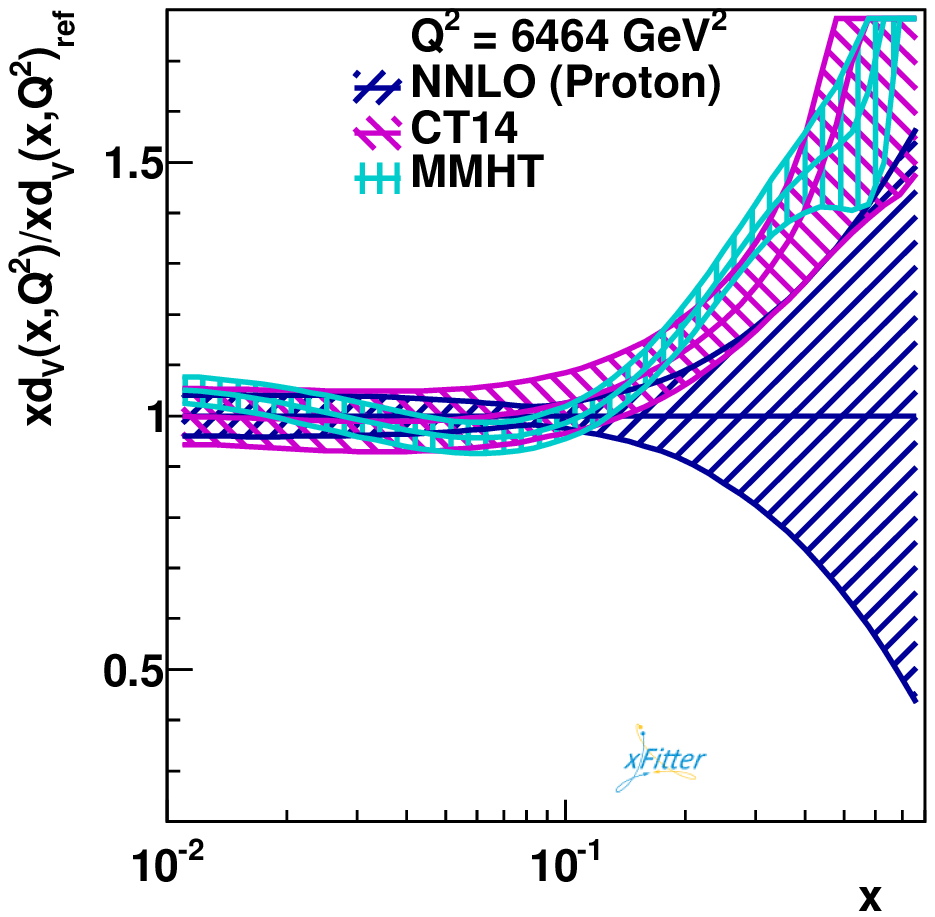}

\includegraphics[width=0.315\textwidth]{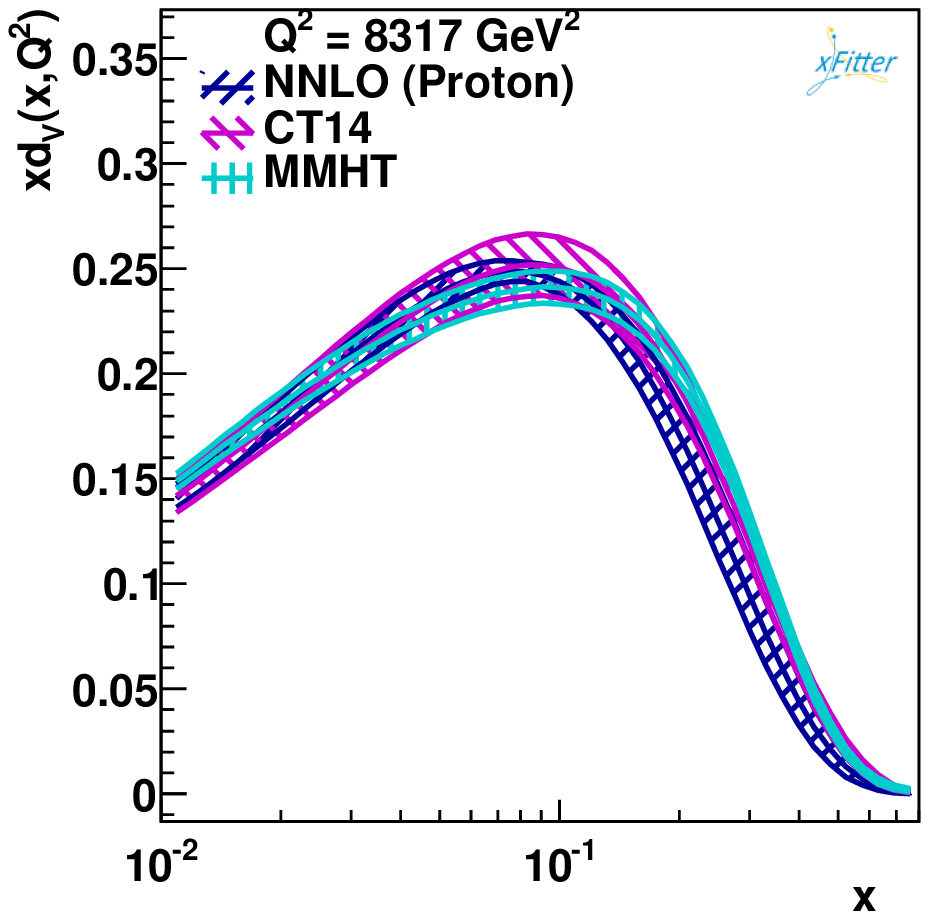}
\includegraphics[width=0.315\textwidth]{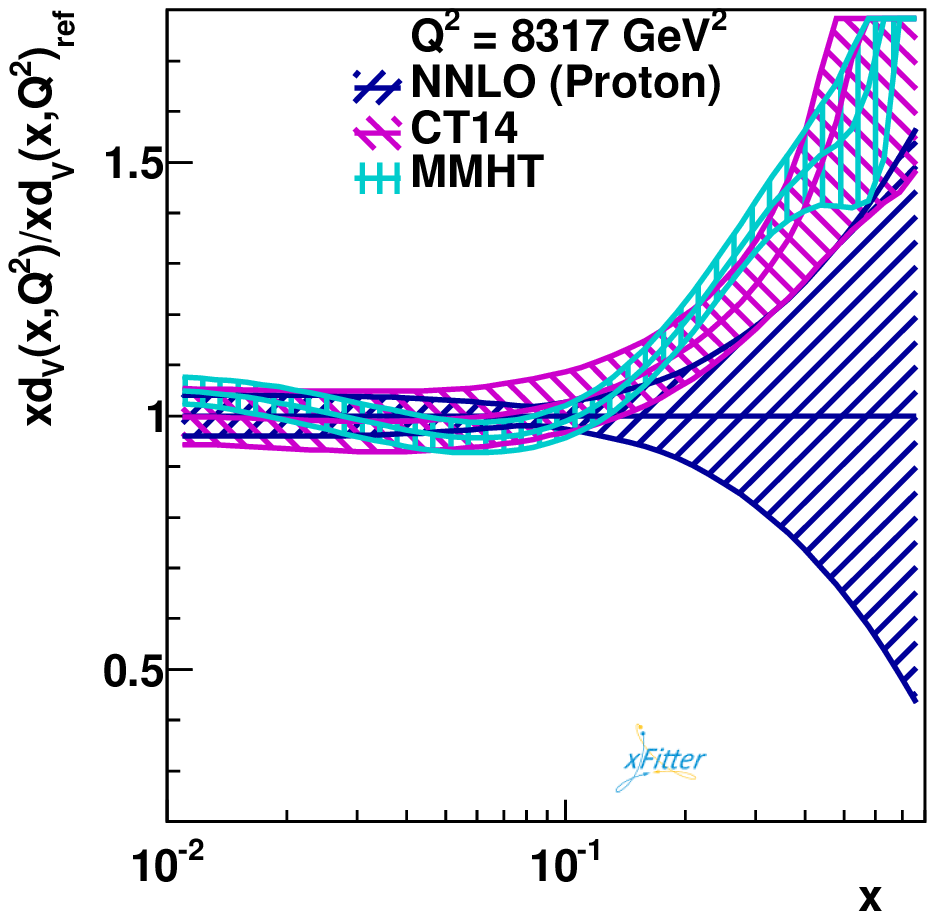}

  \end{center}
  \vspace{-1.15cm}
\caption{$xd_v$  valence PDF results at different values of $Q^2$ = 4, 100, $M_W^2$, and $M_Z^2$ GeV$^2$ obtained with our QCD fits to the DIS neutrino-nucleon data, which have been compared with the results obtained by CT14 \cite{Dulat:2015mca} and MMHT \cite{Harland-Lang:2014zoa} as a function of $x$ at the NNLO (left panel) and the ratio of ${xd_v}/({xd_v})_{ref}$ (right panel) with respect to NNLO (proton). We show our results only in the range of $x\in[10^{-2},0.8]$, where the data are existed and were applied in the present analysis.}
\label{fig:xdv-NNLO}
\end{figure}

\begin{figure}[h!]
\begin{center}

\includegraphics[width=0.35\textwidth]{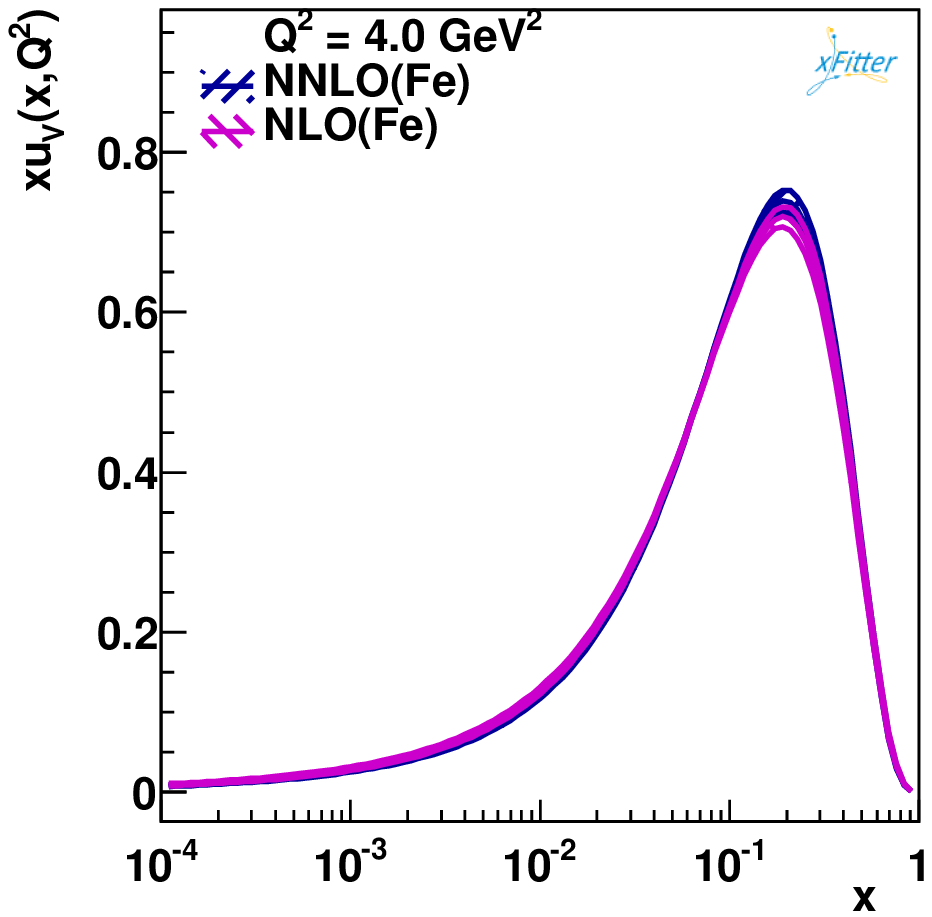}
\includegraphics[width=0.35\textwidth]{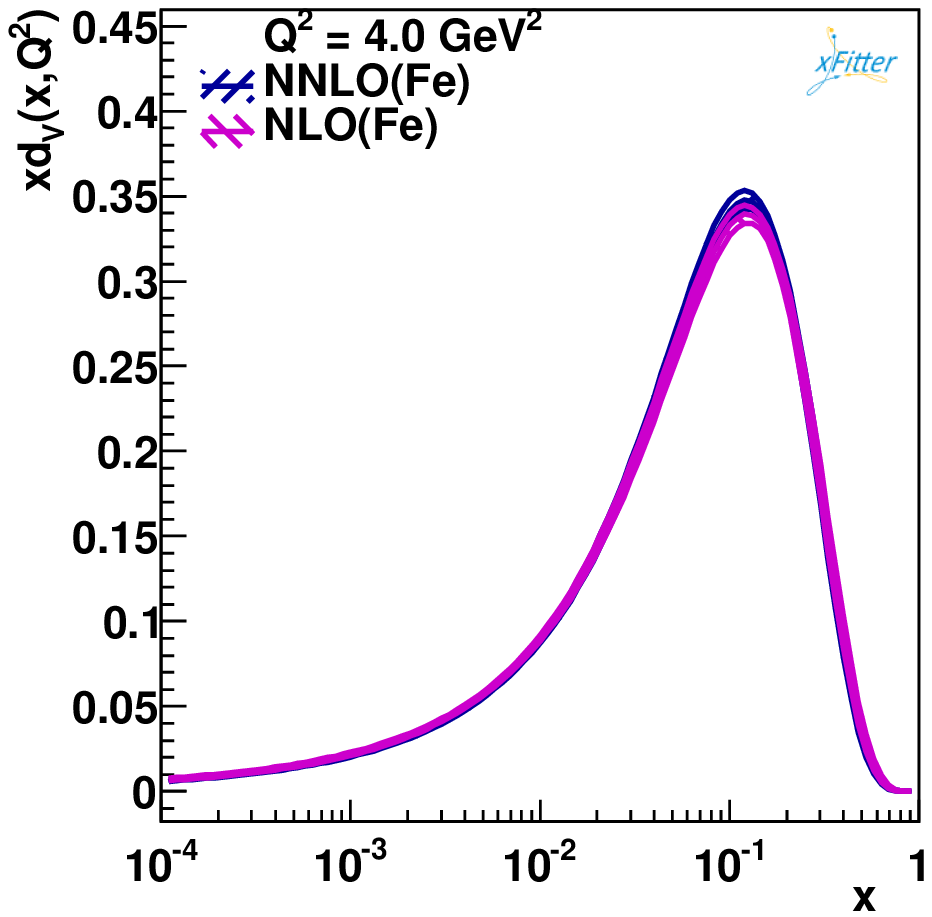}

\includegraphics[width=0.35\textwidth]{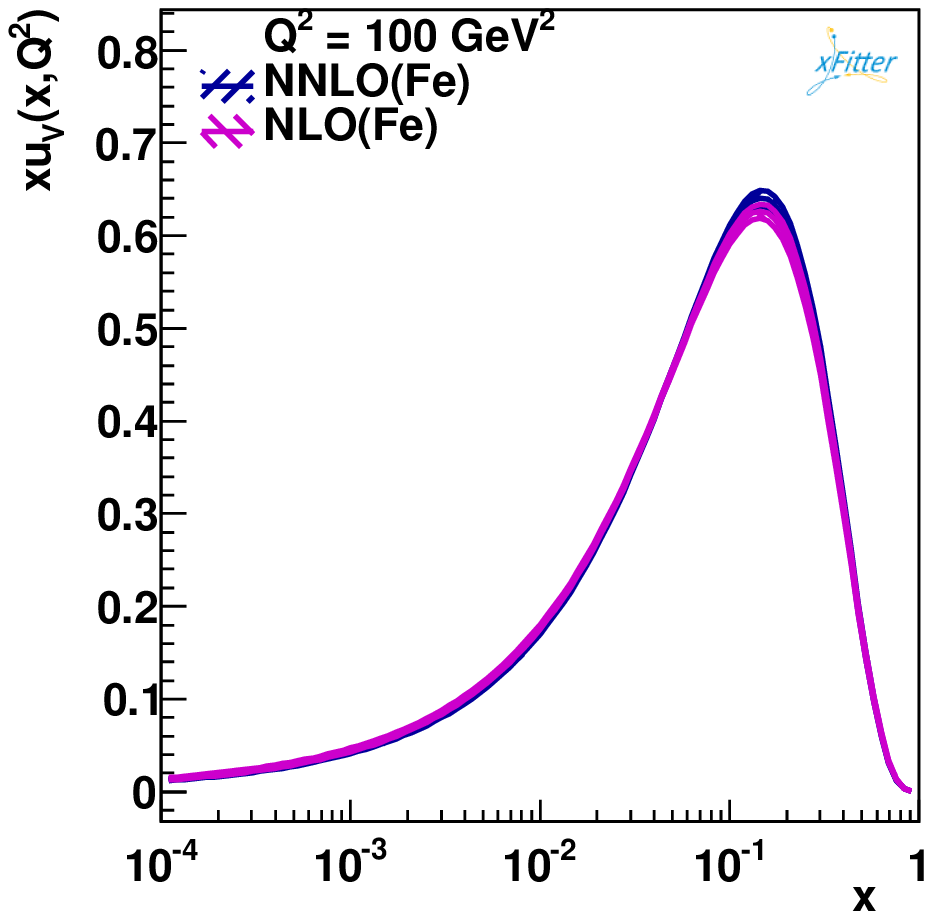}
\includegraphics[width=0.35\textwidth]{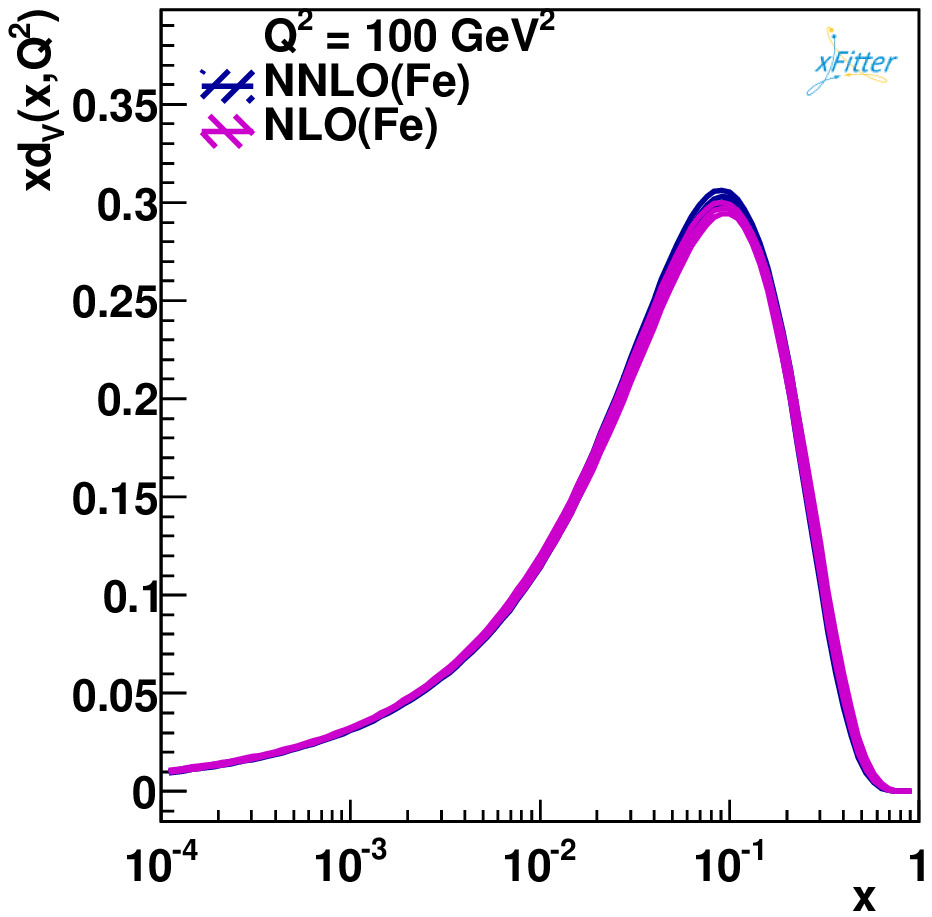}

\includegraphics[width=0.35\textwidth]{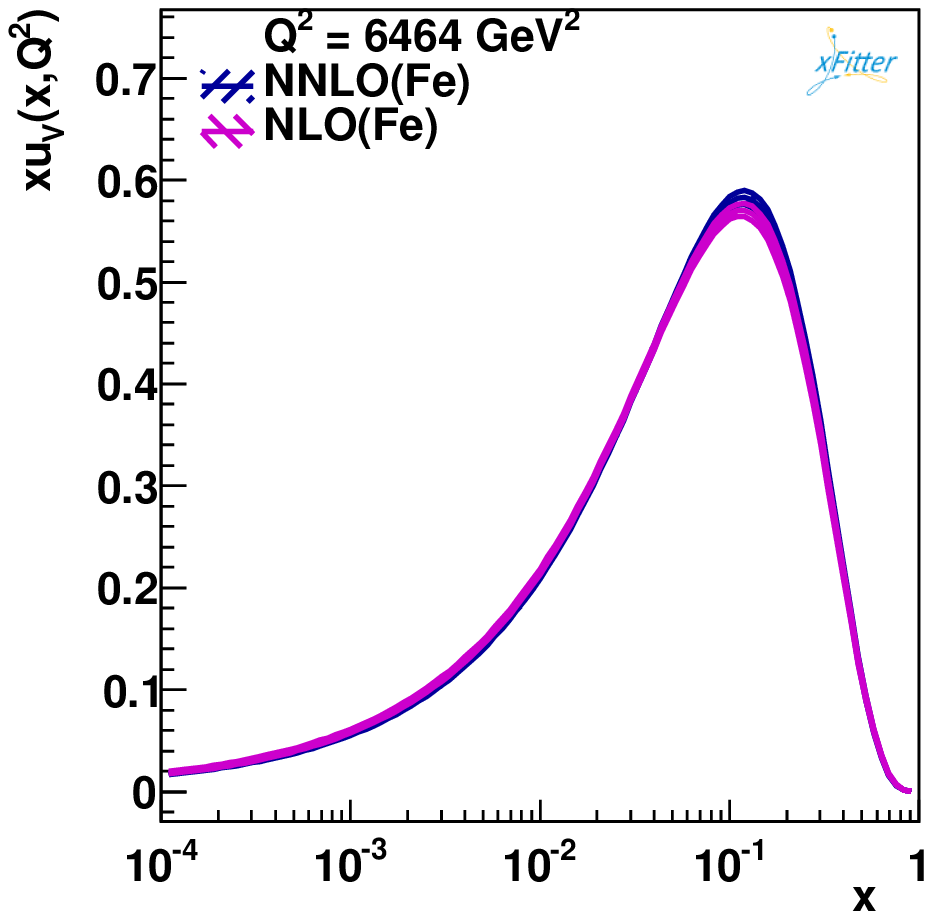}
\includegraphics[width=0.35\textwidth]{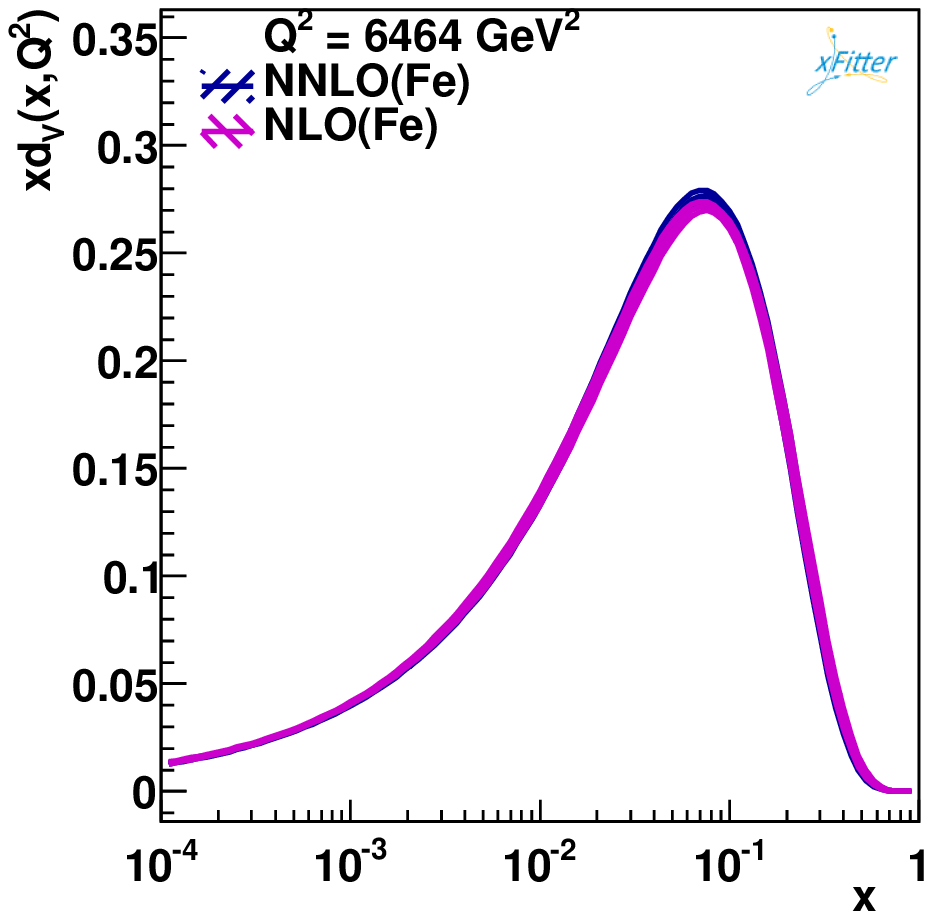}

\includegraphics[width=0.35\textwidth]{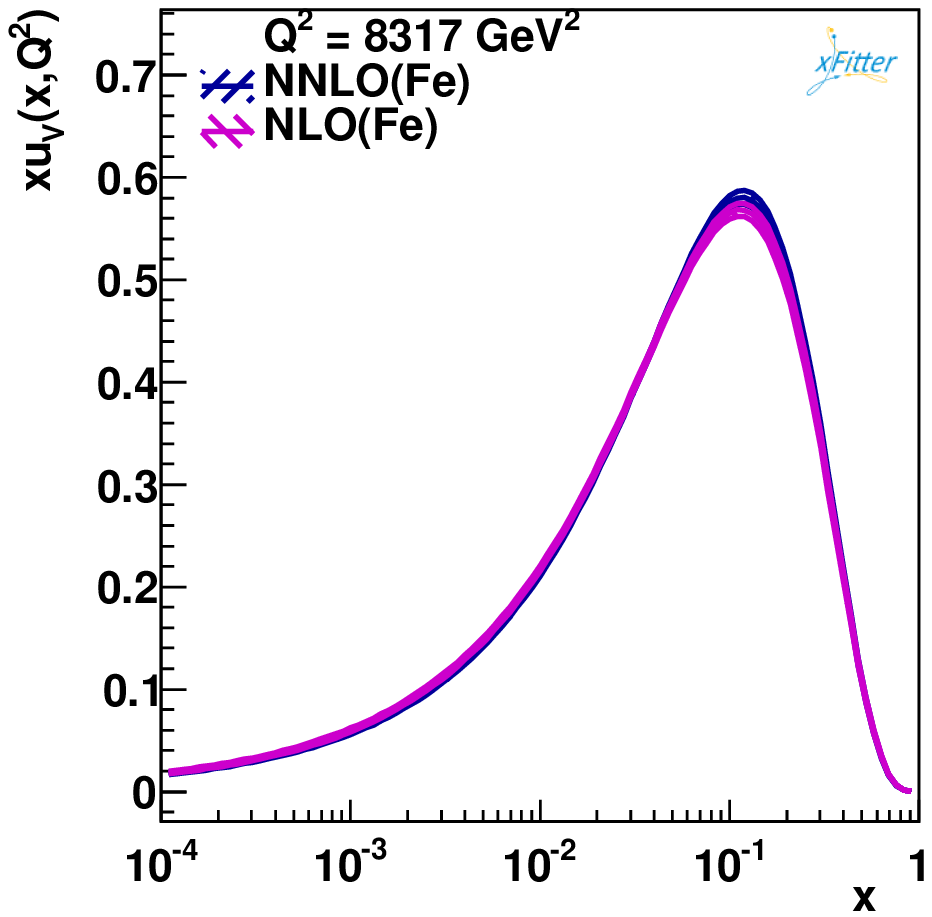}
\includegraphics[width=0.35\textwidth]{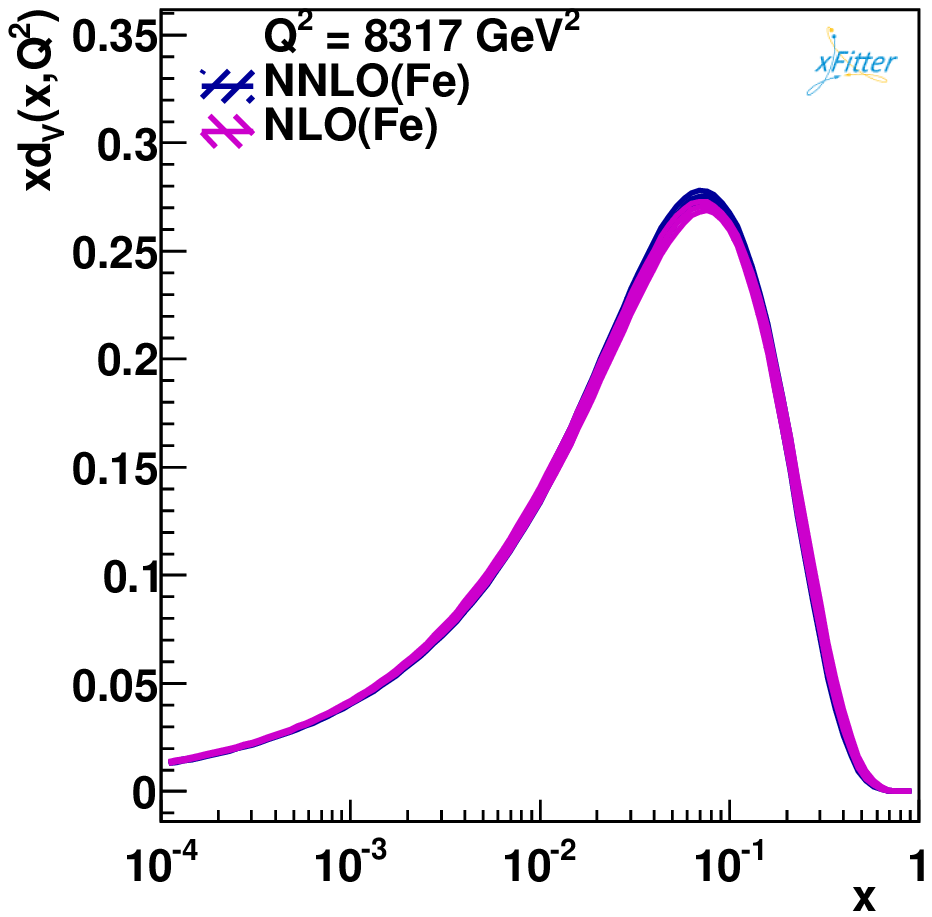}

  \end{center}
  \vspace{-1.0cm}
\caption{ The ${xu_v}^{p/Fe}$  and ${xd_v}^{p/Fe}$ parton density distribution at the NLO and NNLO with their uncertainty bands as a function of $x$ at different values of $Q^2$ = 4, 100, $M_W^2$, and $M_Z^2$ GeV$^2$.}
\label{fig:uv}
\end{figure}

\begin{figure*}[h!]
\begin{center}
\includegraphics[width=1\textwidth]{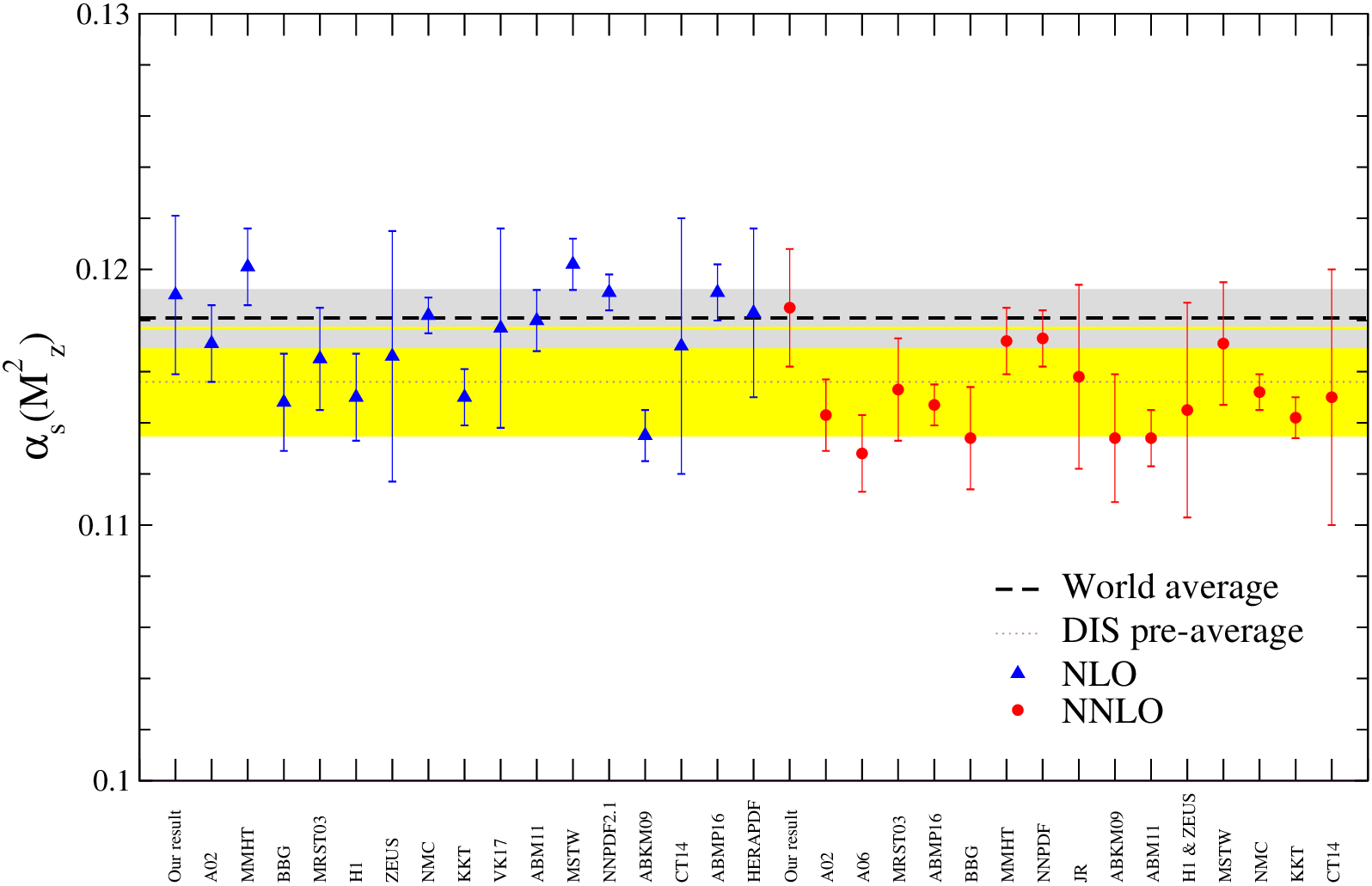}
\end{center}
\caption{ The value of $\alpha_{s}(M_{Z}^{2})$ in comparison with different QCD analyses at NLO  A02 \cite{Alekhin:2002fv}, MMHT \cite{Harland-Lang:2014zoa}, BBG \cite{Blumlein:2006be}, MRST03 \cite{Martin:2003tt}, H1 \cite{Adloff:2000qk}, ZEUS \cite{Chekanov:2001qu}, NMC \cite{Arneodo:1996qe}, KKT \cite{Khorramian:2009xz}, VK17 \cite{Vafaee:2017nze}, ABM11 \cite{Alekhin:2012ig}, MSTW \cite{Martin:2009iq}, NNPDF2.1 \cite{Lionetti:2011pw}, ABKM09 \cite{Alekhin:2009ni}, CT14 \cite{Dulat:2015mca}, ABMP16 \cite{Alekhin:2018pai}, HERAPDF \cite{Abramowicz:2015mha}, and  NNLO  A02 \cite{Alekhin:2002fv}, A06 \cite{Alekhin:2006zm}, MRST03 \cite{Martin:2003tt}, ABMP16 \cite{Alekhin:2018pai}, BBG \cite{Blumlein:2006be}, MMMHT \cite{Harland-Lang:2014zoa}, NNPDF \cite{Ball:2014uwa}, JR \cite{JimenezDelgado:2008hf}, ABKM09 \cite{Alekhin:2009ni}, ABM11 \cite{Alekhin:2012ig}, H1 and ZEUS \cite{Aaron:2009aa}, MSTW \cite{Martin:2009iq}, NMC \cite{Arneodo:1996qe}, KKT \cite{Khorramian:2009xz}, CT14 \cite{Dulat:2015mca} approximations. The dotted line with yellow band indicates the preaverage results of the strong coupling constant $\alpha_{s}(M_{Z}^{2})$  in the DIS subfield \cite{Tanabashi:2018oca}. Also, the grey band and dashed line present the world average value of the strong coupling constant $\alpha_{s}(M_{Z}^{2})$.}{\small   \label{fig:alphamz}}
\label{fig:Alphas}
\end{figure*}

\end{document}